\documentclass[a4paper,english,11pt]{article}
\pdfoutput=1
\usepackage{jheppub}

\usepackage{epsfig,multicol}
\usepackage{amssymb,amsmath,bm, mathrsfs}
\usepackage{epsf}
\usepackage{graphicx}
\usepackage{longtable}
\usepackage{axodraw}
\usepackage{afterpage}
\usepackage{placeins}
\usepackage{color}

\newcommand{\lsim}{
\mathrel{\hbox{\rlap{\hbox{\lower4pt\hbox{$\sim$}}}\hbox{$<$}}}}

\newcommand{\gsim}{
\mathrel{\hbox{\rlap{\hbox{\lower4pt\hbox{$\sim$}}}\hbox{$>$}}}}

\newcommand{\vcb}{|V_{cb}|}

\newcommand{\vub}{|V_{ub}|}

\def\nnb{\nonumber}

\def\eps{\varepsilon}

\newcommand{\tev}{\, {\rm TeV}}
\newcommand{\gev}{\, {\rm GeV}}
\newcommand{\mev}{\, {\rm MeV}}

\newcommand{\Heff}{{\cal H}_\text{ eff}}

\allowdisplaybreaks[2]

\newcommand{\be}{\begin{equation}}
\newcommand{\ee}{\end{equation}}
\newcommand{\bea}{\begin{eqnarray}}
\newcommand{\eea}{\end{eqnarray}}

\newcommand{\bi}{\begin{itemize}}
\newcommand{\ei}{\end{itemize}}
\newcommand{\ord}{{\cal O}}

\renewcommand{\Re}{{\rm Re}}
\renewcommand{\Im}{{\rm Im}}

\newcommand{\Br}{{\rm Br}}

\newcommand{\ts}{\tilde s}
\newcommand{\tc}{\tilde c}

\leftline{TUM-HEP-820/11}
\leftline{FLAVOUR(267104)-ERC-5}
\bigskip

\title{\boldmath $\Delta F=2$ observables and 
$B\to X_q\gamma$ decays in the Left-Right Model: Higgs particles 
striking back}

\author[a]{Monika Blanke}
\author[b,c]{Andrzej J.~Buras}
\author[b]{Katrin Gemmler}
\author[b]{and Tillmann Heidsieck}

\affiliation[a]{Laboratory for Elementary Particle Physics, Cornell University,\\ 142 Sciences Drive, Ithaca,~NY~14853,~USA}
\affiliation[b]{Physik Department, Technische Universit\"at M\"unchen,
James-Franck-Stra{\ss}e, \\D-85747 Garching, Germany}
\affiliation[c]{TUM Institute for Advanced Study, Lichtenbergstr. 2a, D-85747 Garching, Germany}

\emailAdd{mb744@cornell.edu}
\emailAdd{Andrzej.Buras@ph.tum.de}
\emailAdd{Katrin.Gemmler@ph.tum.de}
\emailAdd{Tillmann.Heidsieck@ph.tum.de}

\abstract{{We present a complete study of $\Delta S=2$ 
and $\Delta B=2$ processes in the left-right model (LRM) based 
on the weak gauge group $SU(2)_L\times SU(2)_R\times U(1)_{B-L}$. 
This includes $\varepsilon_K$, $\Delta M_K$,  $\Delta M_q$, $A_{\rm SL}^q$, $\Delta\Gamma_q$ with $q=d,s$
and the mixing induced CP asymmetries $S_{\psi K_S}$ and  $S_{\psi \phi}$. Compared to the Standard Model (SM) these observables 
are affected by tree level
contributions from heavy neutral Higgs particles ($H^0$) as well as new box diagrams with $W_R$ gauge boson and charged Higgs 
$(H^\pm)$  exchanges. We also analyse the 
$B\to X_{s,d}\gamma$ decays that receive important new contributions from the 
$W_L-W_R$ mixing and 
$H^\pm$  exchanges. Compared to the existing literature the novel feature of our analysis 
is the  search for correlations between various observables that could  
help us to distinguish this model from other extensions of the 
SM and to obtain an insight into the structure of the mixing 
matrix $V^{\rm R}$ that governs right-handed currents. Moreover, we perform the 
full phenomenology including both gauge boson and Higgs boson contributions. We find that even for $M_{H^0}\approx M_{H^\pm}\sim\ord(20)\tev$, the tree level $H^0$ 
contributions to $\Delta F=2$ observables are by far dominant and the 
$H^\pm$ contributions to $B\to X_q\gamma$ can be very important, even dominant for certain parameters of the model.
While in a large fraction of the parameter space this  model has to struggle with 
the experimental constraint from $\varepsilon_K$,
we demonstrate that there exist regions in parameter space 
which satisfy all existing $\Delta F=2$, 
$B\to X_{s,d}\gamma$, tree level decays and electroweak precision constraints for scales 
$M_{W_R}\simeq 2-3\tev$ in the reach of the LHC.  We also show that the 
$S_{\psi K_S}$ - $\varepsilon_K$ tension present in the SM can be removed 
in the LRM. Simultaneously ${\rm Br}(B\to X_s\gamma)$ can be brought closer 
to the data.
{However, we point out that with the increased lower bound 
on $M_{W_R}$, the LRM cannot help in explaining the difference between 
the inclusive and exclusive determinations of $\vub$, when all constraints are 
taken into account, unless allowing for large fine-tuning.} {Finally we present a rather complete list of Feynman rules involving quarks, gauge bosons and Higgs particles.}}

}


\begin{document}

\maketitle

\section{Introduction}\label{sec:int}
A very important property of the Standard Model (SM) regarding flavour 
violating processes is the left-handed  structure of the charged current
interactions reflecting
the maximal violation  of parity observed in low energy processes. 
Left-handed charged currents encode  at the level of the Lagrangian the full information about flavour mixing 
and CP violation represented compactly by the CKM matrix. 
Due to the GIM \cite{Glashow:1970gm} mechanism this structure has automatically profound implications for
the pattern of FCNC processes that seems to be 
in remarkable accordance with the  present data within theoretical and experimental uncertainties \cite{Bevan:2011zz,Charles:2004jd}, bearing in 
mind certain anomalies \cite{Lunghi:2008aa,Buras:2008nn,Buras:2010wr,Isidori:2010zz,Lunghi:2010gv,Lenz:2010gu,Bevan:2011zz,Lenz:2011ti,Laiho:2011nz,Lunghi:2011xy}, in particular in CP-violating observables which are discussed below. 

As the SM is expected to be only the low-energy limit of a more fundamental
theory it is conceivable that at very short distance scales parity could 
be a good symmetry implying the existence of
right-handed (RH) charged currents. Prominent examples of such fundamental 
theories are  left-right (LR) symmetric models on which a rich literature exists.
Indeed, LR symmetric models were born 37 years 
ago~\cite{Pati:1974yy,Mohapatra:1974gc,Mohapatra:1974hk,Senjanovic:1975rk,Senjanovic:1978ev}. Early papers mainly cover the 
examinations of two special cases, known as ``manifest`` scenario \cite{Senjanovic:1978ev} and ''pseudo-manifest'' scenario 
\cite{Mohapatra:1977mj,Chang:1982dp,Branco:1982wp,Harari:1983gq}, which are characterised by no spontaneous and fully spontaneous CP violation, 
respectively. The right-handed counterpart of the CKM matrix appears then in a special form being either identical to or the 
complex-conjugate of the CKM matrix up to certain phases as e.g. summarised in \cite{Kiers:2002cz}. The phenomenology of both 
scenarios has widely been  studied in the literature \cite{Beall:1981ze,Ecker:1985rr,Frere:1991db,Barenboim:1996wz}. 
By now in both scenarios strong constraints on the heavy charged gauge boson mass $M_{W_R} \gsim 4\tev$ have been obtained 
\cite{Beall:1981ze,Mohapatra:1983ae,Zhang:2007da} from the constraints on the $K_L - K_S$ mass difference, CP violation in kaon 
decays and the neutron electric dipole moment, making these scenarios {difficult to access in} direct searches at the LHC. In 
addition the ''pseudo-manifest'' scenario has been ruled out by both the appearance of light Higgs triplets \cite{Barenboim:2001vu}
and the correlation of $\varepsilon_K$ and $\sin(2\beta)$ \cite{Ball:1999mb}. This means that the right-handed mixing matrix must be 
different from the CKM matrix in order to reach agreement with experiment. Motivated by this fact more general studies on CP violation 
have been performed in \cite{Langacker:1989xa,Barenboim:1996nd,Kiers:2002cz}. More recent extensive analyses of many observables in 
the LR symmetric framework can be found in e.g.~\cite{Zhang:2007fn,Zhang:2007da,Maiezza:2010ic,Hsieh:2010zr,Crivellin:2011ba}.

Theoretical interest in models with an underlying 
$SU(2)_L \times SU(2)_R$ 
global symmetry has also been motivated by Higgsless 
models~\cite{Csaki:2003zu,Nomura:2003du,Barbieri:2003pr,Georgi:2004iy}. 
Moreover, the recent phenomenological interest in having another look  at the right-handed 
currents in general originated from tensions between inclusive and exclusive
determinations of the elements of the CKM matrix $|V_{ub}|$ and  $|V_{cb}|$. 
As pointed out
and analysed recently in particular in \cite{Crivellin:2009sd,Chen:2008se,Feger:2010qc}, 
the presence of right-handed  currents could either remove 
or significantly weaken some of these tensions, especially in the 
case of $|V_{ub}|$. The implications of these findings for many observables 
within an effective theory approach have been studied in
\cite{Buras:2010pz}. 

Yet an effective theory approach, as interesting as it may be, involves a 
number of unknown couplings that limit the predictive power of the theory and 
in particular does not allow to correlate low-energy high precision 
observables to high energy processes being already explored in a new domain 
of energy at the LHC. In this context we refer to \cite{Nemevsek:2011hz,Grojean:2011vu} where
extensive analyses of LR symmetric models have been performed for early LHC data.
Therefore it is of interest as a preparation for 
new discoveries both through high energy processes and high precision 
experiments in this decade to perform a detailed phenomenological analysis 
in a concrete class of models with right-handed currents, in particular 
models with LR symmetry. As manifest and pseudo-manifest LR models have already been 
ruled out we term the model with extended gauge group but without exact $P$ or $C$ symmetry {\it Left-Right Model} (LRM).

The goal of the present paper is to analyse the well measured FCNC observables
related to the particle-antiparticle mixings $K^0-\bar K^0$ and $B_{d,s}^0-\bar B_{d,s}^0$ in this NP scenario. 
For a RH scale in the reach of the LHC the off-diagonal mixing amplitudes $M_{12}^i$ ($i=K, d, s$) receive important and often dangerous tree level
contributions from neutral heavy scalar particles ($H^0$) present in this model. 
Additionally box diagrams with exchanges of a heavy $W^{'\pm}$ and
charged Higgs 
$(H^\pm)$ exchanges and right-handed couplings of the light $W^\pm$ are present. Similarly 
the $W_L-W_R$ mixing can have a significant impact on
the $B\to X_s\gamma$ decay that often puts a severe constraint 
on extensions of the SM. Also heavy charged Higgs 
$(H^\pm)$ exchanges can contribute and in fact these contributions 
cannot be neglected as often done  in the literature.

We would like to know whether  this class of very interesting models
can be made consistent with all existing data for RH scales as low as $M_{W_R} \simeq (2-3)\tev$, which is still consistent with direct collider searches, 
while 
solving various anomalies observed in the quark sector.

As there have been other analyses of particle-antiparticle mixing \cite{Beall:1981ze,Mohapatra:1983ae,Ecker:1983uh,Gilman:1983ce,Ecker:1985vv,Hou:1985ur,Langacker:1989xa,London:1989cf,Ball:1999yi,Sahoo:2005wb,Kiers:2002cz,Zhang:2007fn,Zhang:2007da} and 
$B\to X_s\gamma$ \cite{Asatrian:1989iu,Asatryan:1990na,Cocolicchio:1988ac,Cho:1993zb,Babu:1993hx,Fujikawa:1993zu,Asatrian:1996as,Frank:2010qv,Guadagnoli:2011id} in LR models, it is mandatory for us to state what is new in 
our paper:

\bi
\item First of all we perform a simultaneous analysis of the most interesting $\Delta F=2$ observables 
in the $K$ and $B_{d,s}$ meson systems in conjunction with the decays $B\to X_s\gamma$ and $B\to X_d\gamma$. 
The analysis  is performed in a general framework of the LRM and hence without making from the beginning particular 
assumptions on the structure of the right-handed mixing matrix or equality of gauge couplings. That is, in contrast to previous analyses we do not assume 
a certain specific form for $V^{\rm R}$ by restricting its parameters, but search for 
its structure  
by using the bounds from tree-level decays, $\Delta F=2$ observables, $B\to X_q\gamma$ decays and imposing constraints from electroweak precision tests.
This strategy differs from the existing literature
 which dominantly considered bounds 
on the parameters of LR models, in particular on the masses of $W^{'\pm}$
and $H^0$.
\item 
In this manner we are led to new structures of the $V^{\rm R}$ matrix that are 
still rather simple and allow to monitor transparently which 
anomalies observed in the quark sector can be solved in these models.
In this context we search for
correlations between various observables. 
\item
We perform the 
full phenomenology including both gauge boson and Higgs boson contributions 
finding that even for $M_{H^0}\approx M_{H^\pm}\sim\ord(20)\tev$, the tree level $H^0$ 
contributions to $\Delta F=2$ observables are by far dominant and the 
$H^\pm$ contributions to $B\to X_q\gamma$ can be very important and even dominant for certain parameters of the model. In this context we include the known 
QCD corrections.
\item
{We analyse the issue of the element $\vub$ in this specific model with 
right-handed currents, pointing out that with the increased lower bound 
on $M_{W_R}$, these models cannot help in explaining the difference between 
the inclusive and exclusive determinations of $\vub$, when all constraints are 
taken into account, unless allowing for large fine-tuning of parameters.}
\item
We investigate a soft lower bound on the heavy Higgs mass.
\item We present a collection of Feynman rules necessary for the analysis of all flavour violating processes, in 
particular $\Delta F=2$ transitions considered in the present paper.  These rules could also be useful for collider physics.
\ei

Our paper is organised as follows. In section \ref{sec:model} we summarise  briefly the main ingredients of the LRM. 
In section \ref{sec:trans} we present the effective Hamiltonians for $K^0-\bar K^0$, $B_{d}^0-\bar B_{d}^0$ and $B_{s}^0-\bar B_{s}^0$ mixings
and we calculate the most interesting observables such as the CP-violating parameter $\varepsilon_K$, the mass differences $\Delta M_K$ and  $\Delta M_q$, the CP-asymmetries $A_{\rm SL}^q$ ($q=d,s$), $S_{\psi K_S}$ and 
$S_{\psi \phi}$ and the width differences $\Delta\Gamma_q$. 
In section~\ref{sec:BSG} we present the analysis of the $B\to X_{s,d}\gamma$ decays 
including CP-violating asymmetries. In section~\ref{sec:tree} we face tree level decays including $B^+\to\tau^+\nu_\tau$. 
In section~\ref{sec:EWP} we summarise the 
existing constraints on the LRM from the electroweak 
precision tests.
In section \ref{sec:strategy} we outline our strategy for the 
numerical analysis. In this context we review the existing anomalies in 
the flavour data and present a simple analytical expression for the 
right-handed matrix $V^{\rm R}$ that allows to see how these anomalies can 
be solved in a correlated manner for different values of $\vub$.
In section~\ref{sec:EWPNUM} we present the bounds on the 
electroweak sector of the LRM from electroweak precision tests. 
In section~\ref{sec:VR4S} a general study of the right-handed mixing matrix $V^{\rm R}$ 
is presented. 
Subsequently in section \ref{sec:num} a detailed numerical analysis of  
particle-antiparticle mixing observables and of $B\to X_{s,d}\gamma$ decays
including tree-level constraints is performed. {In section \ref{sec:higgsmass} we derive a soft lower limit on the heavy Higgs masses.}
In section~\ref{CMODELS} a brief comparison of the LRM with other models is 
presented.
We summarise our results in section \ref{sec:conc}. In the appendices we provide a more detailed description of
the LRM and the symmetry breaking mechanism, the Higgs sector and numerical insights into the structure of LR contributions to $\Delta F=2$ operators.
Furthermore we provide an extensive list of Feynman Rules for the LRM.

\section{Models with Left-Right gauge symmetry}\label{sec:model}

In this section we give a brief description of the LRM. We restrict our presentation to the key properties, which allows us to set our notation.
More details on LR symmetric and asymmetric models can be found e.\,g.\ in 
\cite{Langacker:1989xa,Zhang:2007da,Maiezza:2010ic} and references 
therein.

\subsection{Gauge group and fermion content}

Among the most popular new physics models are LR extensions 
\be\label{eq:gauge-group}
SU(3)_C \times SU(2)_L \times SU(2)_R \times U(1)_{B-L}
\ee
of the SM gauge group, as they allow for a restoration of parity symmetry at high energies. Note that the spontaneous breaking of parity does not have to be connected to the breakdown to the SM gauge group and can take place at some much higher scale. Therefore we do not restrict ourselves to the study of the parity (or alternatively charge conjugation) symmetric case but consider the generic case with independent gauge couplings $g_L,g_R$.

The left-handed fermions are embedded as $SU(2)_L$ doublets and $SU(2)_R$ singlets, while the right-handed fermions are $SU(2)_{L}$ singlets and $SU(2)_R$ doublets:
\begin{gather}
Q_L = \left( \begin{array}{c} u_L \\ d_L\\ \end{array} \right)\sim
\left(3, 2,1,\frac{1}{3}\right)\,,  \;\;\; Q_R = \left( \begin{array}{c}
u_R \\  d_R  \\
\end{array} \right) \sim \left(3, 1,2,\frac{1}{3} \right)\,,  \\
L_L = \left( \begin{array}{c} \nu_L \\ l_L\\ \end{array} \right)\sim
\left(1,2,1,-1\right)\,,  \;\;\; L_R = \left( \begin{array}{c} \nu_R \\ l_R\\
\end{array} \right) \sim \left(1, 1,2,-1 \right) \,,
\end{gather}
with the quantum numbers given in brackets corresponding to the gauge group in \eqref{eq:gauge-group}. We see that instead of the seemingly arbitrary $U(1)_Y$ charges in the SM, the fermionic $U(1)$ charges are now given by their $B-L$ quantum numbers. The resulting electric charges are
\be
Q = T_{3L}+T_{3R}+\frac{B-L}{2}\,.
\ee

\subsection{Higgs sector and spontaneous symmetry breaking}

The spontaneous symmetry breaking of LR models takes place in two steps. 

{\bf Step 1:} At a high scale $\kappa_R\sim\ord(\text{TeV})$ $SU(2)_R\times U(1)_{B-L}$ is broken to the SM hypercharge gauge group $U(1)_Y$: 
\be\label{STEP1}
SU(2)_R\times U(1)_{B-L} \to U(1)_Y\,.
\ee
The details of this breaking are model-dependent. The two simplest possibilities introduce either two scalar doublets or triplets, 
however also more complicated $SU(2)_{L,R}$ representations are phenomenologically viable. 
It turns out that quark flavour phenomenology does not depend on the particular structure of the Higgs 
sector (for more details see appendix \ref{app:goldhiggs}). Hence from now on we concentrate on the 
triplet model, which is appealing in the neutrino sector as it naturally generates TeV scale Majorana masses 
for the right-handed neutrinos \cite{Mohapatra:1979ia,Mohapatra:1980yp,Khasanov:2001tu,Aranda:2009ut}  
through the VEV of $\Delta_R$ defined below. Consequently the light neutrino masses are suppressed by the TeV scale see-saw mechanism. 

In the triplet model, the symmetry breaking
$ SU(2)_R \times U(1)_{B-L} \to  U(1)_Y$
is achieved by a Higgs triplet $\Delta_R$
\be
\Delta_R = \left( \begin{array}{cc} \delta_{R}^+ / \sqrt{2} &
\delta_{R}^{++} \\ \delta_{R}^{0} & -\delta_{R}^+ / \sqrt{2} \end{array} \right) \sim (1,1,3,2)\,,
\ee
which develops a VEV
\be
\langle \Delta_R \rangle = \left( \begin{array}{cc} 0 &
0 \\ \kappa_{R} & 0 \end{array} \right)\,.
\ee
{The most recent experimental direct $W_R$ searches 
 find roughly 
 $M_{W_R}\ge 1.5 - 2\tev$} and this can only be satisfied with $\kappa_R$ much larger than the EWSB scale $v$.
This hierarchy of scales implies that all LR effects can be expanded in powers of the small  dimensionless parameter
\be\label{eq:epsilon}
\epsilon = v/\kappa_R\,.
\ee
Throughout our phenomenological analysis we keep contributions of up to $\ord(\epsilon^2)$, which constitute the leading corrections relative to the SM result.

Keeping up the possibility of straightforwardly incorporating the limits of manifest $P$ or $C$ symmetry  we also introduce an $SU(2)_L$ triplet
\be
\Delta_L = \left( \begin{array}{cc} \delta_{L}^+ / \sqrt{2} &
\delta_{L}^{++} \\ \delta_{L}^{0} & -\delta_{L}^+ / \sqrt{2} \end{array} \right) \sim (1,3,1,2)\,,
\ee
whose VEV
\be
\langle \Delta_L \rangle = \left( \begin{array}{cc} 0 &
0 \\ \kappa_{L}e^{i\theta} & 0 \end{array} \right)\,,
\ee
is constrained to be $\kappa_L\lsim \ord(\text{eV})$ in order not to generate  large Majorana masses for the left-handed neutrinos \cite{Mohapatra:1979ia,Mohapatra:1980yp,Khasanov:2001tu}. 

{\bf Step 2:} The second step of symmetry breaking is then achieved by 
the bidoublet
\be
\phi = \left( \begin{array}{cc} \phi_{1}^0 & \phi_{2}^{+} \\
\phi_{1}^{-} & \phi_{2}^0\\ \end{array} \right)\sim(1,2,2,0)\,,
\ee
whose vacuum expectation value
\be\label{eq:phiVEV}
\langle\phi\rangle = 
\left( \begin{array}{cc} \kappa &0 \\
0 & \kappa' e^{i\alpha} \end{array} \right)\,,
\ee
breaks $SU(2)_L\times SU(2)_R$ to its diagonal subgroup $SU(2)_V$:
\be\label{STEP2}
SU(2)_L\times SU(2)_R \to SU(2)_V.
\ee
This step of spontaneous symmetry breaking takes place at the scale 
\be\label{vev}
v=\sqrt{\kappa^2+\kappa'^2}=174\gev\,.
\ee
Together with the spontaneous breaking 
$SU(2)_R\times U(1)_{B-L} \to U(1)_Y$ in (\ref{STEP1}) at the scale $\kappa_R\gg v$  the VEV of $\phi$ leads to the standard electroweak symmetry breaking (EWSB) 
$SU(2)_L \times U(1)_Y \to U(1)_Q$ so that finally 
\be
SU(2)_L\times SU(2)_R \times U(1)_{B-L} \to U(1)_Q\,.
\ee

{As we see later in \eqref{eq:quarkmasses} in order to obtain a mass splitting between up and down type quarks $\kappa \ne \kappa'$ is required.}
More explicitly requiring the hierarchy $m_b \ll m_t$ to be natural implies $\kappa' \ll \kappa$. However, we do not assume that 
$\kappa/ \kappa'=m_t/m_b$ as done in some papers. In fact such large values of $\kappa/ \kappa'$ are disfavoured by electroweak precision observables, in particular by $A^b_{FB}$ (see section \ref{sec:EWPNUM} for details). {In our analysis we confine the values to $1<\kappa/\kappa'<10$. It should be stressed again that the limit $\kappa = \kappa'$ is not allowed, which can be seen explicitly from the divergent behaviour of several observables.}

For our choice of scalar fields the Higgs Lagrangian is then given by 
\be\label{eq:Higgskin}
\mathcal{L}_{\text{Higgs}}= {\rm Tr}[(D_\mu \Delta_L)^{\dagger}(D^\mu
\Delta_L)]  + {\rm Tr}[(D_\mu \Delta_R)^{\dagger}(D^\mu \Delta_R)] 
+ {\rm Tr}[(D_\mu \phi)^{\dagger}(D^\mu \phi)] + V(\phi, \Delta_L, \Delta_R) \,,
\ee
where the covariant derivatives are
\begin{eqnarray}
D_\mu \phi &=& \partial_\mu \phi + i g_L (\overrightarrow{W}_{L\mu}\cdot
\vec{\tau})\phi - i {g_R} \phi(\overrightarrow{W}_{R\mu}\cdot
\vec{\tau}) \,,\label{eq:covder} \\
D_\mu\Delta_{(L,R)} &=&
\partial_\mu\Delta_{(L,R)} + i {g_{(L,R)}} \left[\overrightarrow{W}_{(L,R)\mu}\cdot
\vec{\tau}, ~\Delta_{(L,R)} \right] + ig'B_\mu \Delta_{(L,R)}  \nonumber \,.
\end{eqnarray}
and the Higgs potential, as being used in our analysis, is given in  appendix \ref{app:potential}.

\subsection{Gauge sector after electroweak symmetry breaking}
After performing the two steps of symmetry breaking the gauge boson mass matrices can be constructed and diagonalised. We summarise our results in appendix \ref{app:gauge-boson-masses}. 
In the process of electroweak symmetry breaking only the gluons and the photon remain massless, while the $W^\pm$ and the $Z$ boson and their heavy counterparts $W'^\pm$ and $Z'$ acquire masses as given in appendix \ref{app:gauge-boson-masses}.
The spontaneous symmetry breaking additionally introduces mixing between the light and heavy gauge bosons. In the case of the light SM like bosons $W^\pm$ this introduces right-handed
couplings at ${\mathcal O}(\epsilon^2)$. The $Z$ boson couplings are also modified with respect to the SM but they do not enter the present analysis.

\subsection{Yukawa interactions and fermion masses}\label{sec:Yuk}
The most general renormalisable Yukawa coupling of the quark fields with our choice of Higgs fields is
given by
\begin{equation}\label{yukawa}
\mathcal{L}_\text{Yuk} = - y_{ij} \overline{Q}_{Li}  \phi  Q_{Rj} - \tilde y_{ij} \overline{Q}_{Li}  \tilde \phi Q_{Rj} + {\rm h.c.}\,,
\end{equation}
where $\tilde \phi = \sigma_2 \phi^*  \sigma_2$ and flavour indices $i,j=1,2,3$. Note that the quantum numbers of the other scalar fields in the 
theory preclude their direct coupling to quarks. \footnote{In the triplet model $\Delta_{L,R}$ couple to the left- and right-handed leptons respectively, generating Majorana mass terms $M_{\nu_{L,R}}\sim\kappa_{L,R}$.}
The resulting fermion mass matrices read
\be
(M_u)_{ij} = v (Y_u)_{ij}\,,\qquad (M_d)_{ij} = v (Y_d)_{ij}\,,
\ee
where
\be\label{eq:quarkmasses}
(Y_u)_{ij}=y_{ij}c+\tilde y_{ij}s e^{-i\alpha}\,,\qquad
(Y_d)_{ij}=y_{ij} s e^{i\alpha}+\tilde y_{ij}c\,,
\ee
and $s=\kappa'/v$ and $c=\kappa/v$.
These matrices are diagonalised by the bi-unitary transformations
\bea
M_u^\text{diag} &=& U_L^\dagger M_u U_R\,,\\
M_d^\text{diag} &=& D_L^\dagger M_d D_R\,,
\eea
where $U_{L,R}, D_{L,R}$ are unitary matrices connecting the flavour and mass eigenstates of quarks.

As we discuss in more details below, the extended Higgs sector with respect to the SM leads to flavour changing neutral Higgs couplings already at the tree level. While the flavour violating couplings of the light SM-like Higgs are highly suppressed and therefore irrelevant for the study of 
$\Delta F=2$ observables and
$K$ and $B$ decays, the new heavy Higgses lead to dangerously large effects in FCNC observables. {In particular the structure of the Yukawa coupling of the Higgs bidoublet $\phi$ in \eqref{yukawa} leads to couplings of the down-type quarks proportional to the up-type quark mass matrix. Since up and down masses are not diagonalised simultaneously, this leads to flavour changing couplings already at the tree level, see table \ref{tab:H01} for details.}

\subsection{Parameter counting}
Having introduced the LR model, let us now count the parameters present in the theory.

The gauge sector is parametrised by the gauge couplings 
\be\label{couplings}
g_s\,,\quad g_L\,,\quad g_R,\quad g'\,,
\ee
i.\,e.\ one additional parameter relative to the SM.

The Higgs potential, see appendix \ref{app:potential}, introduces several
 new  parameters.  However we see below that for our  phenomenological considerations effectively only four  parameters appear to be relevant. Setting $\kappa_L=0$ these are 
\be\label{vevs}
v=\sqrt{\kappa^2+\kappa'^2}\,,\quad  s=\kappa'/v\,,\quad \kappa_{R},\quad  M_H\,,
\ee
where the first three parametrise the  Higgs VEVs \footnote{The phase $\alpha$ in the Higgs potential appears in the analytic expressions below. However we eventually set it 
to zero as the factor $\exp(i\alpha)$ always multiplies $V^{\rm R}$ and cancels out in all expressions for FCNC processes. On a more technical note: the factor 
$\exp(i\alpha)$ always multiplies $V^{\rm R}$, which is a unitary matrix with six phases. Therefore we are always able to absorb $\alpha$ through a redefinition of all phases. Recall that
a unitary matrix cannot have more than six independent phases.}
and $M_H$ is at leading order the common mass of the heavy Higgses $H^0_1$, $H^0_2$ and $H^\pm$
(for more details see appendix \ref{app:goldhiggs}). Note that that the SM VEV $v$ is given in terms of 
$\kappa$ and $\kappa'$. 

In the most general case, the Yukawa couplings $y_{ij}$ and $\tilde y_{ij}$ are arbitrary complex matrices, i.\,e.\ contain each 
9 real parameters and 9 phases. However, not all of these parameters are physical but some can be removed by unitary transformations 
under the flavour symmetry $SU(3)_{Q_L}\times SU(3)_{Q_R}$. Finally we are left with the six quark masses and two mixing matrices in the LH and 
RH sectors respectively:
\be
V^{\rm L}=U_L^\dagger D_L\,, \quad V^{\rm R}=U_R^\dagger D_R\,.
\ee
Adopting the standard CKM phase convention, where the 5 relative phases of the quark fields are adjusted to remove 5 complex phases from the CKM matrix $V^{\rm L}$, we have no more
freedom to remove the 6 complex phases from  $V^{\rm R}$.  In the standard CKM basis  $V^{\rm R}$ can be parametrised as follows \cite{Buras:2010pz}
\be
V^{\rm R} = D_U V_0^{\rm R}D^\dagger_D\,, 
\ee
where $V_0^{\rm R}$ is a ``CKM-like'' mixing matrix, containing only three real mixing angles and one non-trivial phase.
The diagonal matrices $D_{U,D}$ contain the remaining CP-violating phases. Choosing the standard parametrisation for $V_0^{\rm R}$ we have
\be
V_0^{\rm R} = 
\left( 
\begin{array}{ccc}
\tc_{12}\tc_{13}&\ts_{12}\tc_{13}&\ts_{13}e^{-i\phi}\\ -\ts_{12}\tc_{23}
-\tc_{12}\ts_{23}\ts_{13}e^{i\phi} &\tc_{12}\tc_{23}-\ts_{12}\ts_{23}\ts_{13}e^{i\phi}& 
\ts_{23}\tc_{13}\\ 
\ts_{12}\ts_{23}-\tc_{12}\tc_{23}\ts_{13}e^{i\phi}&-\ts_{23}\tc_{12} -\ts_{12}\tc_{23}\ts_{13}e^{i\phi}&\tc_{23}\tc_{13}
\end{array}
\right)\,,
\label{eq:Vtrgen}
\ee
and 
\be
D_U={\rm diag}(1, e^{i\phi^u_2}, e^{i\phi^u_3})\,,  \qquad 
D_D={\rm diag}(e^{i\phi^d_1}, e^{i\phi^d_2}, e^{i\phi^d_3})\,.
\label{eq:Dphases}
\ee

\section{\boldmath $\Delta F=2$ transitions}\label{sec:trans}

\subsection{Preliminaries}
In what follows we use conventions and notation from our papers on various extensions of the SM. An easy comparison with the results for $\Delta F=2$ 
observables in the SM, the Littlest Higgs model with T-parity (LHT) \cite{Blanke:2006sb}, 
the Randall-Sundrum scenario with custodial protection (RSc) \cite{Blanke:2008zb}
and the SM4 \cite{Buras:2010pi} is facilitated in this manner.

In the LR models the effective Hamiltonian for $\Delta F=2$ observables
is constructed by evaluating three classes of diagrams:
\begin{itemize}
\item
The standard box diagrams with quarks and $W_LW_L$, $W_RW_R$ and 
$W_LW_R$ exchanges~\footnote{In what follows in order to make the expressions more transparent 
it is useful to denote $W$ and $W'$ 
by  $W_L$ by $W_R$, respectively, even if they differ by $\ord(\epsilon^2)$
corrections.}. Among the NP contributions involving $W_R$ 
only the latter matter.
\item
Box diagrams with  charged Higgs $H^\pm$ and gauge boson  exchanges. As in  
LR models $H^\pm$ have masses in the multi TeV range,  among 
NP contributions involving $H^\pm$ only the ones with $H^\pm$ and $W_L$ matter.
\item
Tree level neutral heavy Higgs exchanges. These contributions are problematic 
unless the masses of new neutral Higgs particles are significantly larger than 
 the one of $W_R$.  
In first approximation the masses of $H^\pm$ are equal to the masses of the neutral Higgs bosons in question.
\end{itemize}

While in the SM only one operator contributes to each $\Delta F=2$ transition, 
in the model in question there are 8 such operators of dimension 6. 
Consequently the renormalisation group (RG) QCD analysis becomes more 
involved and due to the LR structure of the new operators
QCD corrections play a much more  important role in 
new physics contributions than in the SM contributions.

In what follows, after listing all contributing operators we 
summarise the effective Hamiltonian for $\Delta F=2$ transitions. 
To this end we give the formulae for 
the Wilson coefficients at the matching scales and we  summarise 
the RG QCD corrections and the results for the hadronic matrix elements. 
Subsequently we give the final formulae for the basic mixing amplitudes in 
terms of Wilson coefficients at the high scale and the effective parameters 
$P_i$ that encode perturbative and non-perturbative QCD effects.
{Then we discuss in detail the general and special anatomy of LR contributions and compare the operator structure to the one found in other NP scenarios.}
We end this section with listing the relevant observables. 

\subsection{Local operators}
The contributing operators 
can be split into 5 separate sectors, according to the chirality
of the quark fields they contain. 
For definiteness, we shall first consider  operators responsible for the
$K^0$--$\bar{K}^0$ mixing. 
The operators belonging to the first
three sectors (VLL, LR and SLL) read \cite{Buras:2000if} :
\bea 
Q_1^{\rm VLL}(K) &=& (\bar{s}^{\alpha} \gamma_{\mu}    P_L d^{\alpha})
(\bar{s}^{ \beta} \gamma^{\mu}    P_L d^{ \beta})\,,
\nnb\\[4mm] 
Q_1^{\rm LR}(K) &=&  (\bar{s}^{\alpha} \gamma_{\mu}    P_L d^{\alpha})
(\bar{s}^{ \beta} \gamma^{\mu}    P_R d^{ \beta})\,,
\nnb\\
Q_2^{\rm LR}(K) &=&  (\bar{s}^{\alpha}                 P_L d^{\alpha})
(\bar{s}^{ \beta}                 P_R d^{ \beta})\,,
\nnb\\[4mm]
Q_1^{\rm SLL}(K) &=& (\bar{s}^{\alpha}                 P_L d^{\alpha})
(\bar{s}^{ \beta}                 P_L d^{ \beta})\,,
\nnb\\
Q_2^{\rm SLL}(K) &=& (\bar{s}^{\alpha} \sigma_{\mu\nu} P_L d^{\alpha})
(\bar{s}^{ \beta} \sigma^{\mu\nu} P_L d^{ \beta})\,,
\label{normalK}
\eea
where $\sigma_{\mu\nu} = \frac{1}{2} [\gamma_{\mu}, \gamma_{\nu}]$ and
$P_{L,R} = \frac{1}{2} (1\mp \gamma_5)$5), and summation over the colour indices
$\alpha, \beta = 1,2,3$ is understood. The operators belonging to the
two remaining sectors (VRR and SRR) are obtained from $Q_1^{\rm VLL}$ and
$Q_i^{\rm SLL}$ by interchanging $P_L$ and $P_R$. In the SM only the 
operator $Q_1^{\rm VLL}(K)$ is present.
The operators relevant for  $B_q$ ($q = d,s$) are obtained by replacing 
in (\ref{normalK})
 $s$ by $b$ and
$d$ by $q$.


\subsection{Effective Hamiltonian}
The effective Hamiltonian for $\Delta F=2$ transitions can be written 
in a general form as follows
\be\label{Heff-general}
\Heff^{\Delta F=2} =\frac{G_F^2M^2_{W_L}}{4\pi^2}
\sum_iC_i(\mu)Q_i\,,
\ee
where $Q_i$ are the operators given in (\ref{normalK}) and
$C_i(\mu)$ their Wilson coefficients evaluated at a scale $\mu$ which 
we specify below. In what follows we collect the Wilson coefficients 
of these operators separating the contributions from box diagrams  
with $W_L$ and $W_R$ exchanges from  box diagram charged Higgs $H^\pm$
contributions and   tree level neutral Higgs $H_{1,2}^0$
contributions so that 
\be
C_i=\Delta_{\rm Box}C_i+\Delta_{H^\pm}C_i+\Delta_{H^0}C_i\,.
\ee
These coefficients depend sensitively on the elements of the matrices 
$V^{\rm L}$ and $V^{\rm R}$ through  \cite{Ecker:1985vv}
\be
\lambda_i^{AB}(K)=V_{is}^{A*}V_{id}^{B}\,,\qquad 
\lambda_i^{AB}(B_q)=V_{ib}^{A*}V_{iq}^{B}\,,
\ee
where $A,B=L,R$, $q=d,s$ and $i=u,c,t$. 

\subsubsection{Wilson coefficients from gauge boson box diagrams}

\begin{figure}[h!]
\centering
\includegraphics[width=1.0\textwidth]{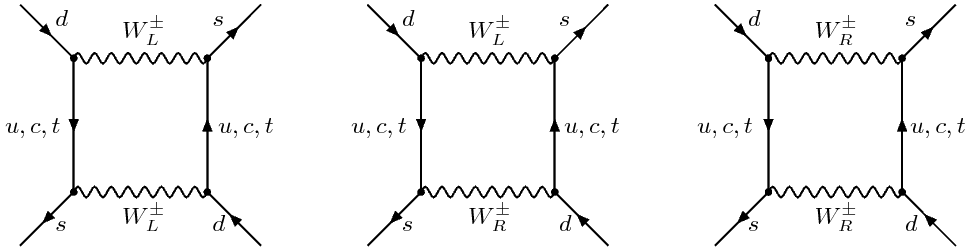}
\caption{Feynman diagrams for contributing gauge boson box diagrams}\label{fig:gauge_box}
\end{figure}

Calculating the diagrams in figure \ref{fig:gauge_box}, we find the following contributions
to the Wilson coefficients relevant for the $K^0-\bar K^0$ system at 
the relevant matching scales $\mu_W=\ord(M_W,m_t)$ and $\mu_R=\ord(M_{W_R})$
\begin{align}
\Delta_{\rm Box}C_1^{\rm VLL}(\mu_W,K) &=\,\sum_{i,j=c,t}\lambda_i^{\rm LL}(K)\lambda_j^{\rm LL}(K)S_{\rm LL}(x_i,x_j)\,,\label{VLLK}\\
\Delta_{\rm Box}C_2^{\rm LR}(\mu_R,K)  &=\,\sum_{i,j=u,c,t}\lambda_i^{\rm LR}(K)\lambda_j^{\rm RL}(K)S_{\rm LR}(x_i,x_j,\beta)\,,\label{VLRK}\\
\Delta_{\rm Box}C_1^{\rm VRR}(\mu_R,K) &=\,\sum_{i,j=c,t}\lambda_i^{\rm RR}(K)\lambda_j^{\rm RR}(K) S_{\rm RR}(\tilde x_i,\tilde x_j)\,,\label{VRRK}
\end{align}
where we introduced the ratios
\be
x_i=\left(\frac{m_i}{M_{W_L}}\right)^2, \qquad 
\tilde x_i=\left(\frac{m_i}{M_{W_R}}\right)^2, \qquad
\beta=\frac{M^2_{W_L}}{M^2_{W_R}}, \qquad r=\left(\frac{s_W}{c_W s_R}\right)^2.
\ee 
Note that at $\ord(\epsilon^2)$ there are no corrections to the Wilson coefficient of  the SM $Q_1^{\rm VLL}$ operator.
In the LR symmetric limit $g_L= g_R$ the factor $r$ reduces to $r=1$.

The loop functions are given as follows  
\begin{eqnarray}
S_{\rm LL}(x_i, x_j) & = & F(x_i,x_j)+F(x_u,x_u)-F(x_i,x_u)-F(x_j,x_u)\,,\\
S_{\rm LR}(x_i, x_j, \beta) & = & 2\beta r\sqrt{x_i x_j} \left[ ( 4 + 
x_i x_j \beta ) I_1(x_i, x_j, \beta) - ( 1 + \beta ) I_2(x_i, x_j, 
\beta) \right]\,,\\
S_{\rm RR}(\tilde x_i, \tilde x_j) & = & \beta r^2 S_{LL} (\tilde x_i, \tilde x_j)\,,\\
F(x_i,x_j) & = & \frac{1}{4}\left[ (4+x_i 
x_j)I_2(x_i,x_j,1) - 8 x_i x_j I_1(x_i, x_j, 1)\right]\,,
\end{eqnarray}
with
\begin{eqnarray}
I_1(x_i,x_j,\beta) & = & \frac{x_i \ln(x_i)}{(1-x_i)(1-x_i 
\beta)(x_i-x_j)} + (i\leftrightarrow j) - \frac{\beta 
\ln(\beta)}{(1-\beta)(1-x_i\beta)(1-x_j\beta)}\,,\qquad\\
I_2(x_i,x_j,\beta) & = & \frac{x_i^2 \ln(x_i)}{(1-x_i)(1-x_i 
\beta)(x_i-x_j)} + (i\leftrightarrow j) - 
\frac{\ln(\beta)}{(1-\beta)(1-x_i\beta)(1-x_j\beta)}\,.\qquad
\end{eqnarray}
The remaining coefficients vanish in the absence of QCD corrections 
but as we discuss below they are generated by QCD effects. 
In obtaining the results in (\ref{VLLK}) and (\ref{VRRK}) we have used the 
unitarity of the matrices $V^{\rm L}$ and $V^{\rm R}$ or equivalently the GIM mechanism to 
eliminate the $\lambda_u^{\rm LL}$ and $\lambda_u^{\rm RR}$ terms.
The GIM mechanism does not apply to the case of LR contributions.
In the case of $B_{q}^0-\bar B_{q}^0$ mixing we just have to replace $K$ by $B_q$.

The results given above were obtained by calculating all box diagram contributions in `t Hooft-Feynman gauge keeping both gauge boson 
and Goldstone boson contributions.  We confirm the results in the literature 
\cite{Ecker:1985vv,Zhang:2007da,Chang:1984hr,Hou:1985ur,Basecq:1985cr}.

While $S_{\rm LL}$ and $S_{\rm RR}$ are gauge independent, this is not the case of 
$S_{\rm LR}$ as pointed out in \cite{Chang:1984hr}. As anticipated in that paper and explicitly 
demonstrated in \cite{Hou:1985ur,Basecq:1985cr} a gauge independent result for $S_{\rm LR}$ 
is obtained by  including vertex 
and self-energy corrections to the tree-level physical Higgs 
$H_1^0$ and $H_2^0$ exchanges that we discuss subsequently. The 
diagrams relevant for the cancellation of the gauge dependence of $S_{\rm LR}$ 
are the ones that include the vertices $H_i^0 G^+G^{'-}$ but for consistency 
all vertex and self-energy corrections 
should be included. Detailed analyses in \cite{Hou:1985ur,Basecq:1985cr}
show that the main role of these additional contributions is the restoration of 
the gauge invariance of $S_{\rm LR}$ without any relevant modification 
of the `t Hooft-Feynman gauge result given above. Consequently we neglect 
these contributions in our analysis. 

\subsubsection{Wilson coefficients from charged Higgs box diagrams}
\begin{figure}[h!]
\centering
\includegraphics[width=0.8\textwidth]{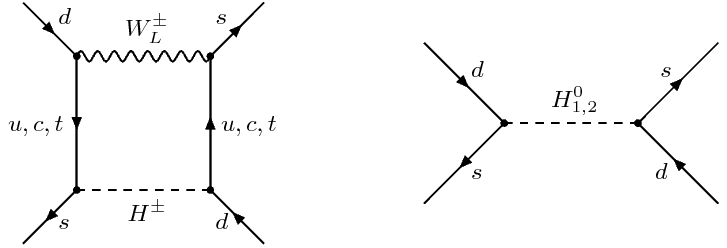}
\caption{Feynman diagrams for Higgs contributions}\label{fig:higgsdiag}
\end{figure}

Among various box diagrams with charged Higgs exchanges the one 
involving $H^\pm$ and $W_L^\pm$ (see left hand side in figure \ref{fig:higgsdiag})
and  the corresponding diagram with $W_L$ replaced by the Goldstone bosons are 
by far the most important ones. The remaining box diagrams involving only 
$H^\pm$ or $H^\pm$ and $W_R^\pm$ can be safely neglected. 
A similar comment applies to box diagrams with heavy neutral Higgs particles.
Calculating the relevant diagram shown in figure \ref{fig:higgsdiag} yields
\be\label{HVLRK}
\Delta_{H^\pm}C_2^{\rm LR}(\mu_H,K)=\sum_{i,j=u,c,t}\lambda_i^{\rm LR}(K)\lambda_j^{\rm RL}(K)S^H_{\rm LR}(x_i,x_j,\beta_H)\,.
\ee
with the master function $S^{H}_{\rm LR}$ defined as
\be\label{SHLR}
S^H_{\rm LR}(x_i, x_j, \beta_H)  =  2u(s)\beta_H\sqrt{x_i(\mu_H) x_j(\mu_H)} \left[x_i x_j  I_1(x_i, x_j, \beta_H) -  I_2(x_i, x_j,\beta_H) \right]\,.
\ee
Here {the function $u(s)$ and $\beta_H$ are given as follows
\be\label{us}
u(s)=\left(\frac{1}{1-2s^2}\right)^2,\qquad \beta_H=\frac{M^2_{W_L}}{M^2_{{H}^\pm}}\,.
\ee
}
We have confirmed the result of \cite{Ecker:1985vv} {except that in that 
paper $s\ll 1$ has been chosen.}
 Our choice of the matching scale is
explained below.  Note that the quark masses in the first factor in (\ref{SHLR}) have to be evaluated at $\mu_H$ as this factor arises from the Yukawa couplings of $H^\pm$. 
The remaining masses have to be evaluated at scales as discussed in the 
context of QCD corrections below. In table \ref{tab:runningmasses} we provide values
of the masses at different high scales.

\subsubsection{Wilson coefficients from tree level Higgs exchange}
Calculating the tree diagrams with neutral Higgs exchanges (right hand side in figure \ref{fig:higgsdiag}), we find 
that the contributions of the Higgs bosons $H_1^0$ and $H_2^0$ to
$C_{1,2}^{\rm SLL}$ and $C_{1,2}^{\rm SRR}$ cancel each other in the limit of 
$M_{H_1^0}=M_{H_2^0}=M_{H}$. Consequently in the case of non-degenerate 
Higgs masses these contributions are $\ord(\epsilon^4)$ and can be 
neglected. On the other hand in the same limit  taking into account the 
overall factor in (\ref{Heff-general}) we find 
\be\label{VLRKH}
\Delta_{\rm H^0}C_2^{\rm LR}(\mu_H,K)=-\frac{16\pi^2}{\sqrt{2} M_H^2 G_F}
u(s)\sum_{i,j=u,c,t}\lambda_i^{\rm LR}(K)\lambda_j^{\rm RL}(K)\sqrt{x_i(\mu_H)x_j(\mu_H)}\,,
\ee
with other Wilson coefficients vanishing at $\mu_{H}=\ord(M_{H})$ in the absence of QCD corrections. The quark masses have to be evaluated at $\mu_H$. 
This result agrees with \cite{Zhang:2007da} {except that in that 
paper $s\ll 1$ has been chosen. As seen in (\ref{us}) e.\,g.\ for $s=0.5$ the 
additional factor $u(s)$ provides an enhancement of a factor of 4.}
 In the case of $B_q^0-\bar B^0_q$ mixing 
$K$ should be replaced by $B_q$. $\Delta_{\rm H^0}C_1^{\rm LR}$ is generated 
by QCD effects as discussed below.

\subsection{QCD corrections and hadronic matrix elements}
The complete analysis  of $\Delta F=2$ processes requires the inclusion of the QCD renormalisation 
group evolution. The local operators have to be evolved from their respective high scales down low energy
scales at which the hadronic matrix elements are evaluated by lattice methods.
A complication arises in the model in question as three rather 
{different high scales $\mu_W \ll \mu_R \ll \mu_H$ are involved}.
Before addressing this problem let us recall a very efficient method 
\cite{Buras:2001ra,Gorbahn:2009pp} for the inclusion of all these QCD effects in the presence of a single
high scale which we denote by $\mu_{\rm in}$.

Instead of evaluating the hadronic matrix elements at the 
low energy scale we can choose to evaluate them at the high scale $\mu_{\rm in}$ at which heavy particles are integrated out. 
Thus the amplitude for $M-\overline{M}$ mixing ($M= K, B_d,B_s$) is given simply by
\be\label{amp6}
A(M\to \overline{M})=\langle\overline{M}|\Heff^{\Delta F=2}|M\rangle =
\frac{G_F^2M^2_{W_L}}{4\pi^2}\sum_{i,a} C^a_i(\mu_{\rm in})\langle \overline{M} |Q^a_i(\mu_{\rm in})|M\rangle\,.
\ee
Here the sum runs over all the operators listed in 
(\ref{normalK}). The 
matrix elements for $B_d-\bar B_d$ mixing are for instance given as 
follows \cite{Buras:2001ra,Gorbahn:2009pp} 
\be\label{eq:matrix}
\langle \bar B_d^0|Q_i^a(\mu_{\rm in})|B_d^0\rangle = \frac{2}{3}m_{B_d}^2 F_{B_d}^2 P_i^a(B_d)\,,
\ee
where the coefficients $P_i^a(B_d)$ 
collect compactly all RG effects from scales below $\mu_{\rm in}$ as well as
hadronic matrix elements obtained by lattice methods at low energy scales.
Analytic formulae for these coefficients are given in \cite{Buras:2001ra} 
while recent applications of this method can be found in 
\cite{Buras:2010mh,Buras:2010zm,Buras:2010pz}. As the 
Wilson coefficients $ C_i(\mu_{\rm in})$ depend directly on the loop functions, 
tree diagram results
and fundamental parameters of a given theory, this formulation is very 
transparent and interesting short distance NP effects are not hidden 
by complicated QCD effects. The numerical values for the coefficients 
$P_i^a(B_q)$ and $P_i^a(K)$ that we require for our analysis 
are given below. 

The question then arises how to generalise this method to the case at 
hand which involves three rather different high scales. There are three  types 
of contributions for which the relevant high energy scales attributed to 
the coefficients quoted above differ from each other:
\begin{itemize} 
\item
The SM box diagrams involving $W_L$ and the SM quarks. Here the scale is chosen to be $\ord(m_t)$ as in \cite{Buras:2001ra}.
\item
{Tree diagrams mediated by neutral heavy Higgs exchanges. In this case we 
take $\mu_H = 15\tev$ as the initial scale for the RG 
evolution.}
\item
The only problematic cases are  the contributions from $W_R$ and $H^\pm$ 
that appear in box diagrams together with much lighter $W_L$ and the 
SM quarks. Here the correct procedure would be to first integrate out 
$W_R$ and $H^\pm$. Subsequently one would construct an effective field theory not involving 
them  as dynamical degrees of freedom. We believe that in view of several unknown parameters in the LR models 
such a complicated analysis would be premature. Therefore we 
 choose $\mu_R$ as the matching scale for box contributions involving $W_R$. 
For diagrams involving Higgs particles we set the high scale to be $\mu_H$.
As the dominant effects from the included RG evolution stem from scales 
below $\mu_t$, this procedure should sufficiently well approximate the true result.
\end{itemize}

Now let us turn to the question of scales in the  quark masses. 
In box contributions we use $m_i(m_i)$ for $i=c\,,\,b\,,\,t$ and $m_i(2\gev)$ for light quarks. An exception are the masses in the overall factor in $H^\pm$ contribution in (\ref{SHLR}), as discussed previously,  which originate in the Yukawa couplings of quarks 
to $H^\pm$. Here similarly to the tree level exchange of heavy neutral Higgs particles (\ref{VLRKH}) quark masses should be evaluated at $\mu_H$.

Having the initial conditions for Wilson coefficients
at a given high scale $\mu_{\rm in}$ we can calculate 
the relevant $M-\overline{M}$ amplitude by means of (\ref{amp6})
provided also the corresponding hadronic matrix elements are known at this scale. 
As seen in (\ref{eq:matrix}) these matrix elements are directly given in terms of the parameters $P_i^a(K)$, $P_i^a(B_d)$ 
and $P_i^a(B_s)$. Explicit expressions for the latter in terms of RG 
QCD factors and the non-perturbative parameters $B_i^a(\mu_L)$ are given in 
equations (7.28)--(7.34) in \cite{Buras:2001ra} with $\mu_L$ denoting the low energy 
scale to be specified below. 

The parameters $B_i^a(\mu_L)$ are subject to considerable uncertainties. 
They can be extracted from the results of \cite{Becirevic:2001xt,Babich:2006bh}.
The parameters $B_i(\mu_L)$ ($i=1,\dots,5$) are given in the basis used by 
Ciuchini et al \cite{Ciuchini:1997bw}.
Both papers provide the values of the parameters $B_i$ in the NDR scheme used in \cite{Buras:2001ra}.
The conversion to our operator basis (\ref{normalK}) is given by
\be
B_1^{\rm VLL}(\mu_L)=B_1^{\rm VRR}=B_1(\mu_L)\,,
\ee
\be
B_1^{\rm LR}(\mu_L)=B_5(\mu_L)\,, \qquad B_2^{\rm LR}(\mu_L)=B_4(\mu_L)\,,
\ee
\be
B_1^{\rm SLL}(\mu_L)=B_2(\mu_L), 
\qquad B_2^{\rm SLL}(\mu_L)=\frac{5}{3}B_2(\mu_L)-\frac{2}{3}B_3(\mu_L).
\ee
The parameters $B_1^{\rm VLL}$ for all meson systems are also known from most recent lattice simulations.
In this case we use the RG invariant parameters $\hat B_1^{\rm VLL}$, usually denoted by $\hat B_K$ and 
$\hat B_{B_q}$, that are already known from the SM analyses. As  these parameters are the same for the VRR contributions,
we can represent the latter as corrections to the SM box function 
$S_0(x_t)$. In this manner the VLL and VRR contributions are governed by 
 meson system dependent functions
\begin{equation} 
S_i=S_0(x_t) + \Delta S_i\, \qquad (i=K,B_d,B_s)\,
\end{equation}
where the $\Delta S_i$ are obtained by evolving the Wilson coefficients from $\mu_R$ 
down to $\mu_t$. 
The formula for $S_i$ is given in (\ref{AJB3}).

{If the unknown $\ord(\alpha_s)$ corrections to the Wilson coefficients of non-standard operators are assumed to be small, our analysis involves only the values of the coefficients
$P_2^{\rm LR}(K)$, $P_2^{\rm LR}(B_d)$ and $P_2^{\rm LR}(B_s)$ calculated at 
$\mu_R=2.5\tev$ in the case of box diagrams and $\mu_{H}=15\tev$ in the case 
of neutral Higgs contributions.} To obtain these values we only need the values 
of $B_4$ and $B_5$.
In the case of the $K$ system for $\mu_L=2$~GeV we have \cite{Babich:2006bh} (see table 16 in 
that paper)
\be
B_4 =  0.810(41)(31)\,, \quad B_5 =  0.562(39)(46)\,.
\ee
Within the uncertainties we can take the same values for $B_d$ and 
$B_s$ systems. In this case for $\mu_L=\mu_b=4.6$GeV we have \cite{Becirevic:2001xt}
\be
B_4 = 1.14(3)(6)\,, \quad B_5 = 1.79(4)(18)\,.
\ee
with the asymmetric errors corrected using \cite{D'Agostini:2004yu}.
In the case of box diagram contributions ($\mu_R=2.5\tev$) we find
\be\label{PI1}
P_2^{\rm LR}(K)= 73(4)(3)\,, \quad P_2^{\rm LR}(B_q)= 4.57(54)(25)\,.\quad\quad {(\rm Box)}
\ee
For the Higgs contribution ($\mu_{H}=15\tev$) we find
\be\label{PI2}
P_2^{\rm LR}(K)= 88(5)(3)\,, \quad P_2^{\rm LR}(B_q)= 5.54(65)(30)\,. \quad\quad {(\rm Higgs)}
\ee
For this calculation we have used the values of
$m_s$ and $m_d$ provided by the lattice-averaging-group. We collect them in table \ref{tab:runningmasses}.
The values relevant for VLL and VRR operators are also given in 
table~\ref{tab:runningmasses} and the final formulae for the mixing observables 
where all these matrix elements enter are presented below.

\subsection{Final expressions for mixing amplitudes}\label{sec:mixing_amplitudes}
We now summarise the expressions for the mixing amplitudes $M_{12}^q$, defined 
in terms of the effective Hamiltonian  
by
\be
2m_{B_q}\left(M_{12}^q\right)^\ast=\langle\bar B_q^0|\Heff^{\Delta B=2}|B_q^0\rangle\,.
\label{eq:3.23new}
\ee
We decompose them first as follows
\be\label{AJB1}
M_{12}^q=(M_{12}^{q})_{\rm SM}+(M_{12}^{q})_{\rm RR}+(M_{12}^{q})_{\rm LR}\equiv
\overline{(M_{12}^{q})}_{\rm SM}+(M_{12}^{q})_{\rm LR}\,.
\ee
Then 
\be\label{AJB2}
\overline{(M_{12}^{q})}_{\rm SM}=\frac{G_F^2}{12\pi^2}F^2_{B_q}\hat B_{B_q}m_{B_q}
M^2_W\left[\lambda_t^{{\rm LL}*}(B_q)\right]^2\eta_B S_q^*(B_q)\,,
\ee
where
\be\label{AJB3}
S_q(B_q)=S_0(x_t)+\frac{\tilde\eta_B}{\eta_B}
\frac{\Delta_{\rm Box}C_1^{\rm VRR}(\mu_R,B_q)}{\left[\lambda_t^{\rm LL}(B_q)\right]^2}\,.
\ee
Here $\eta_B$ is the known SM QCD correction and $\tilde\eta_B/\eta_B \sim 0.95$ describes the QCD evolution from $\mu_R$ down 
to $\mu_W$ and is therefore the same for the $K^0-\bar K^0$ system. 
$S_0(x_t)$ is given in (\ref{S0}) below.

For the LR contribution we first combine the Higgs contributions 
in (\ref{HVLRK}) and (\ref{VLRKH})
into
\be
\tilde\Delta_{\rm Higgs}C_2^{\rm LR}(\mu_R,B_q)=\Delta_{\rm H^0}C_2^{\rm LR}(\mu_H,B_q) + \Delta_{\rm H^+}C_2^{\rm LR}(\mu_H,B_q)\,.
\ee
Then 
\begin{eqnarray}\label{AJB4}
(M_{12}^{q})_{\rm LR}&=&\frac{G_F^2 M^2_W}{12\pi^2}F^2_{B_q}m_{B_q}\left[(\Delta_{\rm Box}C_2^{\rm LR}(\mu_R,B_q))^*P_2^{\rm LR}(\mu_R)+ \right.\\
\nonumber &&\left.(\tilde\Delta_{\rm Higgs}C_2^{\rm LR}(\mu_H,B_q))^*P_2^{\rm LR}(\mu_H)\right]\,.
\end{eqnarray}
In these expressions $\mu_R=\ord(M_{W_R})$ and $\mu_H=\ord(M_{H})$.
In the case of $K^0-\bar K^0$ system $B_q$ should be replaced by $K$ and 
$\eta_B$ by $\eta_2$. Moreover one should add the known contributions from $cc$ and $ct$ box diagrams 
to $(M_{12}^{K})_{\rm SM}$ so that
\begin{eqnarray}
(M_{12}^{K})_{\rm SM}&=& \frac{G_{\rm F}^2}{12 \pi^2} F_K^2 \hat B_K m_K M_W^2
\left[ [\lambda_{c}^{{\rm LL}*}(K)]^2 \eta_1 S_0(x_c) + 
[\lambda_{t}^{{\rm LL}*}(K)]^2 \eta_2 S_0(x_t) + \right.\\
\nonumber &&\left.
2 \lambda_{c}^{{\rm LL}*}(K) \lambda_{t}^{{\rm LL}*}(K) \eta_3 S_0(x_c, x_t) \right],
\label{eq:M12K}
\end{eqnarray}
where $F_K$ is the $K$-meson decay constant and $m_K$
the $K$-meson mass. Here
\bea\label{S0}
S_0(x_t)& \equiv& S_{\rm LL}(x_t,x_t)=\frac{4x_t-11x^2_t+x^3_t}{4(1-x_t)^2}-
 \frac{3x^3_t \ln x_t}{2(1-x_t)^3}\,,\\
\label{BFF1}
S_0(x_c)&\equiv& S_{\rm LL}(x_c,x_c)\approx x_c\,,\\
\label{BFF}
S_0(x_c, x_t)&\equiv& S_{\rm LL}(x_t,x_c)\approx 
x_c\left[\ln\frac{x_t}{x_c}-\frac{3x_t}{4(1-x_t)}-
 \frac{3 x^2_t\ln x_t}{4(1-x_t)^2}\right]\,.
\eea
In the last two expressions we have kept only linear terms in $x_c\ll 1$, 
but of course all orders in $x_t$.

\subsection{General anatomy of LR contributions}\label{sec:lr_gen_anatomy}
In order to see the importance of different NP contributions we rewrite the 
dominant LR contribution in (\ref{AJB4}) by separating contributions of quark mixing
matrices from the loop integral and QCD running. We obtain
\be\label{AJB5}
(M_{12}^{q})_{\rm LR}=\frac{G_F^2 M^2_W}{12\pi^2}F^2_{B_q}m_{B_q}
\sum_{i,j=u,c,t}\Lambda_{ij}(B_q)^* R_{ij}(B_q)\,,
\ee
where we defined
\begin{align}
\nonumber R_{ij}(B_q) &=\,S_{\rm LR}(x_i,x_j,\beta)P_2^{\rm LR}(B_q,\mu_R)\\
&+\, S^{\rm H}_{\rm LR}(x_i,x_j,\beta_H) P_2^{\rm LR}(B_q,\mu_H) \label{Rij}\\
\nonumber &-\,\frac{16\pi^2}{\sqrt{2} M_H^2 G_F} u(s) \sqrt{x_i(\mu_H)x_j(\mu_H)} P_2^{\rm LR}(B_q,\mu_H)\,,\\
\Lambda_{ij}(B_q) &= \lambda_i^{\rm LR}(B_q)\lambda_j^{\rm RL}(B_q)\,.
\end{align}
Here we indicated that the QCD factors $P_2^{\rm LR}$ are the ones for the $B_q$ system.
In the case of $K^0-\bar K^0$ system $B_q$ should be replaced by $K$. Choosing $M_{W_R}=2.5\tev$, $M_H=16\tev\cdot u(s)^{1/4}$ and the central 
values for the factors $P_i^a$ given above, we find for the matrices 
$\hat R(K)$ and $\hat R(B_q)$ the results collected in appendix \ref{app:num_details_deltaf2}, for two different choices of $s$.
This hierarchical structure of the matrix $R_{ij}$ has an impact on the 
resulting structure of the mixing matrix $V^{\rm R}$.
We discuss this in more detail in section \ref{sec:lr_spe_anatomy}.

\begin{figure}
\centering
\includegraphics[width=0.78\textwidth]{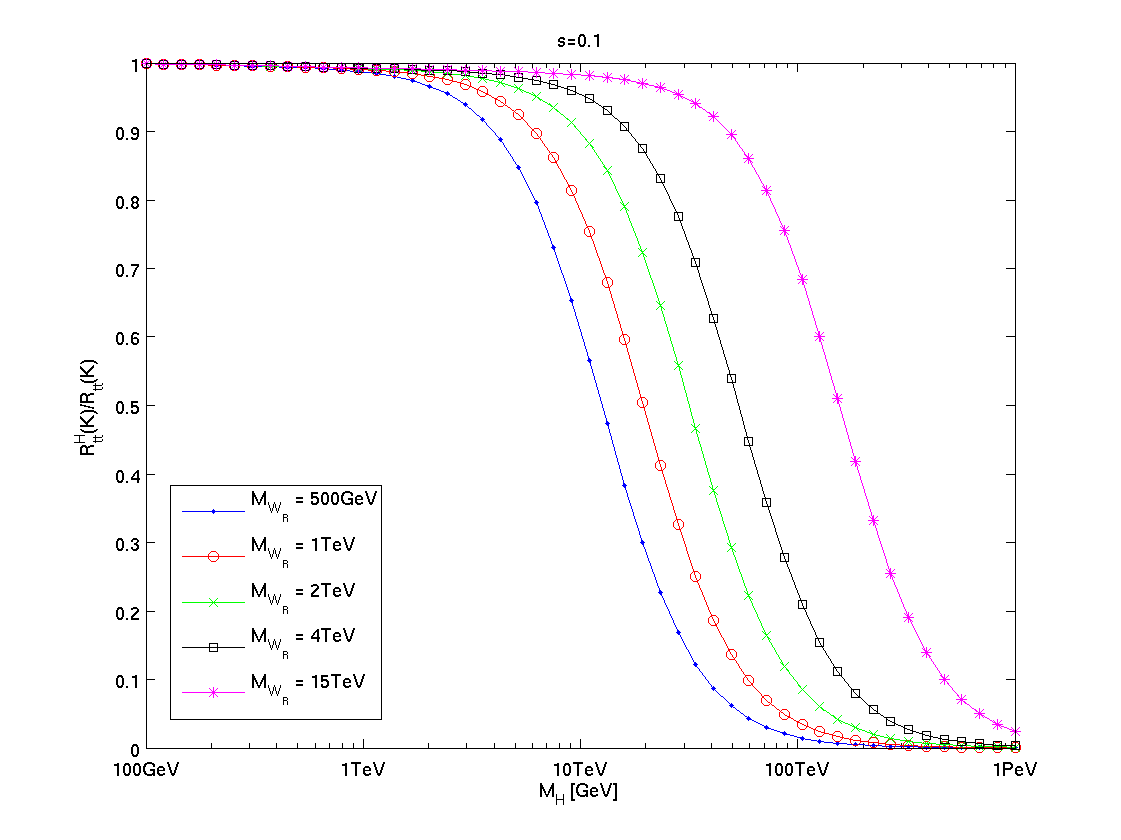}
\includegraphics[width=0.78\textwidth]{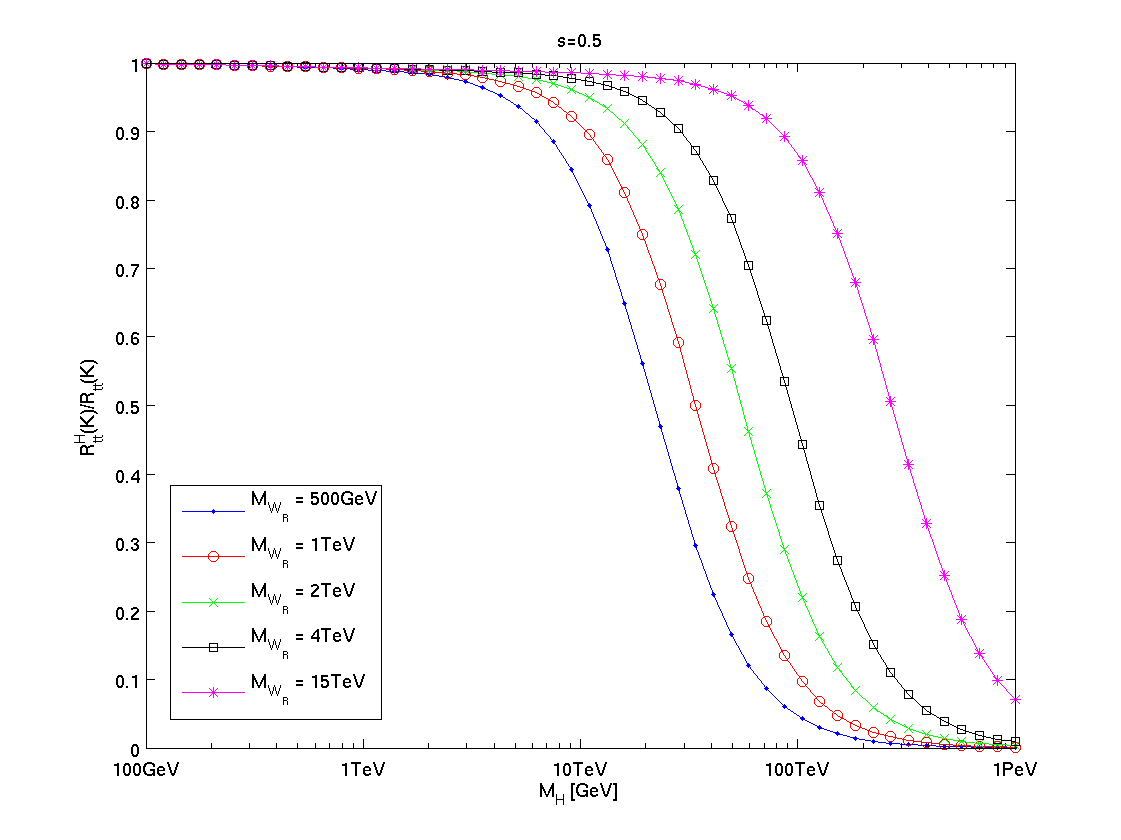}
\caption{The relative importance of the neutral Higgs contribution $R^H_{tt}$ 
(see \ref{Rij}) to $R_{tt}(K)$ as a function of $M_H$ for different $W_{R}$ masses for $s=0.1$ and $s=0.5$}\label{fig:lrcontrib}
\end{figure}
We observe that for fixed $\lambda_i^{\rm LR}\lambda_j^{\rm RL}$ the neutral Higgs 
$H^0$ contributions are by far dominant, followed by gauge boson contributions 
and rather small charged Higgs $H^\pm$ contributions. This shows that 
for $M_{W_R}\ge 2.5\tev$ the neglect of neutral Higgs contributions for 
masses $M_H$ even as high as $16\tev$, as done often in the literature, 
totally misrepresents the LR story in $\Delta F=2$ transitions. In 
figure \ref{fig:lrcontrib} we show the relative importance of the neutral Higgs 
contributions to $R_{tt}(K)$ as a function of $M_H$ for different values of $M_{W_R}$ {and two different choices for $s$.
We observe that even for $M_{W_R} = 400\,{\rm GeV}$, which is excluded 
already for many years, and small $s$ the neutral Higgs contributions 
for $M_H < 20\,{\rm TeV}$ account for at least $20\%$ of the total value.
For realistic $W_R$ masses $M_{W_R} > 2\,{\rm TeV}$ the neutral Higgs
contributions become only negligible for masses beyond $100\,{\rm TeV}$. 
For large values of $s\sim 0.5$ the heavy Higgs contribution becomes even more important.
As both $M_{W_R}$ and $M_H$ are $\kappa_R$ dependent decoupling the heavy Higgs contribution would require 
a non-perturbative coupling $\alpha_3$ in the Higgs sector.}

{\subsection{Special anatomy of LR contributions}\label{sec:lr_spe_anatomy}

As can be seen from the formulae in appendix \ref{app:num_details_deltaf2} the matrices $\hat R(q)$ have
a very special pattern. We first restrict our discussion to the $K$ system. Similar arguments hold
for the $B$ system but the actual hierarchies change. For now we are only interested in the order of magnitude of the elements
and especially in their relative size
\begin{equation}\label{eqn:FK}
\hat R(K) \sim 
(-1)\left(
\begin{array}{ccc}
  10^{-9} & 10^{-6} & 10^{-4}\\
  10^{-6} & 10^{-4} & 10^{-2}\\
  10^{-4} & 10^{-2} & 10^{1}
\end{array}
\right)\,.
\end{equation}
First of all we notice the wide spread of sizes as well as the expected hierarchy
towards the third generation. The relative shift for changing one generation is
about two-to-three orders of magnitude. 
Before summation the elements of $\hat R(K)$ get dressed with elements 
of $\hat\Lambda(K)$ as explicitly given in (\ref{AJB5}).
Here we define $\hat\Lambda(K)$ as the matrix form of $\Lambda_{ij}(K)$. 

Barring fine-tuned cancellations among the various contributions, each of the LR contributions to $\Delta M_K$ and in particular $\eps_K$ has to be suppressed well below the SM contribution in order to obtain agreement with the data. Thus 
we can estimate rough upper bounds on some of the elements of   $\hat\Lambda(K)$
\begin{eqnarray}
\label{eq:LttK}|\Lambda_{tt}(K)|	 &\lsim& 10^{-5} \,,\\
|\Lambda_{ct,tc}(K)| & \lsim & 10^{-3} \,,\\
\label{eq:LccK}|\Lambda_{cc}(K)| &\lsim & 10^{-1}\,,
\end{eqnarray}
while no useful bound can be obtained on the remaining elements of $\hat\Lambda(K)$.

On the other hand for a generic right-handed matrix $V^{\rm R}$ the matrix $\hat\Lambda(K)$ would exhibit the hierarchy implied by the structure of the CKM matrix only 
\be\label{eq:natLambda}
|\hat\Lambda(K)| \sim \begin{pmatrix}
 10^{-1} & 10^{-2} & 10^{-3} \\
 1 & 10^{-1} & 10^{-2} \\
 10^{-2} & 10^{-2} & 10^{-4}
\end{pmatrix}\,.
\ee 
Recall that
\be\label{eq:LijK}
\Lambda_{ij}(K) = V^{\text{L}*}_{is} V^\text{R}_{id} V^{\text{R}*}_{js} V^\text{L}_{jd} \,.
\ee

Comparing the entries of \eqref{eq:natLambda} with the bounds obtained in 
\eqref{eq:LttK}--\eqref{eq:LccK} we see that the $ct$, $tc$ and $tt$ elements need an additional suppression by one order of magnitude. This can only be achieved with the help of an appropriate hierarchy in  $V^{\rm R}$, see \eqref{eq:LijK}. Explicitly we find the constraints
\be
|V_{td}^\text{R}||V_{ts}^\text{R}|\lsim 10^{-1} \,,\qquad |V_{td}^\text{R}||V_{cs}^\text{R}|\lsim 10^{-1} \,,\qquad  |V_{cd}^\text{R}||V_{ts}^\text{R}|\lsim 10^{-1} \,.
\ee
Following an analogous procedure for the $B_d$ system we obtain 
\be
|V_{td}^\text{R}||V_{tb}^\text{R}|\lsim 10^{-2} \,,
\ee
while no relevant constraint can be obtained from the $B_s$ system.

Making the plausible assumption that the diagonal elements of $V^{\rm R}$ and in particular   $|V_{tb}^\text{R}|$ are $\ord(1)$ we obtain
\be
|V_{td}^\text{R}| \lsim 10^{-2}\,.
\ee
This bound agrees surprisingly well with the one obtained from the explicit numerical analysis in section \ref{sec:vr_gen_num}. We stress however that we have performed here a very rough estimate, keeping only the orders of magnitude. It is interesting to see that already this naive estimate allows us to understand certain patterns in our numerical analysis.

}

\subsection{Comparison of the operator structure in various models}\label{sec:comops}
It is instructive to compare the operator structure in the effective 
Hamiltonian for \linebreak $\Delta F=2$ transitions in the specific LR model considered 
in the present paper with the other specific models containing RH 
currents considered by us like RSc models \cite{Blanke:2008zb} or some supersymmetric 
flavour models \cite{Altmannshofer:2009ne} and in particular in
the effective theory approach for the RH currents in \cite{Buras:2010pz}. To this end let us 
note that when $\ord(\alpha_s)$ corrections at the high matching scale are neglected the 
following dynamics is responsible for the structure of the effective 
$\Delta F=2$ Hamiltonian:
\begin{itemize}
\item
A tree level exchange of a colourless gauge boson with LH and RH couplings 
generates the operators $Q_1^{\rm VLL}$, $Q_1^{\rm VRR}$ and $Q_1^{\rm LR}$. This is an 
example of $Z^\prime$ models and  gauge flavour models \cite{Grinstein:2010ve}.
\item
A tree level exchange of  a gauge boson carrying colour generates the operators
$Q_1^{\rm VLL}$, $Q_1^{\rm VRR}$, $Q_1^{\rm LR}$ and $Q_2^{\rm LR}$. An example is the 
tree-level exchange of the KK-gluon in RS models.
\item
A tree level exchange of a colourless Higgs scalar generates the 
operators  $Q_1^{\rm SLL}$, $Q_1^{\rm SRR}$ and $Q_2^{\rm LR}$ but as we have 
seen above at $\ord(\epsilon^2)$ only the last operator contributes
in the model considered.
\item
A tree level exchange of a Higgs scalar carrying colour generates the operators
$Q_{1,2}^{\rm SLL}$, $Q_{1,2}^{\rm SRR}$ and $Q_{1,2}^{\rm LR}$.
\item
Finally box diagrams with internal charged gauge bosons or $H^\pm$ 
carrying both LH and RH couplings generate the operators
$Q_1^{\rm VLL}$, $Q_1^{\rm VRR}$ and $Q_2^{\rm LR}$ at $\ord(\epsilon^2)$ .
\end{itemize}

With this classification in mind it is evident that  
the effective $\Delta F=2$ Hamiltonian at the matching scale in \cite{Buras:2010pz} 
corresponds to a tree level exchange of a colourless gauge boson. 
Clearly with RH currents present in this model also the operator  
$Q_2^{\rm LR}$ is generated through box diagrams, but in the 
presence of QCD corrections this operator is also generated from 
$Q_1^{\rm LR}$ generated by the tree level gauge boson exchange in 
question. An implicit assumption in \cite{Buras:2010pz} 
was that this is the dominant 
mechanism for the generation of  $Q_2^{\rm LR}$.

In the models considered by us there are no flavour changing 
neutral gauge boson exchanges at tree level and the leading 
mechanism for the generation of  $Q_2^{\rm LR}$ are box diagrams. 
The latter can also generate  $Q_1^{\rm LR}$ at $\ord(\epsilon^4)$ 
but this effect is smaller than the QCD mixing  generating 
$Q_1^{\rm LR}$ from $Q_2^{\rm LR}$ that we include in our paper. Thus the 
structures of the $\Delta F=2$ Hamiltonians considered here 
and in  \cite{Buras:2010pz} are in a sense complementary to each other.

\subsection[Basic formulae for $\Delta F=2$ observables]{\boldmath Basic formulae for $\Delta F=2$ observables}
We collect here the formulae that we used in our numerical analysis.
The mixing amplitude $M_{12}$ can be decomposed into SM and NP part
$(i=K,d,s)$
\be
M^i_{12}=\left(M_{12}^i\right)_\text{\rm SM}+\left(M_{12}^i\right)_\text{NP}\,,
\ee
and is related for $i=K$ to the relevant effective Hamiltonian through
\be
2m_K\left(M_{12}^K\right)^\ast=\langle\bar K^0|\Heff^{\Delta S=2}|K^0\rangle\,,
\label{eq:3.23}
\ee
with analogous expressions for $q=d,s$. A general formula for the r.\,h.\,s.\ is 
given in (\ref{amp6}). The $K_L-K_S$ mass difference is then given by
\be
\Delta M_K=2\left[\Re\left(M_{12}^K\right)_\text{\rm SM}+\Re\left(M_{12}^K\right)_\text{NP}\right]\,,
\label{eq:3.34}
\ee
and the CP-violating parameter $\varepsilon_K$ by
\be
\varepsilon_K=\frac{\kappa_\eps e^{i\varphi_\eps}}{\sqrt{2}(\Delta M_K)_\text{exp}}\left[\Im\left(M_{12}^K\right)_\text{\rm SM}+\Im\left(M_{12}^K\right)_\text{NP}\right]\,,
\label{eq:3.35}
\ee
where $\varphi_\eps = (43.51\pm0.05)^\circ$ and $\kappa_\eps=0.94\pm0.02$ \cite{Buras:2008nn,Buras:2010pza} takes into account that $\varphi_\eps\ne \pi/4$ and includes long distance effects in $\Im \Gamma_{12}$ and $\Im M_{12}$. 
The value of $\kappa_\eps$ given here has been calculated within the SM 
using the data on $\varepsilon'/\varepsilon$ that could also contain NP contributions. As analysed in \cite{Buras:2009pj} these 
effects do not have  a significant impact on our analysis.
For the mass differences in the $B_{d,s}^0-\bar B_{d,s}^0$ systems we have
\be
\Delta M_q=2\left|\left(M_{12}^q\right)_\text{\rm SM}+\left(M_{12}^q\right)_\text{NP}\right|\qquad (q=d,s)\,.
\label{eq:3.36}
\ee
Let us then write \cite{Bona:2005eu}
\be
M_{12}^q=\left(M_{12}^q\right)_\text{\rm SM}+\left(M_{12}^q\right)_\text{NP}=\left(M_{12}^q\right)_\text{\rm SM}C_{B_q}e^{2i\varphi_{B_q}}\,,
\label{eq:3.37}
\ee
where
\begin{align}
& (M_{12}^d)_\text{\rm SM}=\big|(M_{12}^d)_\text{\rm SM}\big|e^{2i\beta}\,, &&\beta\approx 22^\circ\,,
\label{eq:3.38} \\
& (M_{12}^s)_\text{\rm SM}=\big|(M_{12}^s)_\text{\rm SM}\big|e^{2i\beta_s}\,,&&\beta_s\simeq -1^\circ\,.
\label{eq:3.39}
\end{align}
Here the phases $\beta$ and $\beta_s$ are defined through
\be
V_{td}=|V_{td}|e^{-i\beta}\quad\textrm{and}\quad V_{ts}=-|V_{ts}|e^{-i\beta_s}\,.
\label{eq:3.40}
\ee
We find then
\be
\Delta M_q=(\Delta M_q)_\text{\rm SM}C_{B_q}\,,
\label{eq:3.41}
\ee
and
\bea
S_{\psi K_S} &=& \sin(2\beta+2\varphi_{B_d})\,,
\label{eq:3.42} \\
S_{\psi\phi} &= & \sin(2|\beta_s|-2\varphi_{B_s})\,,
\label{eq:3.43}
\eea
with the latter two observables being the coefficients of $\sin(\Delta M_d t)$ and $\sin(\Delta M_s t)$ in the 
time dependent asymmetries in $B_d^0\to\psi K_S$ and $B_s^0\to\psi\phi$, respectively. Thus in the presence of 
non-vanishing $\varphi_{B_d}$ and $\varphi_{B_s}$ these two asymmetries do not measure $\beta$ and $\beta_s$ 
but $(\beta+\varphi_{B_d})$ and $(|\beta_s|-\varphi_{B_s})$, respectively.
At this stage a few comments on the assumptions leading to expressions (\ref{eq:3.42}) and (\ref{eq:3.43}) are 
in order. These simple formulae follow only if there are no weak phases in the decay amplitudes for 
$B_d^0\to\psi K_S$ and $B_s^0\to\psi\phi$ as is the case in the SM and also in the LHT model, where due to 
T-parity there are no new contributions to decay amplitudes at tree level so that these amplitudes are
dominated by SM contributions \cite{Blanke:2006sb}. In the model discussed in the present paper new contributions 
to decay amplitudes with non-vanishing weak phases are present at tree level. However, as we demonstrate in 
section~\ref{sec:tree} these contribution can be totally neglected when calculating $S_{\psi K_S}$ and 
$S_{\psi\phi}$. 

Now in models like the LHT model and SM4, the only operators contributing to the amplitudes $M_{12}^K$ and $M_{12}^q$ 
are the SM ones, that is $Q_1^{\rm VLL}$ \cite{Blanke:2006sb,Buras:2010pi}.
Consequently the new phases $\varphi_{B_d}$ and $\varphi_{B_s}$ have purely perturbative character related to the 
fundamental dynamics at short distance scales.
The situation in the LR model in question is different. As now new operators contribute to the $M_{12}^q$ amplitudes, 
the parameters $C_{B_q}$ and $\varphi_{B_q}$ in (\ref{eq:3.37}) are complicated functions of fundamental short distance 
parameters of the model and of the non-perturbative parameters $B_i$ present in $P_i^a(K)$ and $P_i^a(B_d)$. 
Thus the test of the LR models considered with the help of particle-antiparticle mixing and related CP-violation is 
less theoretically clean than in the case of new physics scenarios in which only the operator $Q_1^{\rm VLL}$
contributes. 

Next, we give the expressions for the width differences $\Delta\Gamma_q$ and the semileptonic CP-asymmetries $A_\text{SL}^q$
\bea
\frac{\Delta\Gamma_q}{\Gamma_q}&=& -\left(\frac{\Delta   M_q}{\Gamma_q}\right)^\text{exp}\,\left[\text{Re}\left(\frac{\Gamma^q_{12}}{M^q_{12}}\right)^\text{\rm SM}\frac{\cos{2\varphi_{B_q}}}{C_{B_q}}+
\text{Im}\left(\frac{\Gamma^q_{12}}{M^q_{12}}\right)^\text{\rm SM}\frac{\sin{2\varphi_{B_q}}}{C_{B_q}}\right]\,,
\label{eq:3.44}\\
A_\text{SL}^q&=&\text{Im}\left(\frac{\Gamma^q_{12}}{M^q_{12}}
\right)^\text{\rm SM}\frac{\cos{2\varphi_{B_q}}}{C_{B_q}}-
\text{Re}\left(\frac{\Gamma^q_{12}}{M^q_{12}}\right)^\text{\rm SM}\frac{\sin{2\varphi_{B_q}}}{C_{B_q}}\,.
\label{eq:3.45}
\eea
Theoretical predictions of both $\Delta\Gamma_q$ and $A_\text{SL}^q$ require the non-perturbative calculation of the off-diagonal matrix element $\Gamma_{12}^q$, the absorptive part of the $B_q^0-\bar B_q^0$ amplitude 
as well as perturbative QCD calculations. The latter are known at the 
NLO level \cite{Beneke:1998sy,Beneke:2002rj,Beneke:2003az,Ciuchini:2001vx,Ciuchini:2003ww}. The most recent results read \cite{Lenz:2011ti,Lenz:2011zz}
\begin{gather}
\Re\left
(\frac{\Gamma_{12}^d}{M_{12}^d} \right)^\text{\rm SM} = -5.3(10)\cdot10^{-3}\,,\qquad
\Re\left
(\frac{\Gamma_{12}^s}{M_{12}^s} \right)^\text{\rm SM} = -5.0(10)\cdot10^{-3} \,,\label{eq:r2}\\
\Im\left
(\frac{\Gamma_{12}^d}{M_{12}^d} \right)^\text{\rm SM} = -4.1(6)\cdot
10^{-4}\,,\qquad \Im\left
(\frac{\Gamma_{12}^s}{M_{12}^s} \right)^\text{\rm SM} = 1.9(3)\cdot 10^{-5}\,.\label{eq:r1}
\end{gather}

\begin{table}[htbp]
\renewcommand{\arraystretch}{1.4}
\centering
\begin{tabular}{|l|c|c|c|}
\hline 
{observable} & experimental value & SM prediction\\
\hline \hline
$ \phi_s = -2(\beta_s+\varphi_{B_s})$ & $\in [-1.04,-0.04]$ (CDF \cite{Giurgiu:2010is}) & -0.0363(17) \cite{Charles:2004jd}\\
 & $-0.55^{+0.38}_{-0.36}$  (D0 \cite{Abazov:2011ry}) & \\
 & +0.13(18)(7) (LHCb \cite{Raven:1378074}) &  \\
{  $ \frac{\Delta \Gamma_{d}}{\Gamma_d}$  } & $0.011(37)$ \cite{Asner:2010qj} & 0.0042(8) \cite{Lenz:2011ti}\\
{  $ \Delta \Gamma_{s}$  } & $0.075(35)(1)\,\text{ps}^{-1}$ (CDF \cite{Giurgiu:2010is}) & $0.087(21)\,\text{ps}^{-1}$ \cite{Lenz:2011ti} \\
& $0.163^{+0.065}_{-0.064}\,\text{ps}^{-1} $ (D0 \cite{Abazov:2011ry}) &    \\
 & $0.123(29)(8)\,\text{ps}^{-1} $ (LHCb \cite{Raven:1378074}) & \\
\hline
{  $A_\text{SL}^d$  } & $-0.12(52) \%$  \cite{Abazov:2011yk}
& $ -0.041(6)\%$ \cite{Lenz:2011zz} \\
{ $A_\text{SL}^s$  } & $-1.8(11) \%$ \cite{Abazov:2011yk}
& $0.0019(3)\%$ \cite{Lenz:2011zz} \\
{ $A_\text{SL}^b$  } & $-0.79(20)\%$ \cite{Abazov:2011yk} & $-0.020(3)\%$ \cite{Lenz:2011zz} \\
\hline
\end{tabular}
\caption{\label{tab:DeltaF2SM&exp} Theoretical and experimental values of a number of observables related to $B_{s,d}-\bar B_{s,d}$ mixing.}
\renewcommand{\arraystretch}{1.0}
\end{table}

Finally, we recall the existence of a correlation between $A_\text{SL}^s$ and $S_{\psi\phi}$ that has been pointed out in~\cite{Ligeti:2006pm} and which has been investigated model-independently in \cite{Blanke:2006ig} and in the context of the LHT model in~\cite{Blanke:2006sb}. This correlation follows analytically 
from (\ref{eq:3.43}) and (\ref{eq:3.45}) when the SM phase $\beta_s$ and the first term in (\ref{eq:3.45}) are neglected:
\be
A_\text{SL}^q=
\text{Re}\left(\frac{\Gamma^q_{12}}{M^q_{12}}\right)^\text{\rm SM}
\frac{S_{\psi\phi}}{C_{B_q}}\,.
\label{eq:3.45a}
\ee
In \cite{Lenz:2011zz} it has been pointed out recently that 
the accuracy of a similar correlation that uses $\Delta M_s$ and 
$\Delta\Gamma_s$ \cite{Grossman:2009mn} instead of $\Gamma^s_{12}$ and $M^s_{12}$ is very poor both 
for small and large NP phase $\varphi_{B_s}$. The approximate formula 
(\ref{eq:3.45a}) is instead very accurate for large  $S_{\psi\phi}$. 
In order to improve the accuracy also for small values of this 
asymmetry, in our numerical analysis as in our previous 
analyses in the context of other extensions of the SM, we find such correlation 
numerically by using (\ref{eq:3.43}) and (\ref{eq:3.45}) without making any approximations.

\subsection{Summary}

In summary, in this section, we have calculated the NP contributions in the LR model 
in question to the amplitudes $M_{12}^K$ and $M_{12}^q$.
We have then given formulae for $\Delta M_K$, $\Delta M_q$, $\varepsilon_K$, $S_{\psi K_S}$,
$S_{\psi\phi}$, $\Delta \Gamma_q$ and $A^q_\text{SL}$ in a form
suitable for the study of the size of the NP contributions.
The numerical analysis of these observables is presented in
section~\ref{sec:num}. While particle-antiparticle mixing in LR models has  already been discussed in the literature, our analysis goes beyond these papers as in addition to 
the  full renormalisation group analysis and inclusion of all important effects, 
we search for correlations between various observables that have not been studied by other authors. Most importantly, our philosophy
in performing phenomenology differs from the one used in most papers. Instead 
for looking for bounds on the $W_R$ and Higgs masses we investigate whether 
the LRM can solve certain anomalies present in the flavour data while being 
consistent with electroweak precision tests and the data for tree level 
charged currents. Moreover, we search for the oases in the large space of 
parameters in which the matrix $V^{\rm R}$ takes  special forms that are dictated 
by the data.

\section{\boldmath The decays $B\to X_{s,d}\gamma$} \label{sec:BSG}
\subsection{Preliminaries}
The $B\to X_s\gamma$ decay in a model with $SU(2)_L \times SU(2)_R \times U(1)$ 
symmetry has been analysed by many authors in the past
\cite{Asatrian:1989iu,Asatryan:1990na,Cocolicchio:1988ac,Cho:1993zb,Babu:1993hx,Fujikawa:1993zu,Asatrian:1996as,Frank:2010qv,Guadagnoli:2011id}. There are 
basically two classes of contributions:
\begin{itemize}
\item
First, the ones resulting from the mixing between $W_L$ and $W_R$ that imply 
RH couplings of the SM $W^\pm$ to quarks. In the SM the LH structure of 
these couplings requires the chirality flip, necessary for $b\to s\gamma$
transition to occur, only through the mass of the initial or the final state 
quark. Consequently the amplitude is proportional to $m_b$ or $m_s$. 
In contrast in LR models the RH couplings allow the chirality flip on the 
internal top quark line resulting in an enhancement factor $m_t/m_b$ 
of the NP contribution relative to the SM one at the level of the amplitude.
This is the contribution mostly studied in the literature. The relevant 
LO QCD corrections have been analysed 
for the first  time 
within the effective field theory framework by Cho and Misiak \cite{Cho:1993zb} and have been 
checked since then by many authors, in particular by Bobeth et al. 
\cite{Bobeth:1999ww}, where 
also NLO QCD corrections to the matching conditions at $\mu_H$ 
have been calculated.
In what follows we adopt their results but include NP corrections 
at the LO, while taking into account the known NNLO corrections within the SM. 
\item
The second contribution comes from charged Higgs exchanges. Although in the LR 
models the masses of $H^\pm$ are $\ord(\kappa_R)$ and numerically significantly larger than few TeV, as pointed out 
in \cite{Babu:1993hx} and also analysed in \cite{Fujikawa:1993zu,Asatrian:1996as,Frank:2010qv}, 
the corresponding amplitude is also enhanced by $m_t/m_b$ in 
contrast to the MSSM where it is proportional to $m_b$ or $m_s$.
Moreover, it does not suffer from the suppression  through 
$W_L-W_R$ mixing as is the case of the gauge contribution. As we will see for charged Higgs 
masses even above $10\tev$ this contribution cannot be neglected and in fact it can be dominant for certain ranges of parameters. This should 
be contrasted with $\Delta F=2$ processes where it is generally subleading.
\end{itemize} 

In the next two sections, we summarise the results 
for the Wilson coefficients of the dipole operators 
for these two classes of contributions at the relevant matching scales 
for the SM and NP. Subsequently we include RG QCD corrections to 
these coefficients and present 
the final formula for the branching ratio for the $B\to X_s\gamma$ decay. 
We also 
present the formulae for the CP-averaged branching ratio of the 
$B\to X_d\gamma$ decay and direct CP-asymmetries in both decays.

\subsection{Gauge boson contributions}
Adopting the overall normalisation of the SM effective Hamiltonian  
we have
{\begin{equation} \label{Heff_at_mu}
{\cal H}_{\rm eff}(b\to s\gamma) = - \frac{4 G_{\rm F}}{\sqrt{2}} V_{ts}^* V_{tb}
\left[  C_{7\gamma}(\mu_b) Q_{7\gamma} +  C_{8G}(\mu_b) Q_{8G} \right]\,,
\end{equation}}
where $\mu_b=\ord(m_b)$.
The dipole operators are defined as
\begin{equation}\label{O6B}
Q_{7\gamma}  =  \frac{e}{16\pi^2} m_b \bar{s}_\alpha \sigma^{\mu\nu}
P_R b_\alpha F_{\mu\nu}\,,\qquad            
Q_{8G}     =  \frac{g_s}{16\pi^2} m_b \bar{s}_\alpha \sigma^{\mu\nu}
P_R T^a_{\alpha\beta} b_\beta G^a_{\mu\nu}\,. 
\end{equation}
In writing (\ref{Heff_at_mu}) we have dropped
the primed operators that are obtained from (\ref{O6B}) by  replacing $P_{R}$ 
by $P_L$. In the SM the primed operators (RL) are suppressed by $m_s/m_b$ 
relative to the ones in (\ref{Heff_at_mu}). As we demonstrate below they 
can also be neglected in the LRM discussed by us.

The coefficients $C_i(\mu_b)$ are calculated from their initial values at 
high energy scales by means of renormalisation group methods. Before 
entering the discussion of QCD corrections we describe here our treatment 
of LR contributions at high energy scales.
We first decompose the Wilson coefficients at the scale $\mu_W=\ord(M_W)$ 
as the sum of the SM contribution and the NP contributions: 
\be
C_i(\mu_W)=C_i^{\rm SM}(\mu_W)+\Delta^{\rm LR}C_i(\mu_W)\label{cstart}\,
\ee
and similarly for the primed coefficients. For the SM coefficients we have
\begin{equation}\label{c7}
C^{\rm SM}_{7\gamma} (\mu_W) = \frac{3 x_t^3-2 x_t^2}{4(x_t-1)^4}\ln x_t + 
\frac{-8 x_t^3 - 5 x_t^2 + 7 x_t}{24(x_t-1)^3}\equiv C^{\rm SM}_{7\gamma} (x_t)\,,
\end{equation}
\begin{equation}\label{c8}
C^{\rm SM}_{8G}(\mu_W) = \frac{-3 x_t^2}{4(x_t-1)^4}\ln x_t +
\frac{-x_t^3 + 5 x_t^2 + 2 x_t}{8(x_t-1)^3}\,.                               
\end{equation}

The expressions for $\Delta^{\rm LR}C_i(\mu_W)$ in the LRM have been found 
by Cho and Misiak \cite{Cho:1993zb}. The by far dominant contribution 
comes from the induced right-handed part of the $W_L$ vertex. At the time of 
the work of these authors the expected values for $M_{W_R}$ were of the order 
of several hundred GeV and the scale in the LR contributions could be chosen 
to be $\mu_W$. With $\mu_R\gg\mu_W$ one has to take the effect of large 
logarithms $\log(\mu_R/\mu_W)$ into account. While a complete RG analysis 
would be more involved, in the present paper we take such effects only 
approximately into account by simply declaring the result in  \cite{Cho:1993zb}
to be valid not at $\mu_W$ but $\mu_R$. A more involved analysis will be 
presented elsewhere.

Adapting the formulae of Cho and Misiak \cite{Cho:1993zb} to our notations we find then for the LR contributions
\be\label{c7LR}
\Delta^{\rm LR}C_{7\gamma}(\mu_R) =A^{tb} \left[ \frac{3x_t^2-2x_t}{2(1-x_t)^3} \ln x_t 
\;+\; \frac{-5x_t^2+31x_t-20}{12(1-x_t)^2} \right]\,,
\ee
\be\label{c8LR}
\Delta^{\rm LR}C_{8G}(\mu_R)= A^{tb} \left[ \frac{-3x_t}{2(1-x_t)^3} \ln x_t 
\;+\; \frac{-x_t^2-x_t-4}{4(1-x_t)^2} \right]\,,
\ee
where 
\be\label{tb}
A^{tb}= \frac{m_t}{m_b} \; sc\epsilon^2 e^{i\alpha}
\left(\frac{V^{\rm R}_{tb}}{V^{\rm L}_{tb}}\right)+ {\cal O}(\epsilon^4)\,.        
\ee
The Wilson coefficients of the primed 
operators can be obtained from  \eqref{c7LR}, \eqref{c8LR} by replacing $A^{tb}$ with $(A^{ts})^*$,
where
\be\label{ts}
A^{ts}= \frac{m_t}{m_b} \; sc \epsilon^2 e^{i\alpha}
\left(\frac{V^{\rm R}_{ts}}{V^{\rm L}_{ts}}\right) + {\cal O}(\epsilon^4)\,.
\ee
We observe that they are also enhanced by $m_t/m_b$ in contrast to the 
primed operators in the SM. We stress that $m_t(m_t)$ and $m_b(\mu_b)$ 
should be used here.
We now give arguments that in order to obtain the leading 
$\ord(\epsilon^2)$ corrections to the branching ratio we only 
have to keep the contributions in (\ref{c7LR}) and (\ref{c8LR}) while 
neglecting the contributions from primed operators and the LL and RR 
contributions from NP. Here the LL stands for pure $P_L$ couplings in the weak 
gauge boson--quark couplings and analogously for RR with $P_L$ replaced 
by $P_R$. The unprimed LR contributions result from the $P_R$ coupling 
in the vertex containing the $b$-quark, while the corresponding primed 
contributions result from the $P_R$ coupling in the vertex containing 
the $s$-quark. This is evident from the couplings $A^{tb}$ and $A^{ts}$
given in (\ref{tb}) and (\ref{ts}), respectively.
Now, the LR contributions presented above have been obtained by including 
only SM internal $W_L$ and top quark exchanges taking into account the 
right-handed couplings of $W_L$ in the vertex containing the $b$-quark. 
These contributions are important 
due to the factor $m_t/m_b$ that is absent in LL and RR contributions. 
Moreover, this LR contribution is $\ord(\epsilon^2)$ at the amplitude 
level and interfering 
with the SM contribution gives also $\ord(\epsilon^2)$ contribution 
to the branching ratio for $B\to X_s\gamma$.
Concerning the primed LR contribution to the decay amplitude while being 
of the same order in $\epsilon$ as the unprimed LR coefficients and also 
enhanced by $m_t/m_b$, 
it does not interfere with the SM contributions and 
consequently enters the branching ratio at the $\epsilon^4$ level. 
Therefore it should be neglected for the sake of consistency. 
Only in the case of a special 
hierarchy of the elements of $V^{\rm R}$ matrix could this suppression be 
compensated by the last factor in $A^{ts}$. However, then also other 
$\ord(\epsilon^4)$  contributions to the rate would have to be included, 
which is beyond the scope of this paper. Finally, as already stressed by Cho and Misiak the  
diagrams with internal $W_R$ 
exchanges give negligible contributions and a similar remark applies 
to LL and RR contributions from NP as one can easily check. Note that in this 
case the LL
and RR contributions are governed by SM loop functions in (\ref{c7}) 
and (\ref{c8}) which are strongly suppressed when 
$x_t=m^2_t/M^2_{W}$ is replaced by
$\tilde x_t=m^2_t/M^2_{W_R}$.

\subsection{Charged Higgs contributions}\label{eq:bsgamma-chargedHiggs}
In presenting the results for charged Higgs contributions we follow
\cite{Babu:1993hx,Frank:2010qv} adjusting their formulae to our notations and 
overall normalisations and keeping the phase $\alpha$. In the numerical analysis we set 
$\alpha=0$. As in the case of the gauge boson contributions one can demonstrate that the primed operators can be neglected. 
We also neglect the $H^\pm$ contribution to $C_{8G}(\mu_H)$. The dominant $H^\pm$ contribution at $\mu_H$ is given as follows \cite{Babu:1993hx} 
\footnote{In \cite{Babu:1993hx} $\tan\beta=\kappa/\kappa'=c/s$ has been used.}
\be\label{c7H}
\Delta^{{\rm H}^\pm}C_{7\gamma}(\mu_H)=-u(s) \left[sc\frac{m_t}{m_b} e^{i\alpha}
\left(\frac{V^{\rm R}_{tb}}{V^{\rm L}_{tb}}\right)A^1_{H^+}(y) +2 s^2 c^2A^2_{H^+}(y)\,\right],
\ee
where {the function $u(s)$ has been defined in (\ref{us}) and } \cite{Babu:1993hx}
\begin{align}
A^1_{H^+}(y)&=\,\left[ \frac{3y^2-2y}{3(1-y)^3} \ln y 
\;+\; \frac{5y^2-3 y}{6(1-y)^2} \right]\,,\\  
A^2_{H^+}(y)&=\,\frac{1}{3}A_{\rm SM}(y)-A^1_{H^+}(y)\,,\\
\quad A_{\rm SM}(y)&=\,-2 C^{\rm SM}_{7\gamma} (y)\,,\quad y=\,\frac{m_t^2}{M_{H}^2}\,.
\end{align}
The last function is given on the r.h.s of (\ref{c7}). In these contributions
we set $m_t=m_t(\mu_H)$. 
{For large $s$ close to the limit $s\to1/\sqrt{2}$ a strong enhancement 
of the $H^\pm$ contribution {through the factor $u(s)$ }is possible. As we discussed in section \ref{sec:Yuk} taking this limit, corresponding to $\kappa' = \kappa$, is phenomenologically not viable.} 
Interestingly,  
the inspection of gauge boson and charged Higgs contribution shows that
provided the element $V_{tb}^{\rm R}$ has only a small phase, these contributions 
always enhance the branching ratio for $B\to X_s\gamma$ which brings the 
theoretical value in (\ref{bsgth}) closer to the data in (\ref{bsgexp}).

\subsection{QCD corrections}
In order to complete the analysis of $B\to X_s\gamma$ we have to include 
QCD corrections which play a very important role in this decay. 
In the SM these corrections are known at the NNLO level. In the LR model 
a complete LO analysis has been done by Cho and Misiak \cite{Cho:1993zb}. On the other hand 
Bobeth et al provide matching conditions to the Wilson coefficients of 
LR operators at $\mu_R$ relevant for a NLO analysis, that is $\ord(\alpha_s)$ 
corrections to the coefficients (\ref{c7LR}) and (\ref{c8LR}). However,  
performing a complete NLO analysis would require additional complicated
calculations and in view of many parameters present in this model, such an involved analysis is certainly 
premature. In view of these remarks we proceed as follows:
\begin{itemize}
\item
For the SM contribution we use the full result at the NNLO \cite{Misiak:2006zs}.
\item
For the LR contribution we use explicit LO formulae which we obtained 
on the basis of  
\cite{Cho:1993zb} and the recent paper \cite{Buras:2011zb}. We 
set the $\mu_b$ scale to the one used for the SM contributions, 
this means $\mu_b=2.5$ GeV. While the NNLO SM contribution is not sensitive 
to this choice, some sensitivity is present in the LO LR contribution but in 
view of several parameters involved this uncertainty is not essential.
We are aware of the approximate nature of this treatment of QCD corrections 
in the NP part but we think that such an approach is sufficient before the discovery of 
$W_R$. 
\end{itemize}

\noindent Thus the basic formula for $C_{7\gamma}(\mu_b)$ used by us reads:
\be
C_{7\gamma}(\mu_b)=C_{7\gamma}(\mu_b)^{\rm SM}+\Delta^{\rm LR}C_{7\gamma}(\mu_b)
+ \Delta^{{\rm H}^\pm}C_{7\gamma}(\mu_b)\,,
\ee
where 
\begin{equation}  
\Delta^{\rm LR} C_{7\gamma}(\mu_b)=
\kappa_7(\mu_R)~\Delta^{\rm LR} C_{7\gamma}(\mu_R) +\kappa_8(\mu_R)~\Delta^{\rm LR}C_{8G}(\mu_R)+  A^{cb}\kappa_{\rm LR}(\mu_R)\,,
\label{eq:DeltaC7effA}
\end{equation}
\begin{equation}  
 \Delta^{{\rm H}^\pm}C_{7\gamma}(\mu_b)= \kappa_7(\mu_H)~\Delta^{{\rm H}^\pm}C_{7\gamma}(\mu_H)\,.
\label{eq:DeltaC7Higgs}
\end{equation}
Here $\Delta^{\rm LR}C_{7\gamma}(\mu_R)$, $\Delta^{\rm LR}C_{8{\rm G}}(\mu_R)$ 
and  $\Delta^{{\rm H}^\pm}C_{7\gamma}(\mu_H)$ can be found 
in (\ref{c7LR}), (\ref{c8LR})  and (\ref{c7H}), respectively.
$A^{cb}$ is given as follows
\be
A^{cb}= \frac{m_c}{m_b} \; sc\epsilon^2  e^{i\alpha}
\frac{V^{\rm R}_{cb}}{V^{\rm L}_{cb}}\,. 
\ee
The term proportional to $A^{cb}$ is absent in 
(\ref{eq:DeltaC7Higgs}) as, being related to 
the mixing with new charged current operators, it is already included in 
(\ref{eq:DeltaC7effA}) and above the scale $\mu_R$ this mixing does 
not take place.

Finally, $\kappa$'s are the NP magic numbers listed in Tab.~\ref{tab:c7magicnumbers} \cite{Buras:2011zb}, calculated taking $\alpha_s(M_Z=91.1876\,\text{GeV})=0.118$ \footnote{We thank Emmanuel Stamou for providing this table.}.
\begin{table}
\centering
\begin{tabular}{|l|r|r|r|r|}
\hline
$\mu_R$                     &       1 TeV   & 2.5 TeV       &       10 TeV & 15 TeV \\
\hline
\hline
$\kappa_7$                  &       0.457   &       0.427   &   0.390  & 0.380  \\\hline
$\kappa_8$                  &       0.125   &       0.128   &   0.130 & 0.130  \\\hline
$\kappa_{\rm LR}$               &       0.665   &       0.778   &  0.953 & 1.005  \\\hline
$\rho_8$                    &       0.504   &       0.475   &  0.439 & 0.429  \\\hline
$\rho_{\rm LR}$                 &       -0.052  &   -0.043  &  -0.025 & -0.019 \\
\hline
\hline
$\tilde{\kappa}_7$          &  0.857   &  0.801   &       0.731 & 0.712  \\\hline
$\tilde{\kappa}_8$          &       0.044   &   0.060   &  0.078 & 0.082 \\\hline
$\tilde{\kappa}_{\rm LR}$       &  0.063   &  0.099   &   0.156 & 0.175  \\\hline
$\tilde{\rho}_8$            &       0.874   &   0.824   &   0.760 & 0.743 \\\hline
$\tilde{\rho}_{\rm LR}$         &       -0.033  &  -0.044  &  -0.056 & -0.058 \\\hline
\end{tabular}
\caption{\it The NP magic numbers for $\Delta C^{\rm LR}_{7\gamma}$ and
$\Delta C^{\rm LR}_{8G}$  at $\mu_b=2.5\gev$ and $\mu_t(m_t)$.}
\label{tab:c7magicnumbers}
\end{table}

\noindent For later purposes we also give
\begin{equation}  
\Delta^{\rm LR} C_{8G}(\mu_b)=\rho_8(\mu_R)~\Delta^{\rm LR} C_{8G}(\mu_R)+  A^{cb}{\rho}_{\rm LR}(\mu_R)\,,
\label{eq:DeltaC8effA}
\end{equation}
with the NP magic numbers $\rho_i$ listed in Tab.~\ref{tab:c7magicnumbers}.

\subsection{The branching ratio}
For the branching ratio we follow the strategy of \cite{Buras:2011zb} 
which used the results of \cite{Misiak:2006ab}. One has then 
\be
\Br(B\to X_s\gamma)=\Br(B\to X_s\gamma)_{\rm SM}+\Delta \Br,
\ee
where
\be\label{master2}
\Delta \Br = R\left[2 \Re(C_{7\gamma}(\mu_b)^{\rm SM}\tilde\Delta^{\rm LR}C_{7\gamma}(\mu_b))+
|\tilde\Delta^{\rm LR}C_{7\gamma}(\mu_b)|^2\right]\,,
\ee
with
\be\label{master3}
\tilde\Delta^{\rm LR}C_{7\gamma}(\mu_b)=\Delta^{\rm LR}C_{7\gamma}(\mu_b)
+ \Delta^{{\rm H}^\pm}C_{7\gamma}(\mu_b)\,.
\ee
Next
\be
R=0.00247\,, \qquad C_{7\gamma}(\mu_b)^{\rm SM}=-0.353\,,
\ee
and 
\be\label{bsgth}
\Br(B\to X_s\gamma)_{\rm SM}=(3.15\pm0.23)\times 10^{-4}
\ee
are extracted from \cite{Misiak:2006ab}.
We refer to \cite{Buras:2011zb} for details. In particular in obtaining the value for 
$C_{7\gamma}(\mu_b)^{\rm SM}$ we have taken the non-perturbative corrections 
into account. Strictly speaking the last term in (\ref{master2}) is of 
$\ord(\epsilon^4)$ and should be dropped together with the contributions 
of primed operators. We recall that experimentally \cite{Asner:2010qj}
\be\label{bsgexp}
\Br(B\to X_s\gamma)_{\rm exp}=(3.55\pm0.26)\times 10^{-4}\,.
\ee
so that $\Delta \Br>0$ is favoured implying $\tilde\Delta^{\rm LR}C_{7\gamma}(\mu_b) < 0$. 
{This is guaranteed for $\text{Re}V^{\rm R}_{tb}>0$.}

\subsection[Dissecting the LR contributions to $\Br(B\to X_s\gamma)$]{\boldmath Dissecting the LR contributions to $\Br(B\to X_s\gamma)$}\label{sec:bsg_anatomy}
\begin{figure}
\centering
\includegraphics[width=0.78\textwidth]{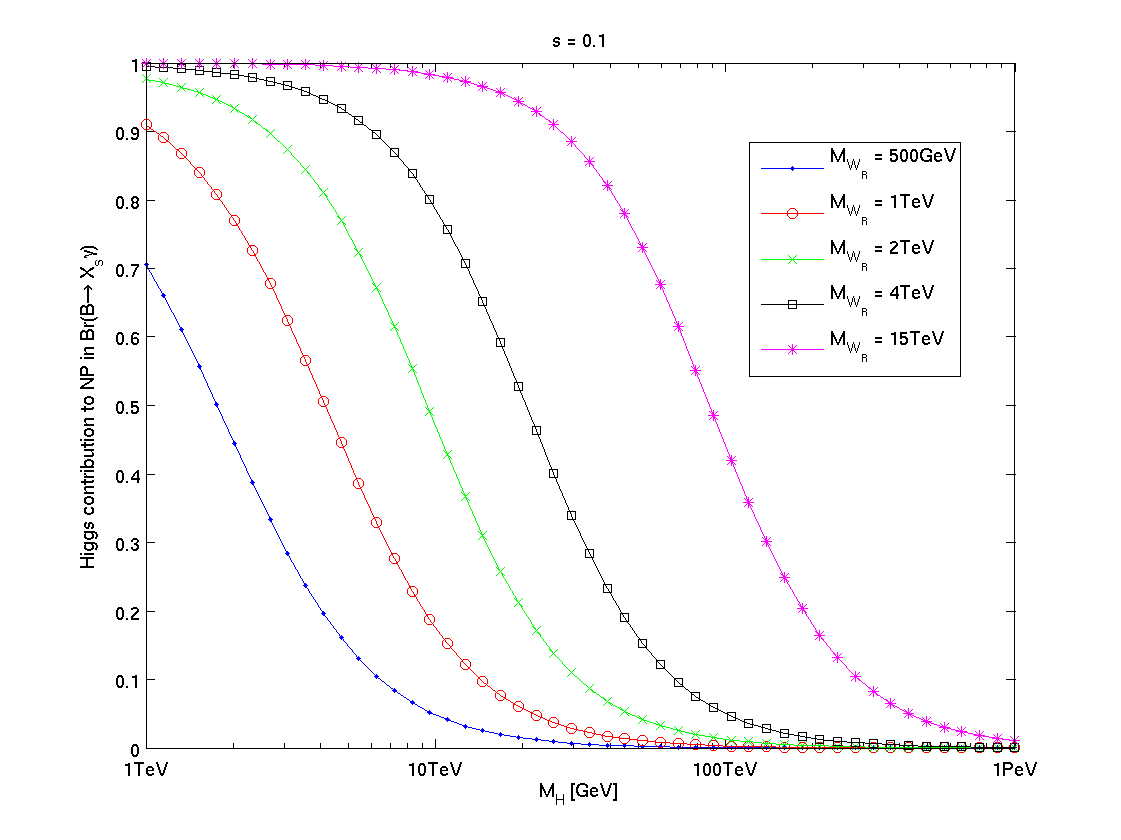}
\includegraphics[width=0.78\textwidth]{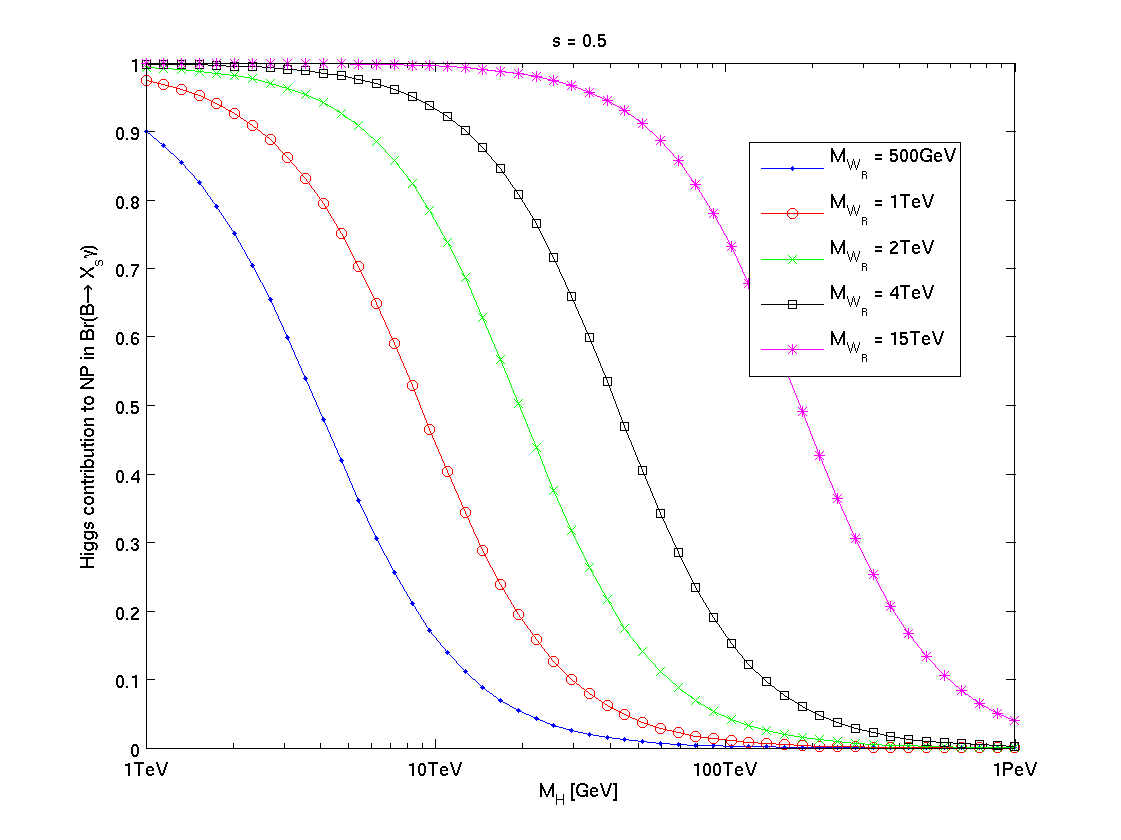}
\caption{The relative importance of the charged Higgs diagrams in the NP contributions to $\Br(B\to X_s\gamma)$ for $s=0.1$ (top panel) and $s=0.5$ (bottom panel) as a function of the Higgs mass for different $W_R$ masses.}\label{fig:bsg_higgs_contrib}
\end{figure}

Let us next calculate the fraction of $H^\pm$ contributions of the full 
NP contribution at the level of the branching ratio. As seen in 
(\ref{master2}) and (\ref{master3}), if we neglect the last term in 
(\ref{master3}) this is simply given by 
$\Delta^{{\rm H}^\pm}C_{7\gamma}(\mu_b)/\tilde\Delta^{\rm LR}C_{7\gamma}(\mu_b)$.
{In figure~\ref{fig:bsg_higgs_contrib}  we plot this  ratio 
as a function 
of $M_H$ for different values of $M_{W_R}$ and two choices of $s$.} For these plots we set $V^{\rm R}$ to be the identity matrix,
making the formulae $V^{\rm R}$ independent. Since the main contributions from charged Higgs and gauge bosons  are  both proportional to $V_{tb}^{\rm R}$
this assumption has a very limited impact.

As in our previous discussion of $\Delta F=2$ contributions in section \ref{sec:lr_gen_anatomy}, the Higgs exchanges (this time $H^\pm$) again turn out to be very important.
{For realistic $W_R$ masses above $2\tev$ and $H^\pm$ masses below $20\tev$, the $H^\pm$ contributions account for at least $20\%$ ($s=0.1$) or $50\%$ ($s=0.5$) of the total NP effect. As expected for the large value of $s=0.5$ the $H^\pm$ contribution is dominant. }

\subsection[The decay $B\to X_d\gamma$]{\boldmath The decay $B\to X_d\gamma$}\label{BDG}
Of considerable interest is also the decay $B\to X_d\gamma$ for which the 
measured CP-averaged branching ratio reads \cite{Asner:2010qj}
\be\label{bdgav}
\langle\Br(B\to X_d\gamma)\rangle=(1.41\pm 0.49)\times10^{-5}\,,
\ee
to be compared with the most recent SM value \cite{Crivellin:2011ba}
\be
\langle\Br(B\to X_d\gamma)\rangle_{\text{\rm SM}}=(1.54^{+0.26}_{-0.31})\times10^{-5}\,.
\ee
While the data agree with the latter estimate there is still significant 
room for NP contributions. Yet already these results can imply 
important constraints on the extensions of the SM. 
Indeed, it was recently pointed out by Crivellin and Mercolli \cite{Crivellin:2011ba} that the contributions of 
primed operators to $\langle\Br(B\to X_d\gamma)\rangle$ could
be significantly constrained implying a bound on  $|V_{td}^{\rm R}|$. This bound 
is approximately 3.5 times stronger than what is found for the best-fit solution in 
\cite{Buras:2010pz}, where this decay has not been considered.
In the notation of the present paper the bound in \cite{Crivellin:2011ba} reads \footnote{Note that this bound does not take into account the charged Higgs contribution.}
\be\label{CMbound} 
sc\epsilon^2 |V_{td}^{\rm R}|\le 1.4\times 10^{-4}\,.
\ee
{As our numerical analysis shows the electroweak precision tests imply 
$sc\epsilon^2\lsim 1\cdot 10^{-3}$. Therefore as long as $|V_{td}^{\rm R}|\le 0.14$ 
the bound in \cite{Crivellin:2011ba} is satisfied in our paper. Moreover as 
we can see in section~\ref{sec:num} the typical values for 
 $|V_{td}^{\rm R}|$ that are consistent with $\epsilon_K$
are below 
\be
|V_{td}^{\rm R}| \le \left\{ \begin{array}{c}
0.047 \quad (s=0.5) \\
0.113 \quad (s=0.1)
\end{array}\right. \,,
\ee
when no fine-tuning constraint is taken into account. Otherwise the allowed ranges for  $|V_{td}^{\rm R}|$ become even smaller. For further details
see section \ref{sec:vr_gen_num}. In all cases $|V_{td}^{\rm R}|$ is well below the bound
in \cite{Crivellin:2011ba}, since e.\,g.\ a large $s$ implies small values for $|V_{td}^{\rm R}|$.}
As we explained before, in our model the contributions of primed 
operators are $\ord(\epsilon^4)$ and consistently neglected in our 
analysis. While a detailed analysis of $\ord(\epsilon^4)$ contributions 
to $\Br(B\to X_d\gamma)$ would be required to assess the numerical importance 
of primed operators in this case, the discussion above indicates that in our model 
they are well below the NP contributions of unprimed operators when all 
additional constraints are taken into account. 

Clearly, the improved data 
on the CP-averaged branching ratio in (\ref{bdgav})
and the finding of \cite{Benzke:2010js} that the CP-averaged branching ratio contains only 
small hadronic uncertainties, invites us to consider this decay as well. 
In analysing it we use, as done in \cite{Crivellin:2011ba}, 
the formulae of  \cite{Hurth:2003dk}. 
In doing this we should emphasise that these formulae strictly speaking apply 
only to cases where the RG evolution below the $\mu_W$ scale is the same as 
in the SM. Due to the presence of new LR current-current operators, additional 
contributions to the Wilson coefficients of the dipole operators are present. 
They are represented by the last term in (\ref{eq:DeltaC7effA}). However, in 
the model considered by us these contributions are very small and can be 
neglected. We find then
\be\label{BDGAV}
\langle\Br(B\to X_d\gamma)\rangle=\frac{\cal{N}}{100}
\left|\frac{V^{{\rm L}*}_{td}V^{\rm L}_{tb}}{V^{\rm L}_{cb}}\right|^2\left[\tilde a+P_7+P_8+P_{78}\right]\,,
\ee
where ${\cal{N}}=2.57\times 10^{-3}$ and 
\be
\tilde a= a +a_{\epsilon\epsilon}|\epsilon_d|^2+a_\epsilon^r \Re(\epsilon_d)\,,
\ee
\be
P_7=a_{77}|R_7|^2+a_7^r \Re(R_7)+a_{7\epsilon}^r \Re(R_7\epsilon_d^*)\,,
\ee
\be
P_8=a_{88}|R_8|^2+a_8^r \Re(R_8)+a_{8\epsilon}^r \Re(R_8\epsilon_d^*)\,,
\ee
\be
P_{78}=a^r_{87}\Re(R_8R_7^*).
\ee

\begin{table}
\centering
\begin{tabular}{|c|c|c|c|c|c|c|c|}
\hline
$a$ & $a_{\epsilon\epsilon}$ & $a_\epsilon^r$  & $a_{77}$ & $a_7^r$ & $a_{7\epsilon}^r$ & $a_{88}$ & $a_8^r$ \\	
\hline\hline  
$7.8221$ & $0.4384$ & $-1.6981$ & $0.8161$ & $4.8802$ & $-0.7827$ & $0.0197$ & $0.5680$\\[1mm]
\hline \hline
$a_{8\epsilon}^r$ & $a^r_{87}$ & $a_7^i$ & $a_8^i$ & $a_\epsilon^i$ & $a_{87}^i$ & $a^i_{7\epsilon}$ & $a^i_{8\epsilon}$\\
\hline\hline  
$-0.0601$ & $0.1923$ & $0.3546$ & $-0.0987$ & $2.4997$ & $-0.0487$ & $-0.9067$ & $-0.0661$\\
\hline   	        	
\end{tabular}
\caption{\it The relevant $a_i$ parameters from \cite{Hurth:2003dk}.}
\label{tab:aipar}
\end{table}

\noindent The values of the $a_i$ parameters are collected in table \ref{tab:aipar} and 
\be
R_7=\frac{C_{7\gamma}(\mu_t)}{C_{7\gamma}^{\rm SM}(\mu_W)}\,,\quad
R_8=\frac{C_{8G}(\mu_t)}{C_{8G}^{\rm SM}(\mu_W)}\,,
\ee
where the $C_{7\gamma}(\mu_t),C_{8G}(\mu_t)$ denotes the 
total Wilson coefficients including both the SM contributions 
without QCD corrections, as given in (\ref{c7}) and (\ref{c8})
and the NP contributions as in
(\ref{eq:DeltaC7effA})
and (\ref{eq:DeltaC8effA}) but evaluated at $\mu_t$ and not $\mu_b$:
\begin{align}  
\Delta^{\rm LR} C_{7\gamma}(\mu_t) &=\, 
\tilde\kappa_7(\mu_R)~\Delta^{\rm LR} C_{7\gamma}(\mu_R) +\tilde\kappa_8(\mu_R)~\Delta^{\rm LR}C_{8G}(\mu_R)+  A^{cb}\tilde\kappa_{\rm LR}(\mu_R)\,,
\label{eq:DeltaC7effAuW}\\
\Delta^{{\rm H}^\pm} C_{7\gamma}(\mu_t) &=\,
\tilde\kappa_7(\mu_H)~\Delta^{{\rm H}^\pm} C_{7\gamma}(\mu_H)\,,
\label{eq:DeltaC7HiggseffAuW}
\end{align}
and
\begin{equation}  
\Delta^{\rm LR} C_{8G}(\mu_t)=\tilde\rho_8(\mu_R)~\Delta^{\rm LR} C_{8G}(\mu_R)+  A^{cb}{\tilde\rho}_{\rm LR}(\mu_R)\,.
\label{eq:DeltaC8effAuW}
\end{equation}
The NP magic numbers $\tilde\kappa_i$ and $\tilde\rho_i$ are 
listed in table~\ref{tab:c7magicnumbers}. Finally 
\be
\epsilon_d=\frac{V^{{\rm L}*}_{ud}V^{\rm L}_{ub}}{V^{{\rm L}*}_{td}V^{\rm L}_{tb}}\,.
\ee
For $B\to X_s\gamma$ one should just replace $d$ by $s$.

\subsection[CP asymmetries in $B\to X_{s,d}\gamma$]{\boldmath CP asymmetries in $B\to X_{s,d}\gamma$}
A very sensitive observable to NP CP violating effects is
represented by the direct CP asymmetry in $b\to s\gamma$, i.e. $A_{\rm CP}(b\to s\gamma)$~\cite{Soares:1991te}, in particular as the perturbative contributions within the SM amount to only $+0.5\%$ \cite{Hurth:2003dk}. The corresponding asymmetry in 
$b\to d \gamma$ transition is much larger but could in principle also provide 
a useful test. These asymmetries are defined by $q=(s,d)$
\be\label{eq:acp_bsg}
A_\text{CP}(b\to q\gamma) \equiv \frac{\Gamma(\overline{B} \to X_{\bar{q}} \gamma) - \Gamma({B} \to X_q \gamma)}{\Gamma(\overline{B} \to X_{\bar{q}}\gamma) + \Gamma({B} \to X_q\gamma)}\,,
\ee
and have been studied in particular in
\cite{Kagan:1998bh,Kagan:1998ym} and more recently in \cite{Hurth:2003dk}, 
where further references can be found. In the context of the LR models 
these asymmetries have been analysed in \cite{Asatrian:1996as}. 
Unfortunately, a recent analysis \cite{Benzke:2010tq} shows that these asymmetries, similar to other direct CP 
asymmetries suffer from hadronic uncertainties originating here in the hadronic 
component of the photon. These uncertainties lower the predictive power 
of these observables and in the case of $b\to s\gamma$ the authors conclude 
that only if experimentally $A_\text{CP}(b\to s\gamma)$ was found 
below $-2\%$, one could consider it as a signal of NP.
In order to get a rough idea whether in the models considered by us the 
perturbative part could be affected strongly by NP, we use the formulae 
in \cite{Hurth:2003dk}, which are compatible with \cite{Kagan:1998bh,Kagan:1998ym}. The 
formulae with hadronic contributions that are rather uncertain can be 
found in \cite{Benzke:2010tq}. We have then
\be
A_\text{CP}(b\to q\gamma)=\frac{\cal{N}}{100}
\left|\frac{V^{{\rm L}*}_{tq}V^{\rm L}_{tb}}{V^{\rm L}_{cb}}\right|^2
\frac{\Im(a_7^i R_7+a_8^i R_8+a_\epsilon^i\epsilon_q+a_{87}^iR_8R_7^*+a^i_{7\epsilon}R_7\epsilon_q^*+a^i_{8\epsilon}R_8\epsilon_q^*)}{\langle\Br(B\to X_q\gamma)\rangle},
\ee
with the values of $a_i$ collected in table \ref{tab:aipar}.

\section{\boldmath Constraints from tree level decays}\label{sec:tree}
\subsection{Preliminaries}

In this section we address the constraints from tree level decays. The possible new tree level contributions arise from the new right-handed couplings  
of the $W_L$ gauge boson, from the exchange of the heavy $W_R$ gauge boson, and from the heavy charged Higgs boson.
For a detailed consideration of leptonic and semi-leptonic decays it is necessary to have a closer look at the lepton sector within the LR model. 
In this context in order to derive the Feynman rules we use the findings of \cite{Mohapatra:1979ia,Mohapatra:1980yp}. After transformation to mass 
eigenstates the light neutrinos are dominated by their left-handed contribution with a small right-handed admixture, while the heavy Majorana neutrinos 
are mainly given by the right-handed neutrinos again modified by a small left-handed contribution. This mixing angle can be constrained by the masses 
of heavy and light neutrinos. Assuming reasonable masses for the light neutrinos and the heavy neutrinos not to be lighter than $100$  GeV, the Yukawa 
couplings have to be very small and cause this mixing to be at most $\ord(10^{-6})$ \cite{Chen:2011de}. In agreement with \cite{Czakon:2002wm} we find these 
mixing effects to be negligible. This is in particular a good assumption taking into account that they have to compete with effects of $\ord(\epsilon^2)$ 
being roughly of $\ord(10^{-3})$. Furthermore tree level decays will only take place into the light neutrinos. As the $W_R$ coupling to light 
neutrinos appears first at $\ord(\epsilon^2)$ and the $W_R$ propagator yields another $\ord(\epsilon^2)$ suppression factor, we conclude that for 
the lepton couplings to charged gauge bosons  at $\ord(\epsilon^2)$ only the SM couplings are relevant for the tree-level decays in question. Concerning the tree level 
contributions of charged Higgs bosons it is sufficient to include only leading order couplings. 
In \cite{Buras:2010pz} a detailed analysis of the constraints on the elements of the matrix $V^{\rm R}$ 
(denoted there by $\tilde V$) and implications for the CKM matrix $V^{\rm L}$ have been presented. As the charged Higgs 
contributions to tree level decays in the present model turn out to be negligible (see below), basically all the results obtained  in \cite{Buras:2010pz}   can be 
taken over by making the following identification:
\be
\varepsilon_L=\ord(\epsilon^4)\,,\qquad
\varepsilon_R=c s \epsilon^2\,, \qquad \tilde V= e^{i \alpha}V^{\rm R}\,,
\ee
where the quantities on the l.h.s are the ones used in \cite{Buras:2010pz}.
For completeness we summarise these results in our notation, extending 
the discussion of the impact of the right-handed currents in tree 
level decays on the mixing induced 
asymmetries $S_{\psi K_S}$ and $S_{\psi\phi}$ and on the CKM phase $\gamma$ as 
extracted from $B\to D$ decays within the SM.
We use our findings from \cite{Buras:2010pz} with updated numerical values. 

\subsection[Constraints on the mixing matrices $V^{\rm L}$  and $V^{\rm R}$]{\boldmath Constraints on the mixing matrices $V^{\rm L}$  and $V^{\rm R}$}

In what follows we use the notation
\be\label{VRVA}
|V_{ij}|_V = \left|V_{ij}^{\rm L} + cs e^{i\alpha} \epsilon^2 V^{\rm R}_{ij} \right|\,,\qquad 
|V_{ij}|_A = \left|V_{ij}^{\rm L} - cs e^{i\alpha} \epsilon^2 V^{\rm R}_{ij} \right|\,,
\ee
for the combinations of left- and right-handed contributions entering vector and axial vector couplings.
As $G_F$ enters tree level decays we note that the correction to the width of the $\mu$-decay in this model are at $\ord(\epsilon^4)$ 
and have no impact on our estimates of $V^{\rm R}$.
\begin{table}
\renewcommand{\arraystretch}{1}\setlength{\arraycolsep}{1pt}
\centering
\begin{tabular}{|l|l|}
\hline
$\Br(\pi\to\mu\nu)=0.9998770(4)$ \hfill  \cite{Nakamura:2010zzi}  
& $f_\pi=129.5(17)\mev$ \\\cline{1-1}
$f_+(0)|V_{us}|^{K\to\pi\ell\nu}=0.2163(5)$ & $f_+(0)=0.9584(44)$ \\
$f_K/f_\pi |V_{us}/V_{ud}|^{K\to\mu\nu}=0.2758(5)$\qquad \cite{Antonelli:2010yf}& $f_K/f_\pi=1.1931(53)$ \\\cline{1-1}
$\Br(D_s\to\tau\nu)=0.0529(28)$  & $f_{D_s}=248.9(39)\mev$ \\
$\Br(B\to\tau\nu)=1.64(34)\cdot 10^{-4}$ & $f_B = 205(12)\mev$\qquad\cite{Laiho:2009eu}\\\cline{2-2}
$F(1) |V_{cb}|^{B\to D^*\ell\nu} = 0.03604(52)$ & $F(1)=0.908(17)$\hfill \cite{Bailey:2010gb}\\\cline{2-2}
$G(1) |V_{cb}|^{B\to D\ell\nu} = 0.0423(15)$ \hfill \cite{Asner:2010qj} & $G(1)=1.074(24)$ \hfill \cite{Okamoto:2004xg} \\\hline
\end{tabular} 
\caption{Values of the most important experimental and theoretical
quantities used as input parameters for the constraints on tree level charged currents. \label{tab:tree-input}}
\renewcommand{\arraystretch}{1.0}
\end{table}

\subsubsection[$u\to d$]{\boldmath $u\to d$}

The study of super-allowed $0^+\to 0^+$ transitions yields the constraint on the $u\to d$ vector current  \cite{Nakamura:2010zzi}
\be
|V_{ud}|_V= 0.97425(22)\,.
\ee
A constraint on the axial component can be obtained from the pion decay $\pi^+ \to \mu^+ \nu$, including the known radiative corrections \cite{Finkemeier:1994ev},  with the result
\be
|V_{ud}|_A =  0.981(13)\,.
\ee

\subsubsection[$u\to s$]{\boldmath $u\to s$}
The vector current $s\to u$ transition can be obtained from $K\to\pi\ell\nu$ decays \cite{Antonelli:2010yf}  with the result
\be
|V_{us}|_V = 0.2257(12)\,.
\ee
The axial $s\to u$ transition can be constrained by combining $K\to\mu\nu$ and 
$\pi\to\mu\nu$. The result reads
\be
|V_{us}|_A =  0.2268(32)\,.
\ee

\subsubsection[$c\to d$]{\boldmath $c\to d$}
The constraint 
\be
|V_{cd}|_V=0.229(25)
\ee
can be obtained from the decays $D\to K\ell\nu$ and $D\to \pi\ell\nu$ \cite{Nakamura:2010zzi}. 
$|V_{cd}|$ can also be obtained from neutrino and anti-neutrino charm production off valence $d$ quarks. 
As interference terms between left- and right-handed quarks are suppressed by $m_d$, we can safely 
neglect them. We can therefore directly apply the constraint to left-handed $c\to d$ transitions 
and get \cite{Nakamura:2010zzi}
\be
|V_{cd}^{\rm L}|=0.230(11)\,.
\ee
Due to the large uncertainties these data do not provide significant constraints on the LR parameter space 
but we  include them in our numerical analysis. The situation should change in 
the future when more precise data 
and lattice inputs will be available. Similar comments apply also to $c\to s$  transitions discussed below.

\subsubsection[$c\to s$]{\boldmath $c\to s$}
From semileptonic $D$ decays and $D_s\to\tau^+\nu$ one finds
\be
|V_{cs}|_V = 0.98(10)\,, \qquad |V_{cs}|_A = 0.978(31)\,.
\ee
\subsubsection[$b\to u$]{\boldmath $b\to u$}
This is a place where the RH currents could enter in a potentially important manner.
If  the SM value for $|V_{ub}|$ is determined from the inclusive semileptonic mode $B\to X_u \ell \nu$, the interference term between LH and RH contributions 
is totally negligible, so that the SM result carries over to the LR model. We have \cite{Nakamura:2010zzi}
\be\label{vub1}
|V^{\rm L}_{ub}| = 4.27(38)\cdot 10^{-3}\,,
\ee
which should be regarded as the true value of this CKM element. 
The vector $b\to u$ transition can be probed by $B\to \pi\ell\nu$. This 
implies the constraint \cite{Nakamura:2010zzi}
\be\label{vub2}
|V_{ub}|_V =3.38(36)\cdot 10^{-3}\,.
\ee

Finally the axial $b\to u$ coupling is determined from $\Br(B\to\tau\nu)$ 
implying the constraint
\be\label{vub3}
|V_{ub}|_A =  4.70(56)\cdot 10^{-3}\,.
\ee

\subsubsection[$b\to c$]{\boldmath $b\to c$}

Similar to the case of $b\to u$ transitions, the inclusive determination of $|V_{cb}|$ in the SM carries over to the LR models, where we have
\be
|V_{cb}^{\rm L}|=41.54(73)\cdot 10^{-3}\,.
\ee
On the other hand
from $B\to D\ell\nu$ and $B\to D^*\ell\nu$ transitions one 
finds the constraints
\be
|V_{cb}|_V =                                    39.4(17)\cdot 10^{-3}\,,
\qquad
|V_{cb}|_A =                                    39.70(92)\cdot 10^{-3}\,.
\ee
The good agreement between the two exclusive determinations suggests a small contribution from right-handed currents and consequently  
LR effects cannot explain the tension between inclusive and exclusive determinations.

\subsubsection[$t\to d,s$]{\boldmath $t\to d,s$}

The $t\to d$ and $t\to s$ transitions cannot be measured from tree level decays, so that no constraint is obtained on $V^{\rm L}_{td,ts}$ and $V^{\rm R}_{td,ts}$.

\subsubsection[$t\to b$]{\boldmath $t\to b$}

In the SM $|V_{tb}|$ can be determined from the ratio of branching ratios $\Br(t\to bW)/\Br(t\to qW) = |V_{tb}|^2$ assuming unitarity of the CKM matrix. 
As interference terms are suppressed by $m_q/m_t$ ($q=d,s,b$), this method can directly be applied to LR models. The best available
constraint stems from D0 \cite{Abazov:2011zk} and reads
\be
|V_{tb}^{\rm L}|=0.95(2)\,.
\ee 
This measurement is $2.5\sigma$ below the SM expectation $V_{tb}^{\rm L}\approx 1$. Since as we see later the tight tree level constraints on other elements on the CKM matrix do not allow for a significant deviation of $V_{tb}^{\rm L}$ from its SM value, this tension persists in the LRM. 
The improved data from the LHC will tell us whether this is a real 
problem.

\subsection{Charged Higgs contributions}

The charged Higgs couplings are given by combinations of the Yukawa couplings of the fermions participating in the interaction. 
Therefore for all processes not involving the top quark the charged Higgs contribution is very small and turns out to be negligible
in all cases. The only decay where this is not completely obvious is $B^+\to \tau^+\nu_\tau$  due to the chiral suppression of the $W^+$ 
contribution. However it turns out  that the $H^+$ contribution is negligible also in that case.
Indeed, the Higgs tree level diagrams appear with a suppression factor of ${m_B^2}/{M_{H}^2}$ and hence in the LR models in which $H^\pm$ is very heavy
this contribution is  negligibly small in comparison to the effects from right-handed charged gauge boson couplings discussed above.

\subsection[A comment on {$Z\to b\bar b$}]{A comment on \boldmath{$Z\to b\bar b$}}
Even if the electroweak precision tests are considered in the next 
section it is useful to discuss already here the impact of new 
contributions on the $Z\to f\bar f$ couplings.
Denoting the effective diagonal couplings of the $Z$ 
to down-type quarks as follows
\be
\mathcal{L}_{\rm eff}^{Z} =   \frac{g}{ c_W} \left( ~g_L^{ii}~\bar d^i_L \gamma^\mu d^i_L +
g_R^{ii}~\bar d^i_R \gamma^\mu d^i_R \right) Z_\mu\,,
\ee
we find (see appendix) at tree level
\bea
(g_L^{ii})&=& \left(-\frac{1}{2}+\frac{1}{3} s_W^2\right) 
-\frac{1}{24}s_R^2c_R^2 \epsilon^2\,,
\\
(g_R^{ii}) &=& \frac{1}{3} s_W^2 - 
\frac{c_R^2}{8}(c_R^2+\frac{1}{3}s_R^2)\epsilon^2\,,
\eea
where the first terms are the SM contributions.
The experimental determination of the effective couplings of the $Z$
bosons to light down quarks resulting from the global fit of 
electroweak data collected by the LEP and the SLD experiments is 
in a very good agreement with the SM. For $b$ quark couplings one 
finds ~\cite{:2005ema}
\bea
(g_L^{bb})_{\rm exp} &=& -0.4182 \pm 0.0015\,,  \\
(g_R^{bb})_{\rm exp} &=& +0.0962 \pm 0.0063\,.  
\eea
While the result for the LH coupling is consistent with the 
SM prediction, there is a large disagreement between data and 
SM expectation in the RH sector:
\be
(\Delta g_R^{bb})_{\rm exp}  
=  (g_R^{bb})_{\rm exp} -  (g_R^{bb})_{\rm SM} = (1.9 \pm 0.6) \times 10^{-2}\,,
\label{eq:Zbb_exp}
\ee 
It is evident from the formulae above that with $\epsilon^2=\ord(10^{-3})$ 
as required by our analysis of electroweak precision constraints
the corrections to $Z\to q\bar q$ are marginal and that while not spoiling 
the agreement of the SM with the data for the remaining couplings, these 
corrections cannot help in removing the anomaly in question. In this 
context we would like to emphasise that the flavour independence of the 
corrections would preclude the solution anyway without spoiling the 
agreement for the remaining couplings. Finally, the sign of  corrections 
being strictly negative would make the disagreement with the data even worse.

\subsection{Impact on CP asymmetries}

\subsubsection[CP asymmetry in $B\to DK$  and the angle $\gamma$]{\boldmath CP asymmetry in $B\to DK$ and the angle $\gamma$}

In the SM the angle 
\be
\gamma = \arg(-V_{ud}V_{ub}^*/V_{cd}V_{cb}^*)\,,
\ee
can be measured by means of  CP asymmetries associated with tree level $B\to DK$ decays. The theoretically cleanest
method has been proposed in \cite{Giri:2003ty}: Studying $B^\pm \to D K^\pm$ together with a Dalitz plot
analysis of a following multibody $D$ decay allows for determining the weak phase $\gamma$ without prior 
assumptions on the hadronic parameters involved. Belle \cite{Belle-Moriond-2011} obtained

\be
\gamma = 77(11)^\circ\,,
\ee
where we combined the several uncertainties in quadrature. The details of the Dalitz plot analysis can be 
found in \cite{Giri:2003ty} -- as they are relevant mostly for the extraction of the hadronic parameters 
from the data we do not repeat them here. The measurement relies on the interference between the two modes
\bea
A(B^-\to D^0K^-)&\equiv&A_B\,,\\
A(B^-\to \bar D^0K^-)&\equiv&A_B r_B e^{i(\delta_B-\gamma_\text{eff})}\,,
\eea
with relative absolute value $r_B$, relative strong phase $\delta_B$ and relative weak phase $\gamma_\text{eff}$. 
In the SM the two amplitudes are governed by the operators $(\bar cb)_{V-A}(\bar su)_{V-A}$ and
$(\bar ub)_{V-A}(\bar sc)_{V-A}$ mediated by the $W$ boson, respectively. Therefore in this approximation $\gamma_\text{eff}=\gamma$.\par
In LR models there are two new contributions to $B^\pm\to D K^\pm$ decays:
\begin{enumerate}
\item
The small RH coupling of $W$ leads to contributions of $(V-A)\otimes (V+A)$ structure. 
Unless the corresponding hadronic matrix elements are strongly enhanced with respect to 
the standard $(V-A)\otimes (V-A)$ ones, this contribution is subleading with respect to 
the one obtained from $W'$ exchange as it is suppressed by an additional factor $sc\le 1/2$. We neglect this contribution in view of the presently large  uncertainties in the experimental determination of $\gamma$, keeping in mind that it should be reconsidered once precise data become available.
\item
Similarly to the case of $S_{\psi K_S}$ (see below) we also have the contributions of $W_R$ 
with purely right-handed couplings, that generate the $(V+A)\otimes (V+A)$ operators. As QCD 
is non-chiral the relevant matrix elements are equal to the SM ones. The LR contribution considered here is
therefore purely perturbative.
\end{enumerate}
The interfering amplitudes then get modified as follows
\bea
A(B^-\to D^0K^-)&=& A(B^-\to D^0K^-)_\text{\rm SM}\left(1+r \frac{M_W^2}{M_{W_R}^2}\frac{V^{\rm R}_{cb} (V^{\rm R}_{us})^*}{V^{\rm L}_{cb} V^{{\rm L}*}_{us}} \right)\,,\\
A(B^-\to \bar D^0K^-)&=& A(B^-\to \bar D^0K^-)_\text{\rm SM}\left(1+r \frac{M_W^2}{M_{W_R}^2}\frac{V^{\rm R}_{ub} (V^{\rm R}_{cs})^*}{V^{\rm L}_{ub} V^{{\rm L}*}_{cs}}\right)\,,
\eea
Here we neglected the QCD running of the LR contribution 
from $M_{W_R}$ to $M_W$.  Then the relative weak phase is given by

\be
\gamma_\text{eff} = \gamma + \arg\left[1+r\frac{M_W^2}{M_{W_R}^2}\left(\frac{V^{\rm R}_{cb} (V^{\rm R}_{us})^*}{V^{\rm L}_{cb} V^{{\rm L}*}_{us}} - \frac{V^{\rm R}_{ub} (V^{\rm R}_{cs})^*}{V^{\rm L}_{ub} V^{{\rm L}*}_{cs}}\right)\right]\,.
\ee

\subsubsection[The impact on $S_{\psi K_S}$ and $S_{\psi\phi}$]{\boldmath The impact on $S_{\psi K_S}$ and $S_{\psi\phi}$}
In the SM the decay $B_d\to J/\psi K_S$ is dominated by one single decay amplitude arising from a tree level $W$ exchange and giving rise to the operator 
$(\bar bc)_{V-A}(\bar cs)_{V-A}$. Using the conventions of \cite{Buras:2005xt} the weak phase of the decay amplitude $A(B_d\to J/\psi K_S)$ is to a 
very good approximation given by
\be
\phi_D = \arg V^{{\rm L}*}_{cb}V^{\rm L}_{cs} \simeq 0\,.
\ee
Together with the phase of $B_d-\bar B_d$ mixing
\be
\phi_M = \arg V^{{\rm L}*}_{tb}V^{\rm L}_{td} \simeq -\beta
\ee
the time dependent CP asymmetry therefore measures \footnote{For the estimates 
of the uncertainties in this relation see \cite{Faller:2008zc}.}
\be
S_{\psi K_S} = \sin [2(\phi_D-\phi_M)] = 
-\sin[2\arg\frac{V^{{\rm L}*}_{tb}V^{\rm L}_{td}}{V^{{\rm L}*}_{cb}V^{\rm L}_{cs}}] = \sin2\beta\,,
\ee
In LR models we have new 
contributions to the $b\to c\bar cs$ tree level decay. First of all the $W$ now has a small right-handed 
coupling proportional to $\epsilon^2\sim \ord(10^{-3})$. Second, also $W_R$ exchanges contribute at the 
$\epsilon^2\sim\ord(10^{-3})$ level. Assuming that the QCD matrix elements do not significantly affect 
the hierarchy of new contributions, the direct $W_R$ exchange is dominant, in particular if $s \ll 1$.
As QCD is a non-chiral theory, we find for the matrix elements governing the decay in question
\be
\langle J/\psi K_S | (\bar bc)_{V-A} (\bar cs)_{V-A} | B_d \rangle = \langle J/\psi K_S | (\bar bc)_{V+A} (\bar cs)_{V+A} | B_d \rangle\,.
\ee
Hence the LR correction to  the decay amplitude can be expressed in terms of short-distance physics only. We find
\be
A(B_d \to J/\psi K_S) = A(B_d \to J/\psi K_S)_\text{\rm SM}\left( 1+ x r \frac{M_W^2}{M_{W_R}^2}\frac{V^{{\rm R}*}_{cb}V^{\rm R}_{cs}}{V^{{\rm L}*}_{cb}V^{\rm L}_{cs}}  \right)\,.
\ee
Here $x$ encodes the RG running from $M_{W_R}$ down to $M_W$ and $r=s^2_W/c_W^2 s_R^2=g_R^2/g_L^2$. Consequently
\be
S_{\psi K_S} = \sin [2(\beta + \varphi_{B_d} + \delta \phi_D)]\,,
\ee
with
\be
\delta \phi_D = \arg \left( 1+ x r \frac{M_W^2}{M_{W_R}^2}\frac{V^{{\rm R}*}_{cb}V^{\rm R}_{cs}}{V^{{\rm L}*}_{cb}V^{\rm L}_{cs}}  \right)\,.
\ee
We estimated $x\simeq 1.1$.

The same arguments apply to $B_s \to J/\psi \phi$, hence we find
\be
S_{\psi\phi} = -\sin[2(\beta_s+\varphi_{B_s}+\delta\phi_D)]\,.
\ee
Note that $\delta \phi_D$ is universal in both transitions.

\subsubsection{Summary}
Using the expressions above we find that the impact of new contributions 
on the determination of $\gamma$ and of new phases $\varphi_{B_s}$ and 
$\varphi_{B_d}$ is negligible. This can be traced back primarily to $W_R$ 
being much heavier than $W_L$.

\section{Electroweak precision constraints }\label{sec:EWP}
\subsection{Preliminaries}
Electroweak precision tests (EWPT) provide very strong constraints on basically any model of NP.
The SM agrees to a high accuracy with the measured values of roughly 40 low-and
high energy observables \cite{Flacher:2008zq,Arbuzov:2005ma,Nakamura:2010zzi}. 
Very often such tests are performed with the help of the famous S-T 
variables \cite{Peskin:1991sw}, which  test the allowed size of 
the oblique corrections, but in order to get the full picture including 
the LEP II data the study of all observables should be favoured.

Since a full analysis of the EWP data is clearly beyond the scope of this work we use
the results of \cite{Hsieh:2010zr} where a full analysis of EWP observables 
in a number of models with $SU(2)_1\times SU(2)_2\times U(1)_X$ gauge symmetry 
has been performed. The model denoted there by LR-T is precisely the model
studied in our paper. Most useful for our study are tables IX and X, where
the authors of \cite{Hsieh:2010zr} provide formulae for corrections
to the most constraining  observables in a given model in terms of the {\it fit parameters}.
These formulae allow then to find allowed regions in the space of the 
{\it fit parameters} that are consistent with the data on EWP data. 
{This analysis is independent of the flavour parameters of the matrix $V^{\rm R}$.
On the other hand the analysis of flavour physics observables is affected by the choice of EW parameters. This is in particular the case for $s$, to which the $b\to s\gamma$ decay and to a certain extent also $\Delta F = 2$ processes are very sensitive. 
Because of its enhancement by charged Higgs contributions $b\to s\gamma$ requires $s<0.64$.
In this sense we are able to separate the EWP analysis from the flavour analysis. We first determine the allowed ranges for EW parameters and then perform the flavour analysis for a certain EW benchmark point, varying only the parameter $s$.}

In what follows, we very briefly describe the main points of 
\cite{Hsieh:2010zr}, simultaneously adjusting their notation to ours.

\subsection{Basic structure of the analysis}
As seen in (\ref{couplings}) and (\ref{vevs}) except for $g_s$, $m_t$, 
$M_H$ and the phase $\alpha$, which we set to zero without loss of generality, 
we have six {\it model parameters}  that 
are relevant for this section. 
{For the present analysis it is convenient to make a change of variables 
and to fix the following input ({\it reference})
parameters
\be\label{refpar}
G_F\,,\quad\quad M_Z\,, \quad\quad \alpha_e\equiv\alpha(M_Z^2)\,,
\ee
which are most precisely measured.} In what follows we use, as in 
\cite{Hsieh:2010zr},
the value of the QED coupling constant in the $\overline{MS}$ scheme:
\be
1/\alpha(M_Z^2)= 127.916\pm0.015\,.
\ee
In the case of the fit parameters that parametrise NP contributions, we choose
\be\label{fitpar}
\epsilon\,, \qquad s_R\,, \qquad s\,,
\ee
which are related to the ones in \cite{Hsieh:2010zr} through
\be\label{dictionary}
\tilde x=\frac{1}{\epsilon^2}\,, \quad c_{\tilde\phi}=c_R\,, 
\quad \sin 2\tilde\beta =2sc\,.
\ee
The goal of \cite{Hsieh:2010zr} is to find the allowed ranges for 
the three parameters in (\ref{dictionary}) by fitting 37 EWP observables 
to the existing data. To this end the SM contributions are included at 
the tree and one-loop level, while NP contributions are taken into account at the tree level. This 
is sufficient for our purposes.
{Before describing the manner in which we use the results of \cite{Hsieh:2010zr},
let us make a few comments on how our choice of reference parameters affects
the $\ord(\epsilon^2)$ corrections in our formulae that 
have been exclusively written in terms of {\it model parameters}.} The latter 
parameters have been distinguished in \cite{Hsieh:2010zr} through 
a {\it tilde}. Expressing these parameters through the parameters 
in (\ref{refpar}) and (\ref{dictionary}) we find that
\begin{itemize}
\item
$G_F$ and $M_Z$, as expected, do not receive any $\ord(\epsilon^2)$ 
corrections. This can be explicitly verified by expressing the model 
parameters in terms of the parameters (\ref{refpar}) and (\ref{dictionary}) 
and inserting in the formula for $M_Z$ given in appendix \ref{app:gauge-boson-masses}. The $\ord(\epsilon^2)$ 
correction cancels when $v$ and $s_W$ are defined by \footnote{Our $v$ is 
by $\sqrt{2}$ smaller than the one used in \cite{Hsieh:2010zr}.}
\be\label{defs}
v^2=\frac{1}{2\sqrt{2}G_F}\,,\qquad s_W^2c_W^2=\frac{\pi\alpha(M_Z^2)}{\sqrt{2}M_Z^2G_F}\,.
\ee
\item
On the other hand $M_W$, $M_{Z'}$ and $M_{W_R}$ are given as follows:
\be\label{MWSM}
M_W=M_Zc_W\left[1+\frac{\epsilon^2}{2}\frac{c^2_W}{c_W^2-s_W^2}(\frac{c_R^4}{4}-2s^2c^2)\right]\,,
\ee
\be\label{MPRIMES}
(M_{W_R})^2 = \frac{e^2 \kappa_R^2}{ c_W^2 s_R^2}\,, \quad
(M_{Z'})^2 = \frac{2 e^2 \kappa_R^2}{ c_R^2c_W^2 s_R^2}\,,
\ee
where we have dropped $\ord(\epsilon^2)$ corrections in the last equation, 
$s_W$ and $c_W$ are defined through (\ref{defs}) and the one-loop SM corrections 
have not been shown in (\ref{MWSM}).
\end{itemize}

The implications for our analysis are as follows:
\begin{itemize}
\item
The formula (\ref{MWSM}) after the inclusion of SM electroweak loop corrections 
together with other observables considered in \cite{Hsieh:2010zr} can 
be used to constrain the fit parameters in (\ref{dictionary}).
However in the present paper we simplify this analysis by requiring that 
the shift in $M_W$ due to NP contributions is consistent within $2\sigma$ 
with the difference between its measured value and its SM prediction
\begin{eqnarray}\label{AJB0}
(\Delta M_W)^{\rm NP}&=&\epsilon^2\frac{M_Z}{2}\frac{c^3_W}{c_W^2-s_W^2}(\frac{c_R^4}{4}-2s^2c^2)= M_W^{\rm exp}-M_W^{\rm SM}\\
\nonumber&=& 0.036(34)~\gev\,,
\end{eqnarray}
where the SM and measured value are taken from \cite{Flacher:2008zq,Baak:2011ze}.
\item  
On the other hand (\ref{MPRIMES}) can be used, as done in \cite{Hsieh:2010zr}, 
to find lower bounds on the masses of these new gauge bosons. In turn 
incorporating the corresponding bounds from collider experiments can 
put additional constraints on (\ref{dictionary}). 
\item
Finally, in case $W^{\pm}_R$ and $Z'$ will be discovered and their 
masses precisely measured, the formulae in (\ref{MPRIMES}) will allow to take 
these values as reference parameters reducing the number of fit 
parameters to one.
\end{itemize}

\subsection{Basic constraints}
Using the dictionary in (\ref{dictionary}) and tables IX and X of \cite{Hsieh:2010zr} we 
find the following expressions for the four observables of interest:
\begin{itemize}
\item $\sigma_{\rm had}$ the partial branching fraction of $Z\to q\bar q$ with the NP correction
\begin{equation}
\delta\sigma_{\rm had}/\sigma_{\rm had,SM} = \left[-1.13 \frac{c_R^2}{4} - 0.142 \frac{c_R^4}{4} + 0.0432 (2s^2c^2)\right]\epsilon^2
\end{equation}
\item the forward-backward asymmetry $A_{\rm FB}(b)$ 
\begin{equation}
\delta A_{\rm FB}(b)/A_{\rm FB,SM}(b) = \left[-30.0 \frac{c_R^2}{4} + 67.6 \frac{c_R^4}{4} - 20.6 (2s^2c^2)\right]\epsilon^2
\end{equation}
\item the weak charge $Q_W({\rm Cs})$ of the caesium-133 nucleus 
\begin{equation}
\delta Q_W({\rm Cs})/Q_{W,{\rm SM}}({\rm Cs}) = \left[-0.855 \frac{c_R^4}{4} - 0.145 (2 s^2c^2)\right]\epsilon^2
\end{equation}
\item the left-handed coupling measured in deep inelastic $\nu$-N scattering 
\begin{equation}
\delta (g_L^{N\nu})^2/(g_{L,{\rm SM}}^{N\nu})^2 = \left[ 0.0219 + 0.478 c_R^2 + 0.210 c_R^4 - 1.42 (4s^2c^2) \right] \epsilon^2
\end{equation}
\end{itemize}
A more detailed description of the observables considered here is presented in e.g. \cite{Hsieh:2010zr,Z-Pole}.

The correct treatment of the constraints is then given by the following formula
\be
|\text{EXP} - \text{\rm SM} (1 + \text{CON})| \leq \sqrt{(\Delta \text{EXP})^2 + (\Delta \text{\rm SM} (1 + \text{CON}))^2}\,,
\ee
where EXP and SM stand for the experimental and SM value, respectively. CON stands for one of the conditions we have listed above. 
Furthermore we use the constraint from the W boson mass as given in 
(\ref{AJB0}), where the SM and experimental values are
taken from \cite{Flacher:2008zq,Baak:2011ze}.

We also incorporate the direct experimental constraints from muon decay of the TWIST Collaboration \cite{Bayes:2011zz}. 
They present two bounds on the ratio of $g_R/g_L$ times the mixing angle and one on $g_L/g_R$ with respect to the $W_R$ mass. 
As $s_W/(c_W s_R)=g_R/g_L$ holds, we can make further simplifications as in the first constraint the factor $g_R/g_L$ cancels. 
In our notation their results read then
\begin{equation}
s c \epsilon^2 <0.020 \qquad \text{and} \qquad \frac{c_W s_R}{s_W} M_{W_R} > 578 \gev 
\end{equation}
at 90\% C.L..

Moreover we take into account the direct experimental bounds on $M_{W_R}$.  While \cite{Aad:2011yg,CMS-PAS-EXO-11-024} report  stringent constraints $>2\tev$ on the mass of heavy $W'$ gauge bosons, these bounds only apply to the case of very light right-handed neutrinos which escape detection. In case of a non-negligible $M_{\nu_R}<M_{W_R}$ the constraints on the masses have been analysed in \cite{CMS-PAS-EXO-11-002}, with the bound on $M_{W_R}$ reaching up to $1.7\tev$ depending on the right-handed neutrino mass, with only $240 \,\text{pb}^{-1}$ of data analysed so far. On the other hand if $M_{\nu_R}>M_{W_R}$ the branching ratio of $W_R$ decaying to leptons is $\epsilon^2$ suppressed and the dominant decay mode is into quarks. In that case the searches for resonances in the dijet mass distribution \cite{Chatrchyan:1370086,Aad:2011fq} apply, yielding the constraint $M_{W_R}>1.5\tev$. We note that all these bounds depend on the ratio $g_L/g_R$ through the production cross-section for $W_R$. In order to keep our analysis independent of the details of the LRM neutrino sector, we will assume $M_R \geq 2 \tev$ throughout our analysis.


Finally for completeness we include the bounds $g_R^2 < 4 \pi$ and $g'^2  < 4 \pi$ in order to guarantee perturbativity of the gauge couplings.

\section{Strategy for the numerical analysis}\label{sec:strategy} 
\subsection{Preliminaries}
Having at hand all the relevant formulae for $\Delta F=2$ processes, 
$B\to X_{d,s}\gamma$ observables, EWP observables and tree level decays 
in the LRM in question, we are ready to perform a numerical analysis 
taking all existing experimental constraints into account. 

The first question which one could ask is whether there are regions in the parameter space of the LRM
for which all observables can be found in their allowed ranges. As the LRM has a large number of parameters, in particular
in the mixing matrix $V^{\rm R}$, it is probable that there are such regions even for an LHC accessible mass $M_{W_R}$. 

The next interesting question is if there are experimentally allowed simplified parametrisations for the RH mixing matrix
with only a few free parameters. There are a few popular yet already excluded matrices of this kind. We however derive
inspiration from those and propose a new and simple matrix.

In general regions in the parameters space fulfilling
\begin{itemize}
\item the fine-tuning of parameters is small,
\item the anomalies in the present data are softened or removed,
\end{itemize}
are particularly interesting. These regions can be considered 
as different 'oases' in the parameter space which are disconnected and
lead in general to different phenomenology.

\subsection{Anomalies in the flavour data}
Before proceeding with outlining our strategy we now review the known flavour anomalies.
\subsubsection[The $\varepsilon_K-S_{\psi K_S}$ anomaly]{\boldmath The $\varepsilon_K-S_{\psi K_S}$ anomaly}
It has been pointed out in \cite{Buras:2008nn,Buras:2009pj} that the SM prediction for $\varepsilon_K$ 
implied by the measured value of $S_{\psi K_S}=\sin 2\beta$, the ratio 
$\Delta M_d/\Delta M_s$ and the value of $|V_{cb}|$ turns out to be too
small to agree well with experiment. This tension between $\varepsilon_K$ and
$S_{\psi K_S}$ has been pointed out from a different perspective in
\cite{Lunghi:2008aa,Lunghi:2009sm,Lunghi:2009ke,Lunghi:2010gv}.
These findings have been confirmed by a UTfit  analysis \cite{Bevan:2011zz}. 
The CKMfitter group having a different treatment of uncertainties finds less significant effects in $\varepsilon_K$ \cite{Lenz:2010gu}.
Indeed taking the experimental value of $S_{\psi K_S}=0.679\pm 0.020$, 
$\vcb=0.0406$, the most recent value of the relevant non-perturbative 
parameter $\hat B_K=0.737\pm0.020$ 
\cite{Antonio:2007pb,Aubin:2009jh,Bae:2010ki,Constantinou:2010qv,Aoki:2010pe} 
resulting from unquenched lattice calculations and including long distance (LD) effects in 
${\rm Im}\Gamma_{12}$ and ${\rm Im}M_{12}$  in the $K^0-\bar K^0$ mixing
\cite{Buras:2008nn,Buras:2010pza} as well as recently calculated NNLO 
QCD corrections to $\varepsilon_K$ \cite{Brod:2010mj,Brod:2011ty} one finds 
\cite{Brod:2011ty}
\begin{equation}\label{epnew}
|\varepsilon_K|_{\rm SM}=(1.81\pm0.28)\cdot 10^{-3},
\end{equation}
visibly below the experimental value $|\varepsilon_K|_{\rm exp}=(2.228\pm0.011)\cdot 10^{-3}$. 
On the other hand $\sin 2\beta\approx 0.85\pm 0.05$ from SM  fits of the Unitarity Triangle (UT) 
is significantly larger than the experimental value $S_{\psi K_S}=0.679\pm 0.020$. This discrepancy 
is  to some extent  caused by the desire to fit $\varepsilon_K$ \cite{Lunghi:2008aa,Buras:2008nn,Buras:2009pj,Lunghi:2009sm,Lunghi:2009ke,Lunghi:2010gv} and 
$\Br(B^+\to\tau^+\nu_\tau)$ \cite{Lunghi:2010gv}. For the most recent 
discussions including up to date numerics see \cite{Lenz:2011ti,Laiho:2011nz,Barbieri:2011ci}. 
As demonstrated in \cite{Buras:2008nn,Buras:2009pj}, whether the NP is required in $\varepsilon_K$ or  $S_{\psi K_S}$ 
depends on the values of $\gamma$, $|V_{ub}|$ and $\vcb$. The phase $\gamma$ 
should be measured precisely by LHCb in the coming years while $\vub$ and 
$\vcb$ should be precisely determined by Belle II \cite{Aushev:2010bq} and SuperB \cite{Bona:2007qt,O'Leary:2010af,Meadows:2011bk} provided 
also the hadronic uncertainties will be under  better control. Here 
we concentrate briefly on $\vub$.

\subsubsection[$|V_{ub}|$ problem]{\boldmath $|V_{ub}|$ problem}

As already mentioned previously there is the tension between inclusive and exclusive determinations of $|V_{ub}|$ with the exclusive ones in the ballpark of $3.4\cdot 10^{-3}$ and the 
inclusive ones typically above $4.0\cdot 10^{-3}$. 
As discussed in \cite{Crivellin:2009sd} an interesting possible solution to 
this problem is the presence of RH charged currents, which selects the inclusive value as the true value (see our discussion in section~\ref{sec:tree}), 
implying again $\sin 2\beta\approx 0.80$ 
\cite{Buras:2010pz}. {Unfortunately in the LRM 
the increased value of the $W_R$ mass precludes this explanation if only points with acceptable fine-tuning are considered. 
Indeed the factor $sc\epsilon^2\le 10^{-3}$ from EWPT data and the 
bound on $|V_{ub}^{\rm R}|$ from FCNC processes as found 
by us, see section \ref{sec:vub_in_lram} for details, imply that the effect 
of RH currents is simply too small. Therefore one cannot state that the 
inclusive value is the favoured one from the point of view of the LRM and 
both the exclusive and inclusive determinations are equally valid.}

\subsubsection{Possible solutions}
As discussed in \cite{Lunghi:2008aa,Buras:2008nn} and subsequent papers of these authors a negative NP phase $\varphi_{B_d}$ in 
$B^0_d-\bar B^0_d$ mixing would solve the  $\varepsilon_K-S_{\psi K_S}$ anomaly,
 provided such a phase 
is allowed by other constraints.  Indeed, this is evident from (\ref{eq:3.42}).
With a negative $\varphi_{B_d}$ 
the true $\sin 2\beta$ is larger than $S_{\psi K_S}$, implying a higher value
on $|\varepsilon_K|$, in reasonable agreement with data and a better UT-fit. This 
solution would favour the inclusive value of $|V_{ub}|$. On the other hand 
as stressed in \cite{Buras:2008nn} a sizeable constructive NP physics 
contribution to $\varepsilon_K$ would not require an increased value 
of $\sin 2\beta$ relative to the experimental value of 
$S_{\psi K_S}$ and the exclusive value of $\vub$ would be favoured in this case.
 Clearly also
in such an analysis $\Delta M_{d,s}$ should be considered.

\subsubsection[Enhanced value of  $S_{\psi\phi}$]{\boldmath Enhanced value of  $S_{\psi\phi}$}
This topic became rather hot in the last years but the situation is still
unclear at present. The Tevatron data for $S_{\psi\phi}$ combined with 
the results for the same sign dimuon asymmetry of $D0$ indicated last spring 
non-standard CP violation 
in the $B_s$-system corresponding to $S_{\psi\phi}\approx 0.8$
\cite{Lenz:2011ti,Laiho:2011nz,Lunghi:2011xy}. On the other hand the 
subsequent measurements of CDF and D0 \cite{Giurgiu:2010is,Abazov:2011ry} and  the first 
more accurate results for $S_{\psi\phi}$ from LHCb \cite{Raven:1378074} 
indicate consistency with 
the SM. Yet, the range 
\be\label{Spsiphiexp}
-0.1 \le S_{\psi\phi}\le 0.4
\ee
summarised recently in \cite{Altmannshofer:2011iv}
leaves still a lot of room 
for sizeable NP contributions. 
Let us hope that the future data from Tevatron and in 
particular from the LHCb, will measure this asymmetry with sufficient 
precision so that we will know to which extent NP is at work here.

\subsection{Addressing the flavour anomalies in the LRM}
Now we want to discuss how the LRM can in principle address the anomalies described above.
As the LRM has a large number of parameters, in particular
in the mixing matrix $V^{\rm R}$, it is obvious that for certain choices of these parameters, 
most, or even all of these anomalies can be removed. Yet, as we will see  the situation is more involved than one might think at first 
sight. In order to better understand this matter we introduce an 
analytic expression for $V^{\rm R}$ valid in a specific oasis
of the parameter space. This matrix is given in terms of only few parameters. Other oases will be considered later on.

In this manner we see that already the present data 
give us some hints on possible structures of the $V^{\rm R}$ matrix. The fact 
that the low energy flavour data can give us such information about 
physics taking place at scales of few TeV is clearly remarkable and 
complementary to direct collider searches for NP at the LHC, where it is 
basically impossible to learn about the structure of $V^{\rm R}$. With improved 
flavour data we should be able to find out which of the oases considered 
by us, if any, is favoured by nature.

\subsubsection[Various scenarios for {$|V_{ub}|$}]{Various scenarios for \boldmath{$|V_{ub}|$}}
In view of the fact that the LRM cannot solve the $\vub$ problem, as mentioned 
above and discussed in section \ref{sec:vr_gen_num}, we introduce
two scenarios for $\vub$. In particular, we investigate the correlations between $\varepsilon_K$, $S_{\psi K_S}$, $S_{\psi\phi}$, $\Delta M_d$ and $\Delta M_s$ 
in these two scenarios for $|V_{ub}|$ setting $\gamma=68^\circ$. These scenarios are defined as 
follows:
\begin{enumerate}
\item {\bf \boldmath Small  $|V_{ub}|$}\\[2mm]
In this scenario we set
\be
|V_{ub}|=3.4\times 10^{-3}\,.
\ee
In this scenario within the SM we find $S_{\psi K_S}\approx 0.675$ in agreement with the 
data but $\varepsilon_K\approx 1.8\times 10^{-3}$ is visibly below the data.
\item {\bf\boldmath Large $|V_{ub}|$}\\[2mm]
In this scenario we set
\be
|V_{ub}|=4.4\times 10^{-3}\,.
\ee
This is the case, considering again the SM, in which $\varepsilon_K$ is consistent with the data while 
$S_{\psi K_S}\approx 0.82$, significantly above the data.
\end{enumerate}

We now illustrate with an example how all these discrepancies 
can be solved with a special form of the matrix $V^{\rm R}$ that depends 
only on four parameters: $\tilde s_{13}$, $\tilde s_{23}$ and two 
phases $\phi_1$ and $\phi_2$, all chosen to be in the first 
quadrant.
We want to stress that these scenarios are only used in our discussion regarding the
simplified matrix introduced in the following section. 
As already discussed in section \ref{sec:tree}
in our general analysis we 
treat the constraints on $\vub$ and $\gamma$ like any other measurement.

\subsubsection[Understanding  anomalies through {$V^{\rm R}$}]{Understanding  anomalies through \boldmath{$V^{\rm R}$}}\label{sec:Matrix}
In this section we want to propose a new simplified RH mixing matrix and subsequently
illustrate how it can be used to solve the flavour anomalies.
This matrix has been found with the goal to have very simple expressions 
for the $\Delta F=2$ observables while satisfying all existing constraints. 
To this end it turned out useful to set $\ts_{12}=0$ in $V_0^{\rm R}$ 
as in this manner, as seen in (\ref{eq:Vtrgen}), 
five entries in this matrix simplified considerably. Moreover, the $\Delta M_K$
constraint has been relaxed in this way. The reduction of the number of phases
to two turned out also to improve the transparency.

The resulting special form of $V^{\rm R}$ is then given by 
\be
V^{\rm R} = 
\left(\begin{array}{ccc}
-\tc_{13}e^{-i\phi_1}   &   0       &\ts_{13}\\ 
-\ts_{23}\ts_{13}e^{i(\phi_2-2\phi_1)} &-\tc_{23}e^{-i\phi_1}& 
-\ts_{23}\tc_{13}e^{i(\phi_2-\phi_1)}\\ 
\tc_{23}\ts_{13}e^{-i\phi_1}& -\ts_{23}e^{-i\phi_2}&\tc_{23}\tc_{13}
\end{array}\right)~.
\label{eq:golden Vr}
\ee

In order to derive simple formulae we restrict the mixing angles to the phenomenologically viable ranges
\be
\ts_{13}\lsim 0.02 \,, \qquad \ts_{23}\lsim 0.2 \,, 
\ee
with the constraints coming from the $B_d$ and $B_s$ system, respectively. Without further assumptions on the phases involved, a much stronger combined constraint
\be
\ts_{13}\ts_{23} \lsim 10^{-5} 
\ee
can be derived from $\eps_K$. In this range of parameters 
we find then  to a very good 
approximation that the terms proportional to $\lambda_t^{\rm LR}\lambda_t^{\rm RL}$ 
dominate as far as $\Delta M_{d,s}$, $\varepsilon_K$, $S_{\psi K_S}$ and $S_{\psi\phi}$ are concerned. 
For $\Delta M_K$ also $\lambda_t^{\rm LR}\lambda_c^{\rm RL}$ is relevant but 
the appearance of the same phase $\phi_1$ in $V^{\rm R}_{td}$ and $V^{\rm R}_{cs}$ 
makes its contribution to $\varepsilon_K$ to be negligible.
We find then
\be\label{AJB1a}
( \Im~M_{12}^K)_{\rm LR}=|R_{tt}(K)|\times\frac{G_F^2 M_W^2}{12\pi^2}F_K^2m_K|V_{td}^{\rm L}|  |V_{ts}^{\rm L}|\tc_{23}
\ts_{13}\ts_{23}\sin(\phi_2-\phi_1-\beta+\beta_s)\,.
\ee

For $B_q^0-\bar B_q^0$ mixing we have
\be\label{CBq}
C_{B_q}e^{2i\varphi_{B_q}}=1-\frac{|R_{tt}(B_q)|}{S_0(x_t)\hat B_{B_q} \eta_B}
\left[\frac{\lambda_t^{RR}(B_q)}{\lambda_t^{LL}(B_q)}\right]^*\,.
\ee
Using the matrix (\ref{eq:golden Vr}) we obtain
\be
\left[\frac{\lambda_t^{RR}(B_d)}{\lambda_t^{LL}(B_d)}\right]^*=\frac{\tc_{23}\ts_{13}V_{tb}^{\rm R}}{|V_{td}^{\rm L}|}e^{i(\phi_1-\beta)}\,,
\qquad
\left[\frac{\lambda_t^{RR}(B_s)}{\lambda_t^{LL}(B_s)}\right]^*=
\frac{\ts_{23}V_{tb}^{\rm R}}{|V_{ts}^{\rm L}|}e^{i(\phi_2-\beta_s)}\,,
\ee
and consequently
\bea\label{AJB2a}
\sin 2\varphi_{B_d} &=& -\frac{|z_d|}{C_{B_d}}\frac{\tc_{23}\ts_{13}V_{tb}^{\rm R}}{|V_{td}^{\rm L}|}\sin(\phi_1-\beta)\,,\\
\label{AJB3a}
\sin 2\varphi_{B_s}&=& -\frac{|z_s|}{C_{B_s}}\frac{\ts_{23}V_{tb}^{\rm R}}{|V_{ts}^{\rm L}|}\sin(\phi_2-\beta_s)\,.
\eea
Here 
\be
z_q=\frac{R_{tt}(B_q)}{S_0(x_t)\hat B_{B_q}\eta_B}\,.
\ee
Setting then $M_{W_R}=2.5\tev$, $s=0.1$ and $M_{H^0}=M_{H^\pm}=16\tev$  we find then 
for central values of other parameters
\be
|R_{tt}(K)|=9.1\,,\qquad |R_{tt}(B_q)|=0.57\,, \qquad |z_d|=0.36\,, \qquad |z_s|=0.34\,.
\ee
These quantities
summarise the dominant 
$S_{\rm LR}$ contribution that governs the box contributions with $(W_L,W_R)$ and $(W_L,H^\pm)$ as well as neutral 
Higgs tree level exchanges. As seen in (\ref{Rij}) it includes also QCD corrections, hadronic 
matrix elements and depends on the masses of $W_R$ and heavy Higgs particles. If only the box 
contributions involving $W_L$ and $W_R$ were taken into account the NP contributions to $\varepsilon_K$
would be reduced universally by $85\%$ allowing for larger values of 
the mixing parameters and phases in these three expressions. The Higgs dominance can be read off figure \ref{fig:lrcontrib}.

The fact that three quantities in question depend only on two new phases 
$\phi_1$ and $\phi_2$ implies interesting correlations in the two
scenarios for $\vub$ that we illustrate now. These correlations depend 
of course on the two positive real values of $\ts_{13}$ and $\ts_{23}$ as we 
 see below.  
\begin{itemize}
\item
In scenario 1, $S_{\psi K_S}$ within the SM agrees well with the data implying 
that $\phi_1\approx\beta$. This in turn implies that $\varepsilon_K$ and 
$S_{\psi\phi}$ are governed by the phases $\phi_2-2\beta$ and  $\phi_2$, respectively. The desire to 
reproduce the experimental value of $\varepsilon_K$ requires 
$\phi_2>2\beta-\beta_s$, and $S_{\psi \phi}$ can be enhanced simultaneously.
\item
In scenario 2, $\varepsilon_K$ in the SM agrees well with the data implying 
that $\phi_1\approx \phi_2-\beta+\beta_s$. Consequently  $S_{\psi K_S}$ and 
$S_{\psi \phi}$ are this time governed by the phases  $\phi_2-2\beta$ and  $\phi_2$, respectively. The desire to 
reproduce the experimental value of $S_{\psi K_S}$  requires again
$\phi_2>2\beta-\beta_s$ which also leads to an enhanced value of $S_{\psi \phi}$.
\end{itemize}

In summary the phases $\phi_1$ and $\phi_2$ should satisfy 
\bea
\phi_1 &\simeq & \left\{
\begin{array}{lr}
\beta\,, &\qquad\text{(scenario 1)}\\
\phi_2 - \beta + \beta_s\,, & \text{(scenario 2)}
\end{array} \right. \\[2mm]
\phi_2 &>& 2 \beta - \beta_s\,.
\eea
As we can see there are relations between the NP phases in the 
$B_d$ and $B_s$ systems but they are more involved than the ones characteristic for 
${\rm 2HDM_{\overline{MFV}}}$  \cite{Buras:2010mh,Buras:2010zm}.

\begin{figure}
\centering
\includegraphics[width=.4\textwidth]{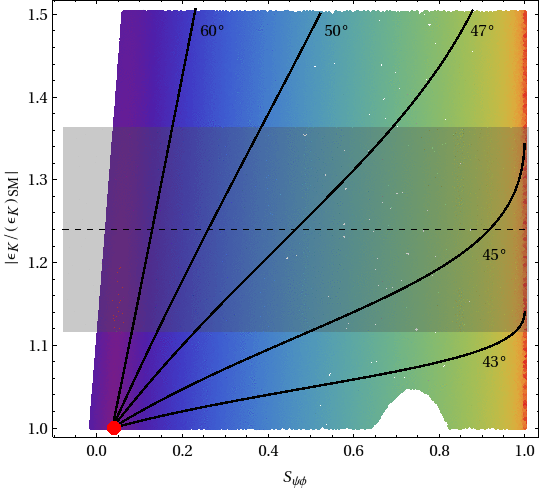}\quad
\includegraphics[width=.048\textwidth]{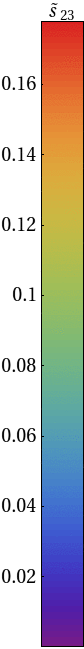}
\caption{Correlation between $S_{\psi\phi}$ and $\eps_K$ in scenario 1, for fixed $\phi_1 = \beta$ and $\tilde s_{13} = 3\cdot 10^{-3}$. The central SM value is indicated by the red dot.\label{fig:S1-Spsiphi-epsK}}
\end{figure}

In figure \ref{fig:S1-Spsiphi-epsK} we show the correlation between $\varepsilon_K$ and $S_{\psi\phi}$ 
in scenario 1. To this end we set $\phi_1 = \beta$ and $\ts_{13}=3\cdot 10^{-3}$ and vary $\ts_{23}<0.2$ and $2 \beta-\beta_s<\phi_2<\pi+2 \beta-\beta_s$, imposing the experimental constraints on $\Delta M_s$ and $\Delta M_d$.
The SM point is reached for $\ts_{23}=0$. The black curves correspond to constant $\phi_2$ values as indicated.   We observe that for not too small $\ts_{23}\gsim 0.05$ and $45^\circ < \phi_2 < 50^\circ$ good agreement with the data for $\eps_K$ can be achieved while at the same time enhancing $S_{\psi\phi}$ w.\,r.\,t.\ its tiny SM value. For larger values of $\phi_2$ a smaller $\ts_{23}$ is required in order to fit the $\eps_K$ data. At the same time the effects in $S_{\psi\phi}$ become much smaller. Within scenario 1 this simple structure for $V^{\rm R}$ can thus solve the $\eps_K$ anomaly. The measured value of $S_{\psi\phi}$ can then be used to measure the values of $\ts_{23}$ and $\phi_2$.

\begin{figure}
\centering
\includegraphics[width=.42\textwidth]{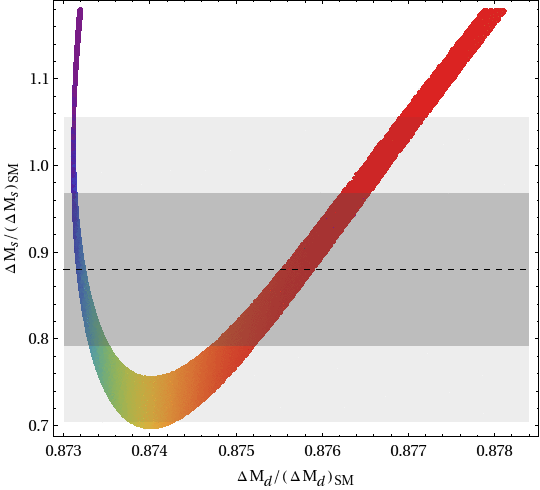}\quad
\includegraphics[width=.044\textwidth]{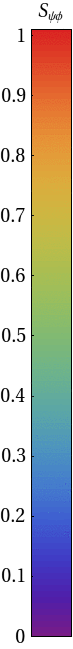}
\caption{Correlation between $\Delta M_d$ and $\Delta M_s$ in scenario 1, for fixed $\phi_1 = \beta$ and $\tilde s_{13} = 3\cdot 10^{-3}$, and imposing the constraint from $\eps_K$.\label{fig:S1-CBd-CBs}}
\end{figure}

In figure \ref{fig:S1-CBd-CBs} we show the correlation between $\Delta M_d$ and $\Delta M_s$, normalised to their SM values, after imposing the constraint from $\eps_K$. Again $\phi_1 = \beta$ and $\ts_{13}=3\cdot 10^{-3}$ are fixed, while $\ts_{23}<0.2$ and $2 \beta-\beta_s<\phi_2<\pi+2 \beta-\beta_s$ are varied. Since the observed values for both mass differences lie below their SM predictions, a suppression is welcome also in this case. We observe that $\Delta M_d$ is very close to its experimental central value and almost constant, since it depends to very good approximation only on $\phi_1$ and $\ts_{13}$ which are fixed. Also $\Delta M_s$ can easily be suppressed and brought closer to the data, which in turn favours moderate enhancements of $S_{\psi\phi}\lsim 0.4$ (light-blue points in the figure) or very large effects $S_{\psi\phi}\sim 1$ (red points) -- the latter being disfavoured by the LHCb data.

As a specific example for how the anomalies are solved in scenario 1, we quote the following parameter point:
\be
\ts_{13} = 3 \cdot 10^{-3}\,,\qquad \ts_{23} = 0.03\,,\qquad \phi_1 = \beta = 21^\circ\,,\qquad \phi_2 = 50^\circ\,.
\ee
With this choice of parameters we find
\be
S_{\psi K_S} \approx 0.67\,, \qquad |\eps_K|\approx 2.2\cdot 10^{-3} \,,\qquad S_{\psi\phi} \approx 0.27\,,
\ee
and
\be
\Delta M_d \approx 0.51\,\text{ps}^{-1}\,,\qquad \Delta M_s \approx 17.4,\text{ps}^{-1}\,
\ee
barring in mind the theoretical and parametric uncertainties that have been omitted here for the sake of simplicity.

Let us stress that this solution to the flavour anomalies is in fact the simplest one possible within scenario 1. Setting any of the four parameters in \eqref{eq:golden Vr} to zero would spoil the relations discussed above.

Let us now turn our attention to scenario 2, in which $\eps_K$ is in good agreement with the data, but $S_{\psi K_S}$ needs to be suppressed w.\,r.\,t.\ its SM value $\approx 0.82$. We show now that although this simple structure of $V^{\rm R}$ can yield some improvement over the SM situation, the tensions cannot be fully resolved in this case.

\begin{figure}
\centering
\begin{minipage}{6cm}
\includegraphics[width=\textwidth]{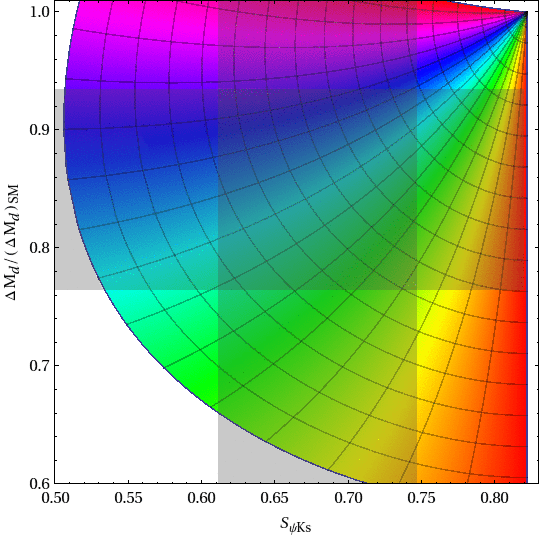}
\end{minipage}\quad
\begin{minipage}{6.2cm}
\includegraphics[width=\textwidth]{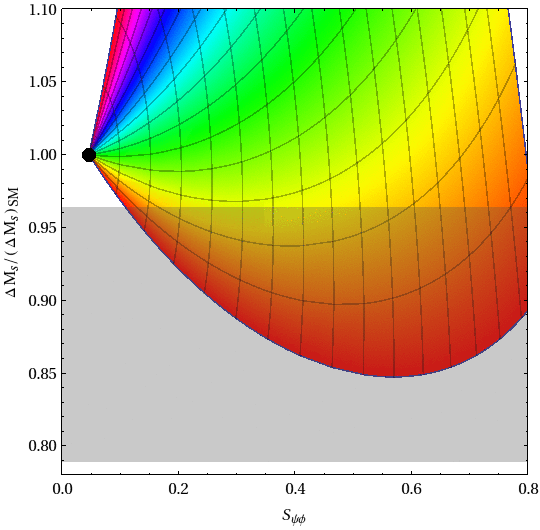}
\end{minipage}\quad
\begin{minipage}{1.cm}
\includegraphics[width=\textwidth]{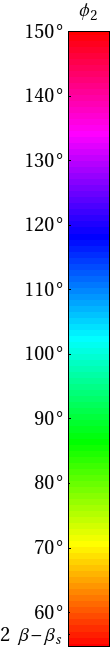}
\end{minipage}
\caption{{\it left:} $\Delta M_d/(\Delta M_d)_\text{SM}$ as a function of $S_{\psi K_S}$ for different values of $\ts_{13}<0.01$ and $2\beta-\beta_s<\phi_2<150^\circ$. {\it right:} $\Delta M_s/(\Delta M_s)_\text{SM}$ as a function of $S_{\psi \phi}$ for different values of $\ts_{23}<0.1$ and $2\beta-\beta_s<\phi_2<150^\circ$. The SM values are displayed as black dots, while the experimental $1\sigma$ regions are indicated by grey bands. \label{fig:S3-matrix}}
\end{figure}

To this end we show in the left panel of figure \ref{fig:S3-matrix} $\Delta M_d/(\Delta M_d)_\text{SM}$ as a function of  $S_{\psi K_S}$ for different values of $\ts_{13}<0.01$ and $2\beta-\beta_s<\phi_2<150^\circ$ and fixed $\phi_1 = \phi_2 -\beta+\beta_s$. We observe that in order to simultaneously fit the data on $S_{\psi K_S}$ and $\Delta M_d$, $\phi_2$ is required to lie roughly in the range $90^\circ \dots 120^\circ$.

In the right panel of figure \ref{fig:S3-matrix} we show $\Delta M_s/(\Delta M_s)_\text{SM}$ as a function of  $S_{\psi \phi}$ for different values of $\ts_{23}<0.1$ and $2\beta-\beta_s<\phi_2<150^\circ$. Here we observe that in order to obtain a suppression of $\Delta M_s$ w.\,r.\,t.\ the SM prediction, as favoured by the data, a much smaller phase $\phi_2 \lsim 70^\circ$ is required. An enhancement of $S_{\psi\phi}$ over its SM value is then obtained automatically. Such a low value for $\phi_2$ would in turn allow only for a mild suppression of $S_{\psi K_S}$ with respect to its SM value. This exercise shows very clearly how important it is to consider 
various observables simultaneously. Each of the panels in figure \ref{fig:S3-matrix} taken alone allows to remove the anomalies in $B_d$ and $B_s$ systems, respectively. 
But when they are taken together a clear tension between the suppression 
of $\Delta M_s$ and suppression of $S_{\psi K_S}$ is present.

In view of the present non-perturbative uncertainties, it is not possible to rule out this structure of $V^{\rm R}$ as a solution to the flavour anomalies within scenario 2, however we note that this scenario is disfavoured since only a slight amelioration of the SM tensions is possible. On the other hand, as seen before, this simple matrix works beautifully in solving all anomalies within scenario 1.

In fact the conclusion is much more general: Within scenario 2 it is not possible to solve the flavour anomalies if  the $tt$ contribution dominates all observables in question. To see this, let us parametrise
\be
V^{\rm R}_{td} = |V^{\rm R}_{td}|e^{-i \phi_1}\,,\qquad V^{\rm R}_{ts} = -|V^{\rm R}_{ts}|e^{-i \phi_2}\,,\qquad V^{\rm R}_{tb} = |V^{\rm R}_{tb}|e^{-i \phi_3}\,.
\ee
{The constraint from $\eps_K$ in \eqref{AJB1a} again implies  $\phi_1\approx \phi_2 - \beta + \beta_s$. In the formulae for $B_{d,s}$ mixing (\eqref{AJB2a}, \eqref{AJB3a}) $\phi_2$ is then replaced by $\phi_2 - \phi_3$. However, since the same combination of phases appears in both meson systems} it cannot help to ameliorate the tension between $B_d$ amd $B_s$ data. Thus in order to solve the flavour anomalies in the LRM in case of a large $|V_{ub}|$ value, charm contributions have to be relevant.

These correlations are reminiscent of the ones found in the context of 
a 2HDM with flavour blind phases (${\rm 2HDM_{\overline{MFV}}}$) 
 \cite{Buras:2010mh,Buras:2010zm}. The difference is in the 
direct contribution to $\varepsilon_K$ in scenario 1 which was absent in the 2HDM in question.

We should also emphasise that with $V^{\rm R}_{tb}$ being close to unity, in 
both scenarios the branching ratio $\Br(B\to X_s\gamma)$ is 
also enhanced over the SM value bringing the central theoretical value
of this branching ratio closer to experiment.

While showing this example we do not claim that the matrix in 
(\ref{eq:golden Vr}) is the only one that is capable of removing 
simultaneously (scenario 1) or ameliorating (scenario 2) all anomalies in a correlated manner, but possibly this 
is the simplest matrix achieving this goal. One can also check that 
moving to different quadrants of the mixing angles and phases would generally 
imply different correlations between the observables in question and 
this fact illustrates how one can determine $V^{\rm R}$ once the data on 
$S_{\psi\phi}$, $S_{\psi K_S}$ and $\vub$ improve.

\subsection{Nominal input parameters}
We have collected the input parameters required for the numerical analysis 
in tables~\ref{tab:input} and \ref{tab:runningmasses}.
The values of other parameters like $P_i^a$ {and the tree level constraints on the CKM matrix} have been given in previous 
sections. At this stage it is important to recall first the theoretical uncertainties 
in the constraints used by us.
Considerable progress in lattice calculations has been made in recent years
reducing the uncertainties in $F_{B_s}$ and $F_{B_d}$ and also in
$\sqrt{\hat B_{B_s}}F_{B_s}$ and $\sqrt{\hat B_{B_d}}F_{B_d}$ down to $5\%$. This 
implies an uncertainty of $10\%$ in $\Delta M_d$ and  $\Delta M_s$  within the SM. The numerical values and their respective errors are given in table \ref{tab:input}. Even bigger progress has been made in the 
case of the CP-violating parameter $\varepsilon_K$, where the decay constant $F_K$ is known with 1\% 
accuracy. Moreover the parameter $\hat{B}_K$ is known within 
3\% accuracy from lattice calculations with dynamical fermions 
\cite{Antonio:2007pb,Aubin:2009jh,Bae:2010ki,Constantinou:2010qv,Aoki:2010pe} 
and an improved 
estimate of long distance contributions to $\varepsilon_K$ reduced this uncertainty down to $2\%$ \cite{Buras:2008nn,Buras:2010pza}. 
Also the calculation of NNLO QCD perturbative corrections to $\varepsilon_K$   \cite{Brod:2010mj,Brod:2011ty} improved the theoretical status of $\varepsilon_K$ by much.
The situation with other $B_i$ parameters describing the hadronic matrix 
elements of other operators is much worse as clearly seen in the errors of 
the factors $P_i^a$ in (\ref{PI1}) and (\ref{PI2}).
Here a significant progress is desired. 

Finally let us recall that the CP-asymmetries $S_{\psi\phi}$ and $S_{\psi K_S}$ 
have rather small hadronic uncertainties 
\footnote{See \cite{Faller:2008zc,Faller:2008gt} for more details 
and references therein.} and the hadronic uncertainties in the {ratio}
$\Delta M_d/\Delta M_s$ amount to roughly 3\%. Also the theoretical uncertainties in the rate $B\to X_s\gamma$ decay are below $10\%$.

On the experimental side  $\Delta M_d$,  $\Delta M_s$ and $\eps_K$ are very 
precisely measured, so that their experimental errors can be neglected for all practical purposes, while
$S_{\psi K_S}$ is known with an uncertainty of $\pm 3\%$. $\Delta M_K$, while very accurately 
measured, is subject to poorly known long distance contributions
and we only require that $(\Delta M_K)_{\text{exp}}$ is reproduced within $\pm 30\%$. Finally the 
rate for  the $B\to X_s\gamma$ decay is known within the accuracy of $10\%$, 
comparable to the theoretical one.

\begin{table}
\renewcommand{\arraystretch}{1}\setlength{\arraycolsep}{1pt}
\centering
\begin{tabular}{|l|l|}
\hline
$G_\mu = 1.16637(1)\cdot 10^{-5}\gev^{-2}$   &  $\eta_1=1.87(76)$ 
\hfill\cite{Brod:2011ty}\\\cline{2-2}
$M_W = 80.399(23) \gev$  & $\eta_3= 0.496(47)$\hfill\cite{Herrlich:1996vf,Brod:2010mj}\\\cline{2-2}
$\alpha(M_Z) = 1/127.9$ & $\eta_2=0.5765(65)$ \hfill\cite{Buras:1990fn} \\\cline{2-2}
$\alpha_s(M_Z)= 0.1184(7) $ & $\eta_B=0.55(1)$ \hfill \cite{Buras:1990fn,Urban:1997gw} \\\cline{2-2}
$\sin^2\hat{\theta}_W = 0.23116(13)$  & $F_K = 156.0(11)\mev$\\
$m_K^0= 497.614(24)\mev$   & $\hat B_K= 0.737(20)$ \\
$\Delta M_K= 0.5292(9)\cdot 10^{-2} \,\text{ps}^{-1}$ \qquad {} &  $F_{B_d} = 205(12)\mev$ \\
$|\eps_K|= 2.228(11)\cdot 10^{-3}$ & $F_{B_s} = 250(12)\mev$\\
$m_{B_d}= 5279.5(3)\mev$ & $\hat B_{B_d} = 1.26(11)$ \\
$m_{B_s} = 5366.3(6)\mev$ \hfill\cite{Nakamura:2010zzi} & $\hat B_{B_s} = 1.33(6)$\\\cline{1-1}
$\Delta M_d = 0.507(4) \,\text{ps}^{-1}$ & $F_{B_d} \sqrt{\hat B_{B_d}} = 233(14)\mev$ \\
$\Delta M_s = 17.77(12) \,\text{ps}^{-1}$ &  $F_{B_s} \sqrt{\hat B_{B_s}} = 288(15)\mev$ \\
$\tau_{B_s} = 1.471(25)\,{\rm ps}$ & $\hat B_{B_s}/\hat B_{B_d} = 1.05(7)$\qquad{}\\
$\tau_{B_d} = 1.519(7)\,{\rm ps}$ & $\xi = 1.237(32)$ \hfill \cite{Laiho:2009eu}\\
$\sin(2\beta)_{b\to c\bar c s}= 0.679(20) $ \hfill\cite{Asner:2010qj} &  \\\hline
$m_c(m_c) = 1.268(9)\gev$\cite{Laiho:2009eu,Allison:2008xk}&  \\
$m_t(m_t) = 163(1)\gev$ & \\
$m_b(2.5\gev) = 4.60(3)\gev$ &\\\hline
\end{tabular}
\caption{Values of the experimental and theoretical quantities used as input parameters. \label{tab:input}}
\renewcommand{\arraystretch}{1.0}
\end{table}

\begin{table}
\centering
\begin{tabular}{|l|c|c|c|c|c|}
\hline
 & $2{\rm GeV}$ & $4.6{\rm GeV}$ & $172{\rm GeV}$ & $2.5{\rm TeV}$ & $15{\rm TeV}$ \\\hline\hline
$m_u(\mu) ({\rm MeV})$ & $2.09(0)(9)$ & $1.74(6)(7)$ & $1.15(8)(5)$ & $0.97(8)(4)$ & $0.88(8)(4)$\\\hline
$m_d(\mu) ({\rm MeV})$ & $4.73(0)(11)$ & $3.94(1)(9)$ & $2.61(2)(6)$ & $2.19(2)(5)$ & $2.00(2)(5)$\\\hline
$m_s(\mu) ({\rm MeV})$ & $93.6(2)(11)$ & $77.9(3)(9)$ & $51.6(4)(6)$ & $43.4(4)(5)$ & $39.5(4)(5)$\\\hline
$m_c(\mu) ({\rm MeV})$ & $1089(7)(0)$  & $907(6)(0)$ & $601(5)(0)$ & $505(4)(0)$ & $460(4)(0)$\\\hline
$m_b(\mu) ({\rm GeV})$ &  -- & $4.074(19)(0)$ & $2.702(14)(0)$ & $2.268(12)(0)$ & $2.068(12)(0)$\\\hline
$m_t(\mu) ({\rm GeV})$ &  -- & -- & $162.3(10)(0)$ & $136.3(9)(0)$ & $124.2(8)(0)$\\\hline
\end{tabular}
\caption{The NLO running quark masses at different scales. The first and the second parenthesis shows the statistical and systematic error, respectively.}\label{tab:runningmasses}
\end{table}

\section{A numerical (re-)analysis of EWPT constraints}\label{sec:EWPNUM}

In this section we revisit the constraints from EWPT following the analysis in \cite{Hsieh:2010zr} and adding a few new constraints as outlined in section \ref{sec:EWP}. 

\subsection{Numerical procedure for the EWP tests}
As discussed in section \ref{sec:EWP} we can reduce the free parameters in the electroweak sector
down to three. We randomly scan over the full ranges of $0 < s < 1/\sqrt 2$, $0.1 \lsim s_R < 1$ and $0 < \epsilon < 0.1$. {For every parameter point we evaluate the $\chi^2$ function corresponding to the constraints discussed in section \ref{sec:EWP}.}

\subsection{Electroweak precision constraints}\label{sec:EWPconstraints}
First of all we want to show the correlations between the three electroweak 
parameters we have chosen in section~\ref{sec:EWP}, that is $\epsilon$, $s_R$ 
and $s$ and in this manner establish the allowed ranges for these parameters 
to be used in the analysis of FCNC processes.
\begin{figure}
\centering
\includegraphics[width=0.78\textwidth]{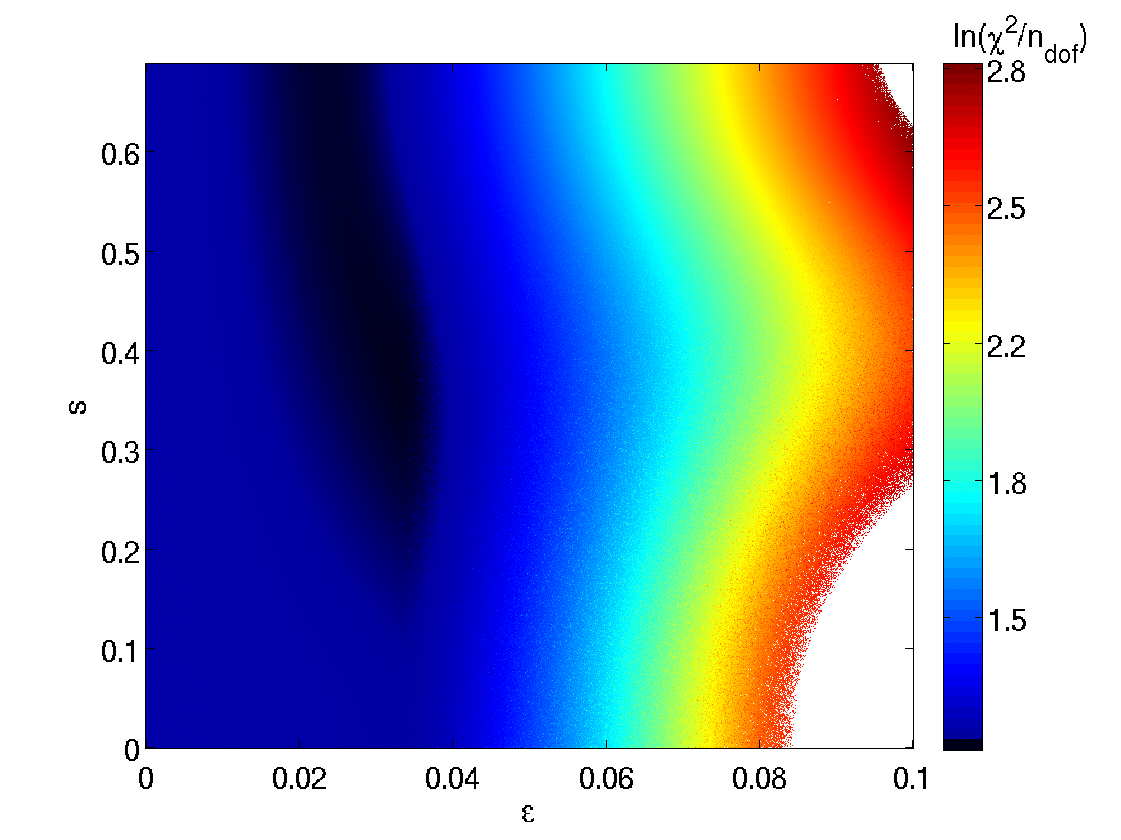}
\caption{In this figure we show $s$ as a function of $\epsilon$. The colour spectrum corresponds to $\ln(\chi^2/n_{\rm d.o.f.})$. }\label{fig:ewpar1}
\end{figure}
In figure \ref{fig:ewpar1} we show the correlation of $s$ and $\epsilon$ while encoding $\ln(\chi^2/n_{\rm d.o.f.})$ as a colour spectrum. As one can see
smaller values of $\epsilon$ are clearly favoured by the data as indicated by the colour gradient. Additionally one can see an area of a 
slightly darker blue at $\epsilon \sim 0.03$ and non-trivial $s$. This area  allows to soften the disagreement with $A^b_{\rm FB}$ as measured at LEP. 
This measurement has to be taken with a grain of salt because the competing SLD experiments did not measure a departure from the SM. We also note 
that values of $\epsilon$ above 0.08 are disfavoured.

\begin{figure}
\centering
\includegraphics[width=0.78\textwidth]{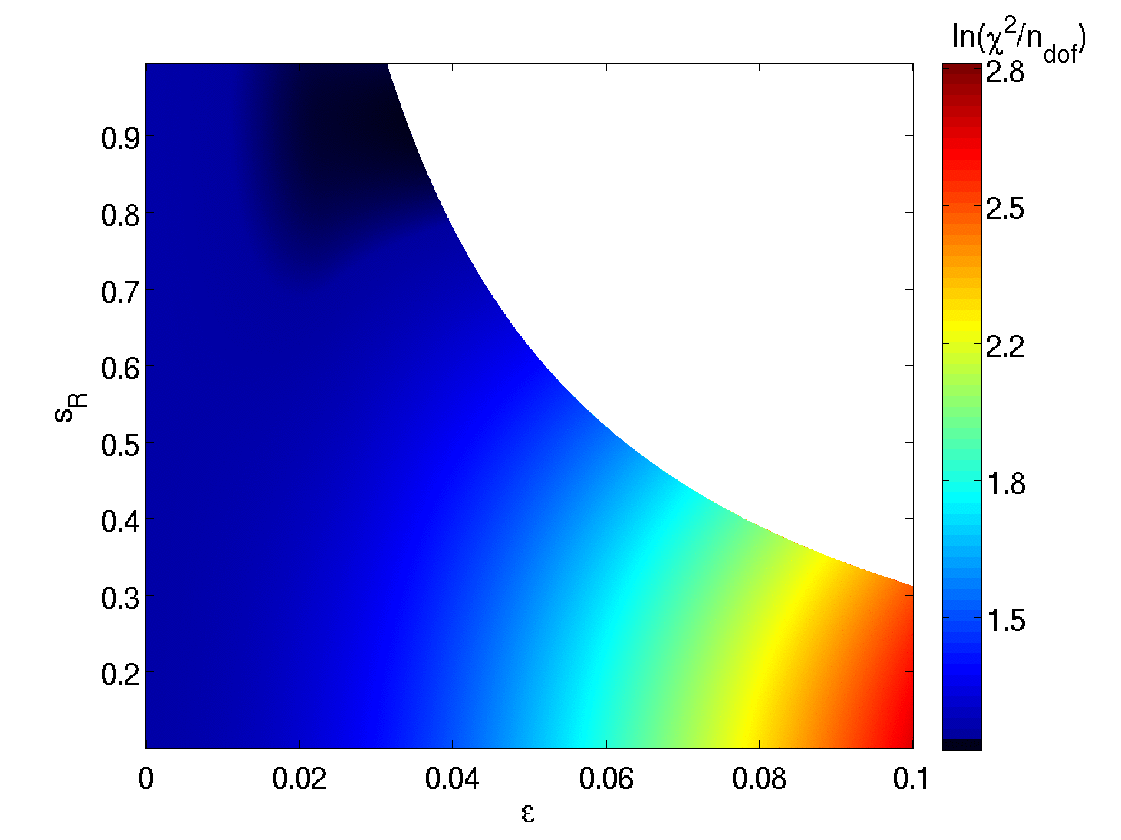}
\caption{In this figure we show $s_R$ as a function of $\epsilon$. The colour spectrum corresponds to $\ln(\chi^2/n_{\rm d.o.f.})$. }\label{fig:ewpar2}
\end{figure}
In figure \ref{fig:ewpar2} we show $s_R$ as a function of $\epsilon$. 
A preference for smaller values of $\epsilon$ can again be observed. For
values of $\epsilon\sim 0.03$ the value of $s_R$ has to be non trivial and even above $0.6$ in order to keep $\chi^2$ low. As in 
figure~\ref{fig:ewpar1} there is a slightly darker region for $\epsilon = 0.01\dots 0.03$ and $s_R = 0.7 \dots 1$ where the $A^b_{\rm FB}$ constraint
would be fulfilled within $2\sigma$.

\begin{figure}
\centering
\includegraphics[width=0.78\textwidth]{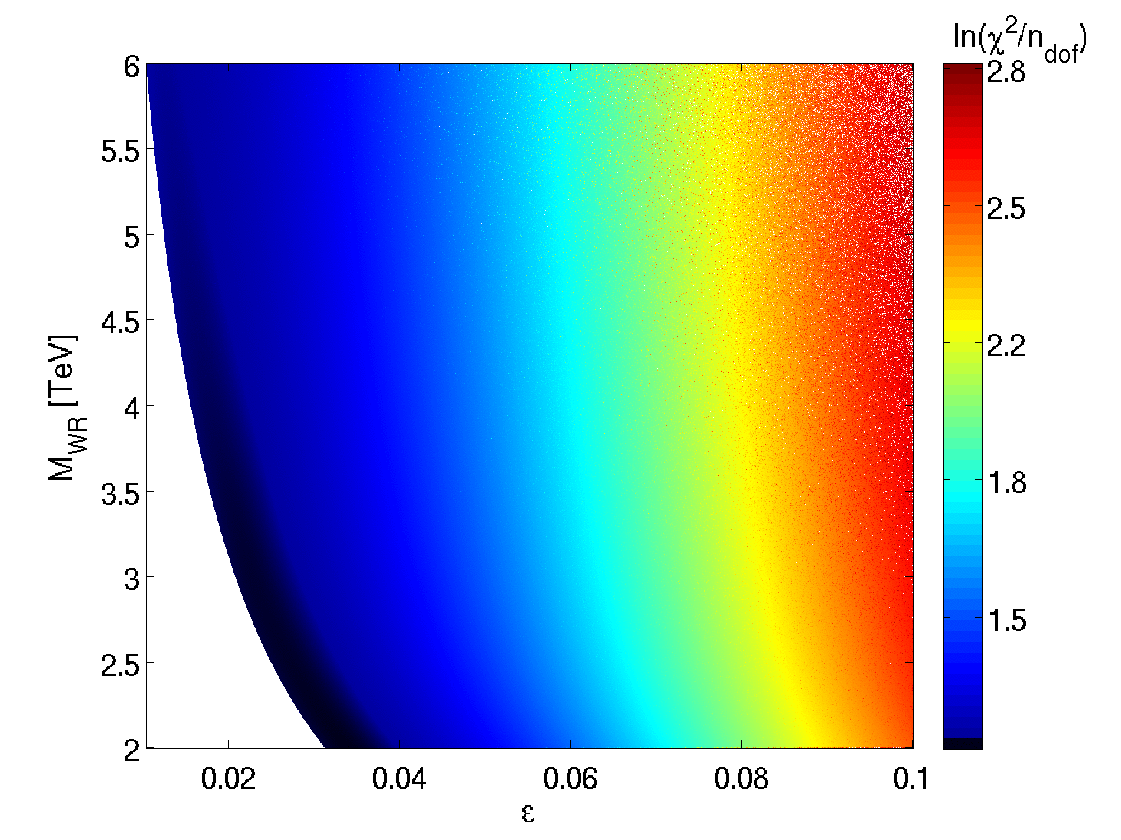}
\caption{In this figure we show $M_{W_R}$ as a function of $\epsilon$. The colour spectrum corresponds to $\ln(\chi^2/n_{\rm d.o.f.})$. }\label{fig:ewpar3}
\end{figure}
Figure \ref{fig:ewpar3} shows $M_{W_R}$ as a function of $\epsilon$. As expected the mass of the heavy gauge boson $W_R$ increases with a decreasing
$\epsilon$. The wide spread is due to other parametric dependencies. Interestingly in this picture the dark blue region implies an upper limit
on $M_{W_R}$. We conclude that in case $A^b_{\rm FB}$ as measured at LEP was true we would be able to put this model under pressure if $M_{W_R}$ is not found
below roughly $7\,{\rm TeV}$.

{For our analysis of $\Delta F=2$ observables we fix two of the electroweak parameters to a point with intermediate values
\begin{align}\label{eq:EWpoint}
  s_R = 0.80\,, && \epsilon &= 0.03\,.
\end{align}
The parameter $s$, which turns out to be very relevant for flavour observables, is varied in the range
\be
0.1 < s < 0.6\,.
\ee
This range of parameter points is contained in the dark blue area in figures \ref{fig:ewpar1}-\ref{fig:ewpar3}.  In this way we are able to incorporate the EWP constraints into our flavour analysis in a quite general manner.
Changing the electroweak reference point \eqref{eq:EWpoint} would affect mostly the masses $M_{W_R}$ and $M_H$. The Higgs mass $M_H$ additionally displays a strong dependence on the parameter $s$. We already studied the effect of changing the masses on $\Delta F=2$ observables and $\Br(B\to X_s\gamma)$ in sections \ref{sec:lr_gen_anatomy} and \ref{sec:bsg_anatomy}, respectively.
In terms of more tangible parameters our choice corresponds to the masses
\be
	M_{W_R} \approx 2.6\,{\rm TeV}\,, \qquad M_{H} \approx \frac{16}{\sqrt{1-2s^2}}\,{\rm TeV}\,, 
\ee
where we choose $\alpha_3 = 8$ in the Higgs potential. We emphasise that increasing the value $M_{H}$ by much can only be done at the price of loosing 
perturbativity in the scalar sector. 

The parameters relevant for our flavour analysis then read
\be
	1.3<\kappa/\kappa'<9.9\,,\qquad r = \frac{g_R^2}{g_L^2} \approx 0.48\,,\qquad 8.9\cdot 10^{-5} < sc\epsilon^2 < 4.3\cdot 10^{-4}\,.
\ee}

Note that using the constraints outlined in section \ref{sec:EWP} we are not able to fulfil $A_{\rm FB}^b$ at below $1.8\sigma$. This can
in part be accounted for by the approximate nature of the constraints used but in our opinion this should not be the dominant effect. This 
suggests that the LRM cannot explain the LEP $A_{\rm FB}^b$ anomaly, but only 
soften it.

\section{A general study of \boldmath{$V^{\rm R}$}}\label{sec:VR4S}

In this section we study the structure of the RH matrix $V^{\rm R}$.
In order to perform our general analysis of $V^{\rm R}$ and of flavour observables we varied all 13 parameters in the $V^{\rm L}$ and $V^{\rm R}$ matrices  in their allowed ranges.
For the parameters of $V^{\rm R}$ this means $0\dots \pi/2$ for the mixing angles and $0\dots 2\pi$ for the phases.
In the case of $V^{\rm L}$ parameters the situation is simpler as due to the smallness of $\varepsilon_K$ and
the tree-level constraints one is able to restrict the allowed ranges beforehand.
We impose all available constraints from $\Delta F=2$, $\Br(B\rightarrow X_q \gamma)$  as well as the 
tree-level measurements of CKM parameters within $2\sigma$. Our choice of electroweak parameters is described in section \ref{sec:EWPconstraints}.

\subsection[Allowed ranges for $V^{\rm R}$]{\boldmath Allowed ranges for $V^{\rm R}$}

{First we  study the matrix $V^{\rm R}$ in full generality, aiming to identify the valid regions of parameter space. To this end we allow the parameter $s$ to vary in the range $0.1\dots 0.6$. Due to the unitarity of the matrix $V^{\rm R}$ it is sufficient to consider the allowed ranges for the absolute values of $V^{\rm R}_{us}$, $V^{\rm R}_{ub}$ and $V^{\rm R}_{cb}$ in order to obtain a complete picture of the possible size of all elements. Therefore in figure \ref{fig:absVR_s} we show the correlations between pairs of these elements of $V^{\rm R}$. The value of $s$ is encoded in colour, where all plots show points with large $s$ in the front layer. Generally the allowed regions grow with decreasing $s$.
First we observe that large regions of the parameter space are already excluded. In addition all three elements can reach their maximal values $\sim 1$ individually, however not simultaneously. The two branches of points are easier to disentangle in the 3D version of these plots, as can be seen in figure \ref{fig:absVR} for $s=0.1$.
\begin{figure}
\centering
\includegraphics[width=0.49\textwidth]{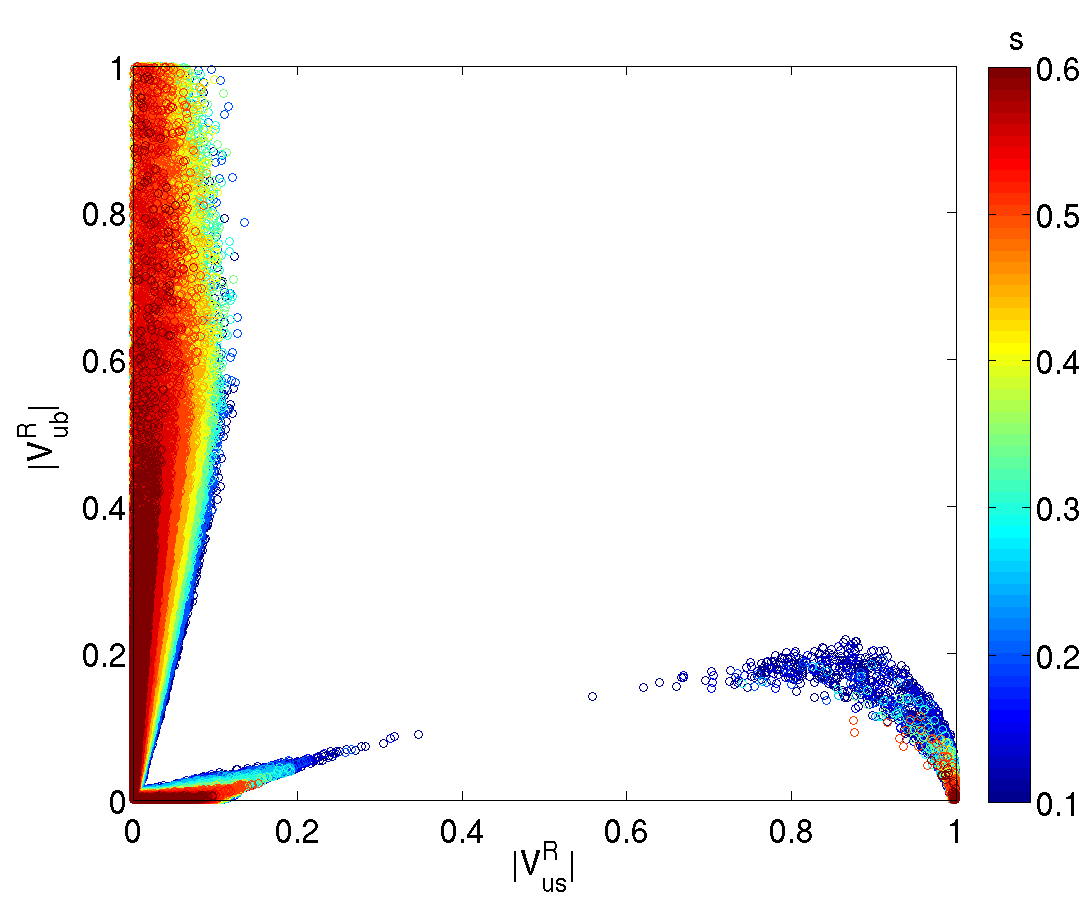}
\includegraphics[width=0.49\textwidth]{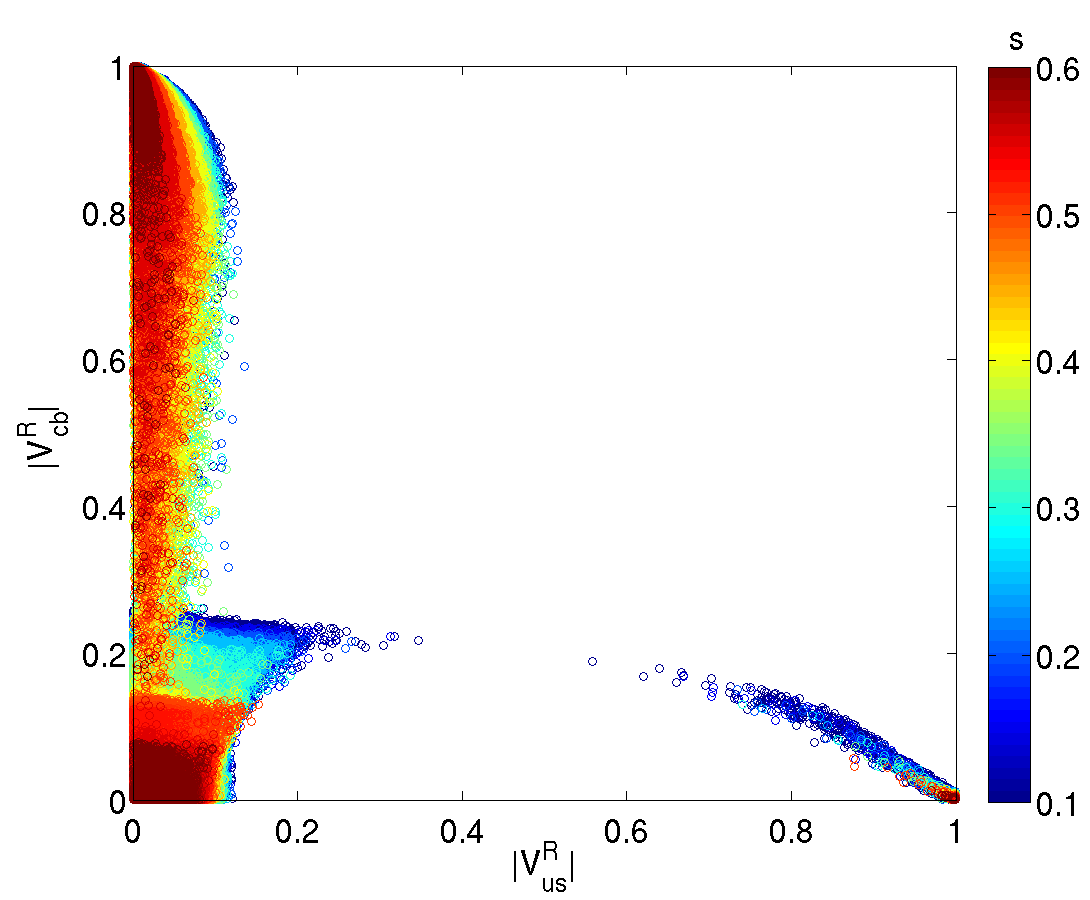}
\includegraphics[width=0.49\textwidth]{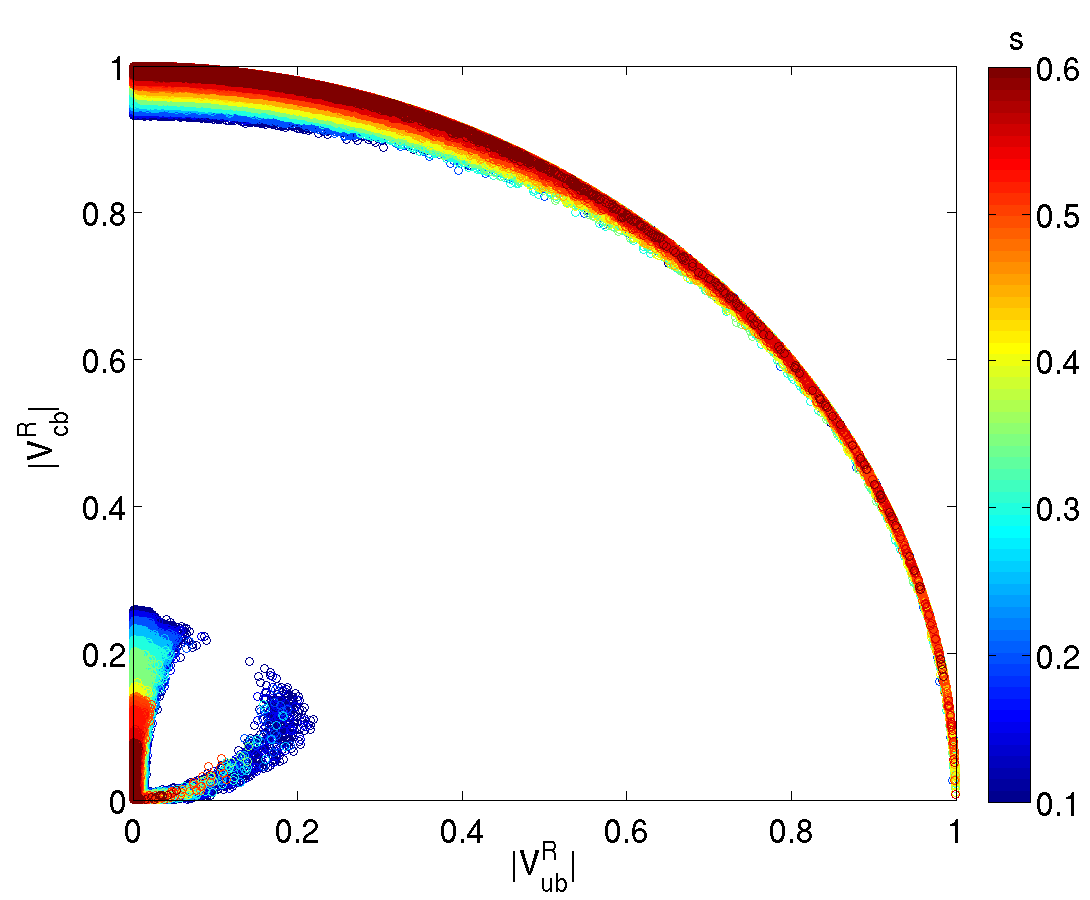}
\caption{$|V^{\rm R}_{us}|$, $|V^{\rm R}_{ub}|$ and $|V^{\rm R}_{cb}|$ as functions of each other with $s$ encoded in colour}\label{fig:absVR_s}
\end{figure}

In order to keep the presentation as transparent as possible, in the following we restrict ourselves to showing results only for fixed values of $s$ when necessary. In the course of our analysis of $\Delta F = 2$ constraints we concentrate on the case $s=0.1$. For the study $\Br(B\to X_{s,d}\gamma)$ large values of $s$ are interesting since they lead to  enhanced effects. In a few cases we consider the $s$ dependence explicitly.
}

\subsection{A fine-tuning study}\label{sec:vr_gen_num}

{We now turn to investigating the fine-tuning necessary for a point to fulfil all the experimental constraints.

 It is well known that models which predict sizeable contributions to the LR $\Delta F = 2$ operators can lead to dangerously large fine-tuning of several observables. For example the RS model with custodial protection has to struggle with huge effects in $\eps_K$ and a potentially large fine-tuning \cite{Csaki:2008zd,Blanke:2008zb}.

First let us define the measure of fine-tuning
\begin{equation}
\Delta^{\rm mod}_{\rm BG} = \frac{1}{N_{\rm Obs}}\sum\limits_{i=1}^{N_{\rm Obs}} \Delta_\text{BG}(O_i) =\frac{1}{N_{\rm Obs}}\sum\limits_{i=1}^{N_{\rm Obs}} \max_j\left(\left|\frac{p_j}{O_i}\frac{\partial O_i}{\partial p_j}\right|\right)
\,,\label{eqn:ft}
\end{equation}
used throughout our analysis.} Here $\Delta_{\rm BG}$ is the well known Barbieri-Giudice (BG) measure of fine-tuning \cite{Barbieri:1987fn}.
Note that the sum should only contain observables which are in fact fine-tuned in some part of the parameter space.
The overall fine-tuning defined in equation (\ref{eqn:ft}) is compatible with the more
sophisticated fine-tuning measure proposed by Athron and Miller \cite{Athron:2007ry}. Since
the numerical application of the Athron-Miller fine-tuning measure requires 
much more computing power than the BG measure while not providing
additional insights we decided to stick with $\Delta^{\rm mod}_{\rm BG}$. {We note that by definition $\Delta_{\rm BG}$ is sensitive only to fine-tuning in terms of cancellations between various contributions, but not to accidentally small parameters.}

\begin{figure}
\centering
\includegraphics[width=0.7\textwidth]{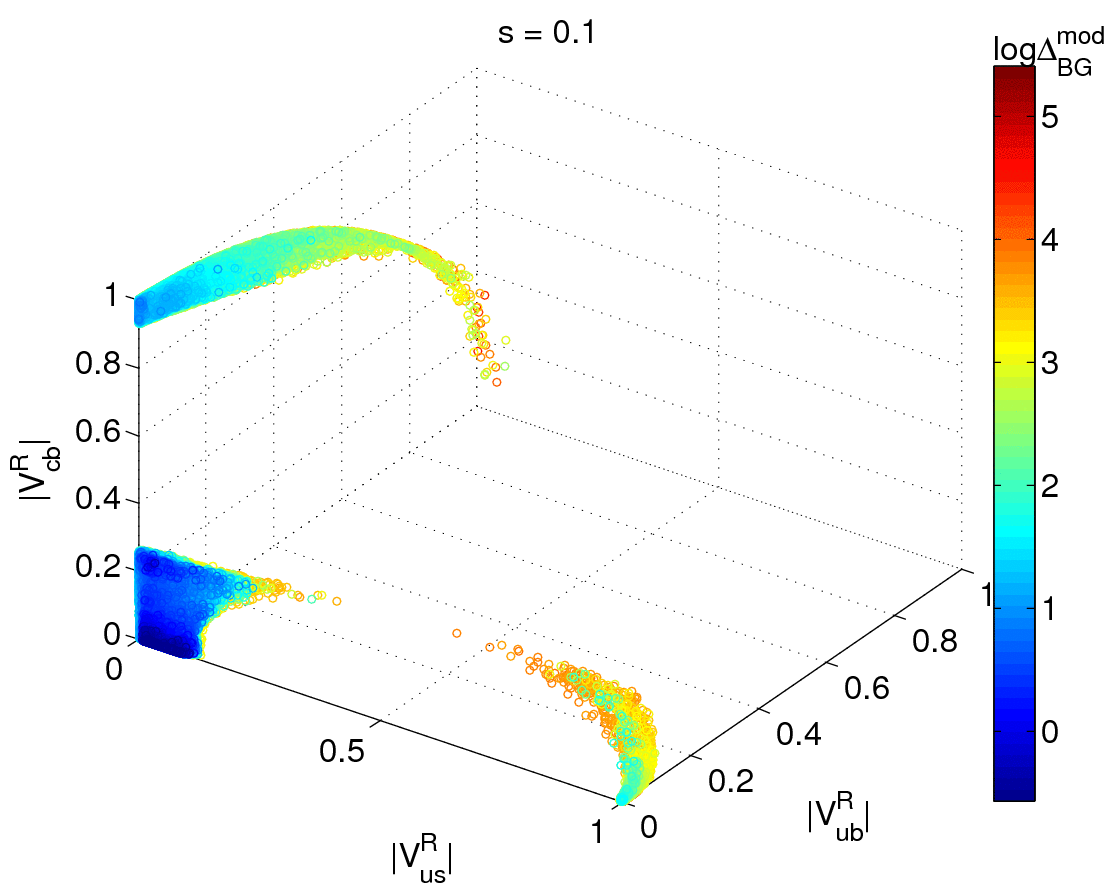}
\caption{The absolute values of $|V^{\rm R}_{us}|$, $|V^{\rm R}_{ub}|$ and $|V^{\rm R}_{cb}|$ with $\log\Delta^{\rm mod}_{\rm BG}$ encoded as the colour.}
\label{fig:absVR}
\end{figure}

In figure \ref{fig:absVR} we show the absolute values of $|V^{\rm R}_{us}|$, $|V^{\rm R}_{ub}|$ and $|V^{\rm R}_{cb}|$ as a three dimensional plot for $s=0.1$. The colour corresponds to 
$\log\Delta^{\rm mod}_{\rm BG}$ and we plotted points with low fine-tuning in front of ones with large fine-tuning. 
This however does not preclude points with high fine-tuning to lie in regions dominated by points with low fine-tuning. 
First of all we observe that the constraints allow for three clearly distinct 
scenarios.
\begin{itemize}
 \item In the first scenario $|V^{\rm R}_{us}|$, $|V^{\rm R}_{ub}|$ and $|V^{\rm R}_{cb}|$ are small. 
 The fine-tuning can be very small and increases slowly with $|V^{\rm R}_{ub}|$.  We call this scenario the ``normal hierarchy'' scenario.
 \item 
We find the second scenario in the case of small $|V^{\rm R}_{us}|$ and large $|V^{\rm R}_{cb}|$. This scenario is not restricted to one corner but allows points along the $|V^{\rm R}_{ub}|$
axis as well. The fine-tuning is in most cases high and increases further with increasing $|V^{\rm R}_{ub}|$. Taking only points with low fine-tuning this scenario corresponds to the so-called ``inverted hierarchy'' scenario.
 \item For small $|V^{\rm R}_{cb}|$, large $|V^{\rm R}_{us}|$ and $|V^{\rm R}_{ub}| < 0.30$ we find the third scenario. This scenario exhibits large fine-tuning for all points, hence it is completely eliminated if we require $\Delta^\text{mod}_{BG} < 10$. 
{We point out that while this scenario is very fine-tuned it is also not rigorously excluded.}
\end{itemize}
{Second we find that while significant regions of the parameter space suffer from large fine-tuning, there exist ranges of parameters in which the fine-tuning is small and all experimental constraints, in particular the ones from $\eps_K$ and $\Delta M_K$, can be satisfied.}

{In the following we restrict ourselves to points in parameter space which exhibit only
a small level of fine-tuning $\Delta^{\rm mod}_{\rm BG} < 10$. In this case we find the following limits:
\be
	|V^{\rm R}_{td}| < 1.2\cdot 10^{-2}\qquad {\rm and}\qquad |V^{\rm R}_{us}| <
\left\{
\begin{array}{c}
 0.18\quad (s = 0.1)\\
 0.13\quad (s=0.5)
\end{array}
\right.
\ee
Note that the constraint on $|V^{\rm R}_{td}|$ is much more stringent than the one on $|V^{\rm R}_{ub}|$. 
The ``normal'' and ``inverted hierarchy'' scenarios introduced above are then defined by
\be
\begin{array}{llc}
|V^{\rm R}_{cb}| & <\, 0.3\,, &\qquad\text{(normal hierarchy)}\\
|V^{\rm R}_{cb}| & >\, 0.9\,. &\qquad\text{(inverted hierarchy)}
\end{array} 
\ee
The first scenario leads to a hierarchical structure of $V^{\rm R}$ with small off-diagonal elements,  while the second one
inverts the hierarchy of the $2,3$ submatrix giving $V^{\rm R}$ a very different
structure. 
}

\begin{figure}
\centering
\includegraphics[width=0.7\textwidth]{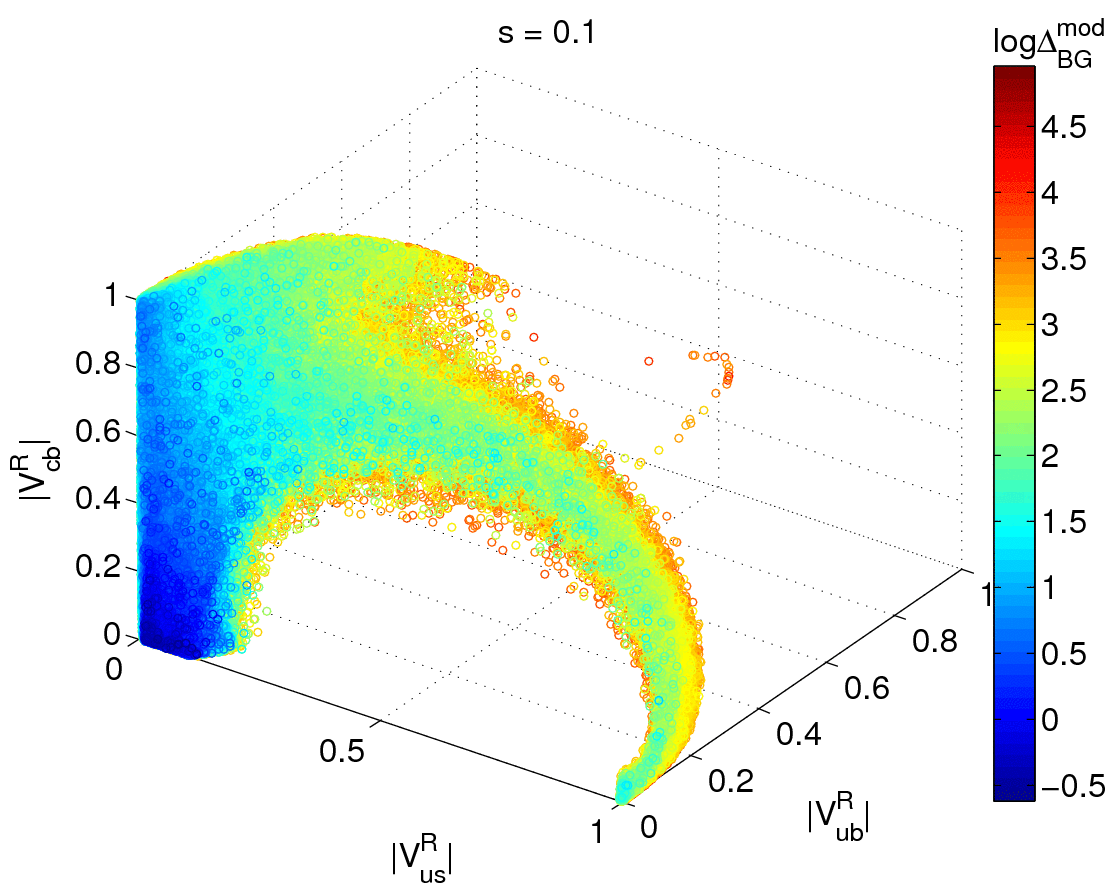}
\caption{The 3D correlation between allowed values of $|V^{\rm R}_{us}|$, $|V^{\rm R}_{ub}|$ and $|V^{\rm R}_{cb}|$,  omitting  the contributions of the heavy Higgses.}\label{fig:absVR_higgs}
\end{figure}
{Let us now consider how these results change if the heavy Higgs contributions are omitted.
In  figure \ref{fig:absVR_higgs} we show the allowed parameter points obtained from a scan analogous to the one that lead to figure \ref{fig:absVR}, taking into account only the gauge boson contributions. We observe that now the points cover a significantly larger region of parameter space, even if only points with low fine-tuning are considered. This drastic change reflects once more the importance of the heavy Higgs particles.}

\subsection{A closer look at the ``normal hierarchy'' scenario}\label{sec:normal_h}

{Now we study in more detail the ``normal hierarchy'' scenario. Recall that we consider only parameter points with small fine-tuning $\Delta^{\rm mod}_{\rm BG} < 10$.
In order to investigate the possible contributions of the matrix $V^{\rm R}$ to flavour processes it is useful to study the relative size of the dominant LR contribution compared to the SM contribution.
From the hierarchical structure of $V^{\rm R}$ in this scenario together with the pattern of the matrix $\hat R$, see appendix \ref{app:num_details_deltaf2}, we expect the $tt$ contribution to dominate.
Therefore it is convenient to define ($q=K, B_d, B_s$)
\be
{\mathscr F}_{t} (q) = \frac{\lambda_t^{\rm LR}(q)\lambda_t^{\rm RL}(q)}
{\lambda_t^{\rm LL}(q) \lambda_t^{\rm LL}(q)} = \frac{\lambda_{t}^{\rm RR}(q)}{\lambda_{t}^{\rm LL}(q)}\,.
\ee

In order to better understand the results obtained in this scenario it is useful to  introduce a Wolfenstein-like parametrisation of the right-handed matrix} by
expanding in $\tilde\lambda = |V^{\rm R}_{us}|$. Defining $\tilde s_{12} \equiv \tilde\lambda$, $\tilde s_{13} = \tilde B \tilde\lambda^2$ and $\tilde s_{23} = \tilde A \tilde\lambda$
and then expanding in $\tilde\lambda$ we arrive at
\be\label{eq:normal-Wolf}
{\footnotesize
\left(
\begin{array}{ccc}
  e^{i \phi_{ud}} \left(1-\frac{\tilde\lambda^2}{2}\right) & e^{i \phi_{us}} \tilde\lambda  & \tilde B e^{i \phi_{ub}} \tilde\lambda ^2 \\
 -e^{i (\phi_{cb}-\phi_{tb}+\phi_{ts}+\phi_{ud}-\phi_{us})} \tilde\lambda  & e^{i (\phi_{cb}-\phi_{tb}+\phi_{ts})} \left(1- \frac{1}{2}\left(\tilde A^2+1\right) \tilde\lambda ^2\right) & \tilde A e^{i \phi_{cb}} \tilde\lambda  \\
  e^{i \phi_{ud}} \left(\tilde A e^{i (\phi_{ts}-\phi_{us})}-\tilde B e^{i (\phi_{tb}-\phi_{ub})}\right) \tilde\lambda ^2 & -\tilde A e^{i \phi_{ts}} \tilde\lambda  & e^{i \phi_{tb}}\left(1 - \frac{\tilde A^2}{2} \tilde\lambda ^2\right)
\end{array}
\right)\,.}
\ee

\begin{figure}
\centering
\includegraphics[width=0.48\textwidth]{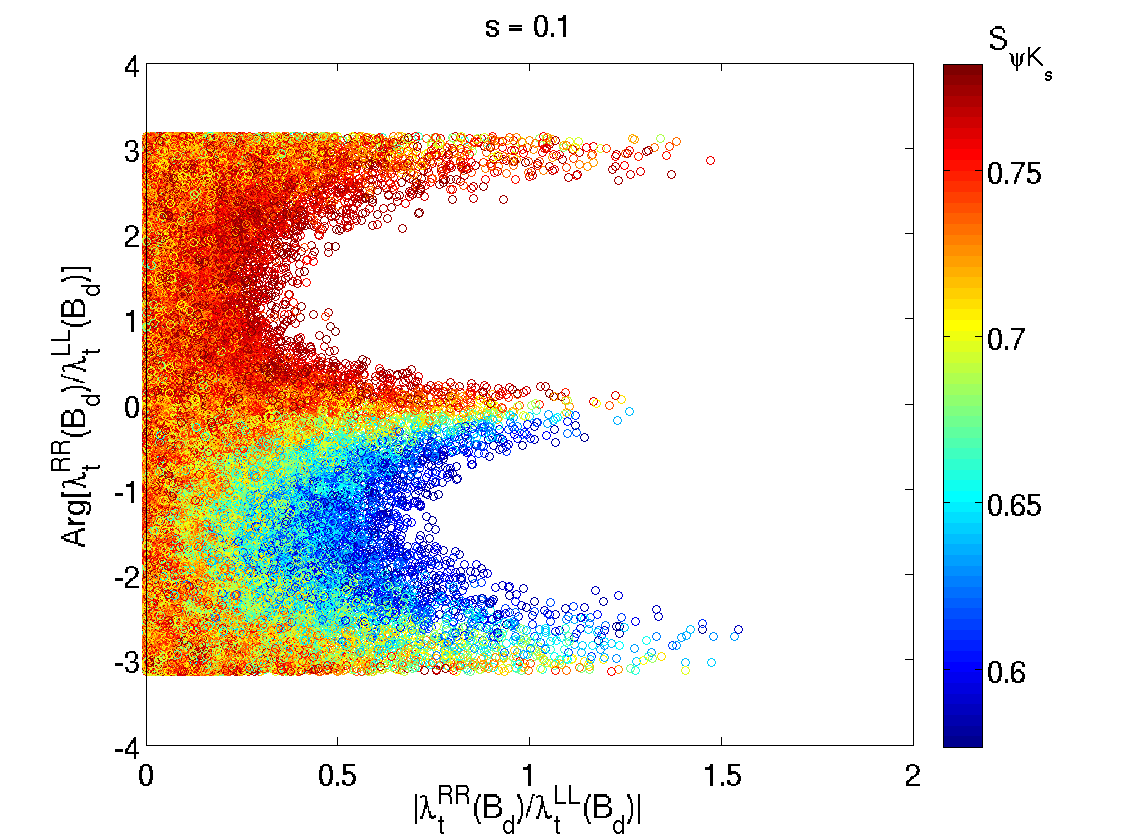}
\includegraphics[width=0.48\textwidth]{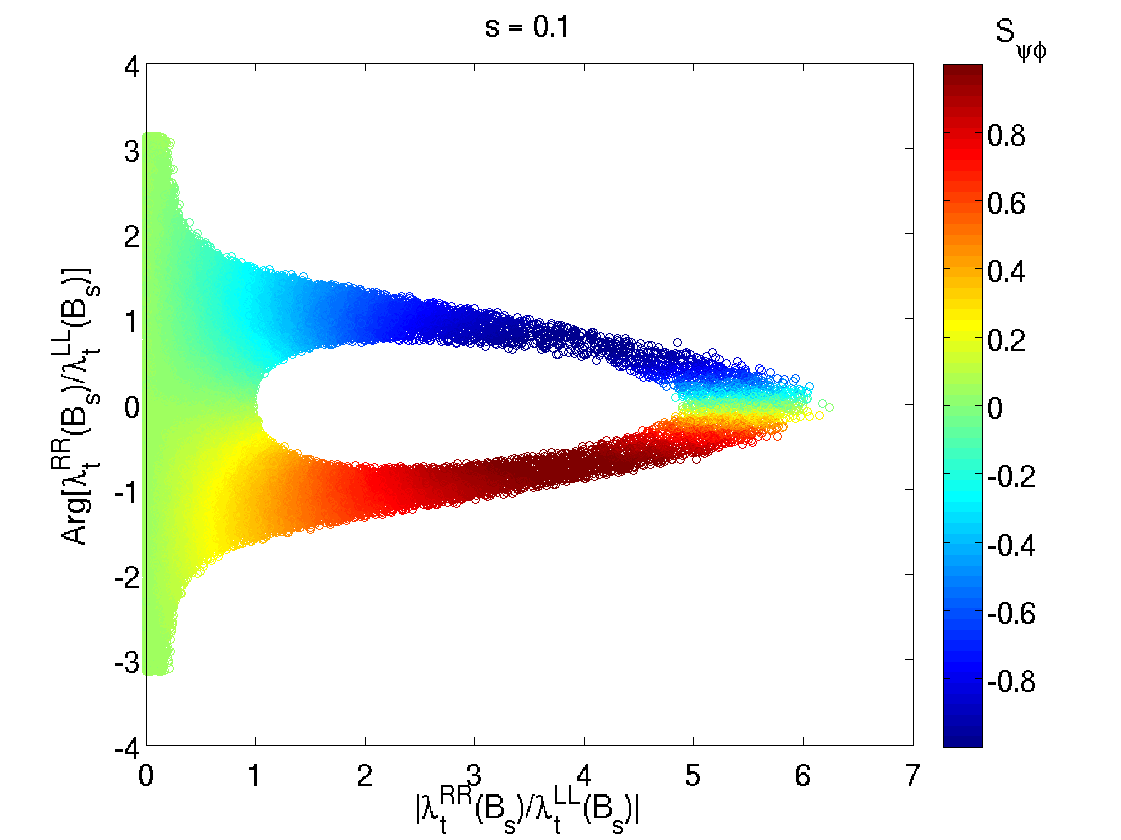}
\caption{$\arg({\mathscr F}_{t}(B_q))$ as a function of $|{\mathscr F}_{t}(B_q)|$ with $B_d$ on the left-hand side and $B_s$ on the right-hand side. The colour coding corresponds to $S_{\psi K_S}$ and $S_{\psi \phi}$ respectively.}\label{fig:s1_lambdaRR_Bq}
\end{figure}
Note that this matrix cannot be reduced to the matrix in $V^{\rm R}$ in  (\ref{eq:golden Vr}) and is valid in a different oasis in the parameter space.
In figure \ref{fig:s1_lambdaRR_Bq} we show the correlation of $\arg\left({\mathscr F}_{t}(B_q)\right)$ and $|{\mathscr F}_{t}(B_q)|$.
On the left hand side of figure \ref{fig:s1_lambdaRR_Bq} we show the situation in the $B_d$ system. In order to gain more information from this
correlation we encode $S_{\psi K_S}$ as the colour of a point. First we notice that the absolute value $|{\mathscr F}_{t}(B_d)|$ only allows enhancements up to $1.5$. The phase $\arg({\mathscr F}_{t}(B_d))$
on the other hand is allowed to be in the whole $[-\pi, \pi]$ range. However there are restrictions on the phase as well. For the maximal allowed enhancement of the absolute value the
phase is restricted to be close to $0$ or $\pm\pi$ and for a phase of $\pm\pi/2$ the absolute value is restricted to be below roughly $0.8$ and $0.5$ for a phase of $\pi\ (-\pi)$ respectively. {The observed shape reminds us of the model-independent constraint on $(M^d_{12})_\text{NP}$. Since the measured values of $\Delta M_d$ and $S_{\psi K_S}$ are in good agreement with the SM prediction (although somewhat on the low side), there is not much room left for NP. Furthermore as $\Delta M_d$ suffers from larger theoretical uncertainties than $S_{\psi K_S}$ the possible size of the NP contribution is largest if its relative phase is close to 0 or $\pm\pi$. 
If the $tt$ contribution was the only contribution to $(M^d_{12})_\text{NP}$, its phase would be equal to $\arg({\mathscr F}_{t}(B_d))$ and we would observe strong peaks for these phases in the plot.
The small ``washout'' indicates  however that other LR contributions cannot be neglected in this case.}

On the right hand side of figure \ref{fig:s1_lambdaRR_Bq} we show the situation in the $B_s$ system. In this plot the colour corresponds to $S_{\psi\phi}$. 
We immediately see that the situation is completely different from the $B_d$ system. The enhancement in the absolute value $|{\mathscr F}_{t}(B_s)|$ is allowed to be up to nearly $7$ while the 
phase $\arg({\mathscr F}_{t}(B_s))$, though allowed to be somewhere in the whole range, is very strongly correlated with the absolute value and $S_{\psi\phi}$. For $|S_{\psi\phi}|$ close to zero
the absolute value is restricted to be smaller than $1.5$ or bigger than $4.7$ while the phase is either roughly free or very close to zero respectively. For big $|S_{\psi\phi}|$ the phase is restricted
to two very specific areas (the blue and red arches in the plot). This might lead to interesting correlations in rare decays of $B_s$ mesons.
{In fact the shape found in the right panel of figure \ref{fig:s1_lambdaRR_Bq} is familiar from the $\Delta M_s$ constraint projected onto the plane $\big(|(M_{12}^s)_\text{NP}|, \arg (M_{12}^s)_\text{NP}\big)$, see e.\,g.\ figure 1 of \cite{Ligeti:2006pm}. This leads us to conclude that the NP amplitude in the present case is totally dominated by the $tt$ contribution, and therefore governed by ${\mathscr F}_{t}(B_s)$. Indeed from the structure of $V^{\rm R}$ in \eqref{eq:normal-Wolf} it is easy to derive that the strong hierarchy in the matrix $\hat R(B_q)$, see appendix \ref{app:num_details_deltaf2}, cannot be overcompensated by hierarchies in the right-handed quark mixing.}

\FloatBarrier 

\section{A brief discussion of flavour observables}\label{sec:num}

{This section is dedicated to the detailed discussion of flavour violating effects in the LRM.
After analysing the possible size of NP effects in the various meson systems, we study the possible enhancements of the $B_s$ mixing phase and effects in $\Br(B\to X_q\gamma)$. Finally we turn our attention to the $|V_{ub}|$ problem, which was recently discussed in the context of RH currents \cite{Crivellin:2009sd,Chen:2008se,Feger:2010qc,Buras:2010pz}.}

\subsection{A summary of possible size of NP effects in the different meson systems}
\begin{figure}
\centering
\includegraphics[width=0.7\textwidth]{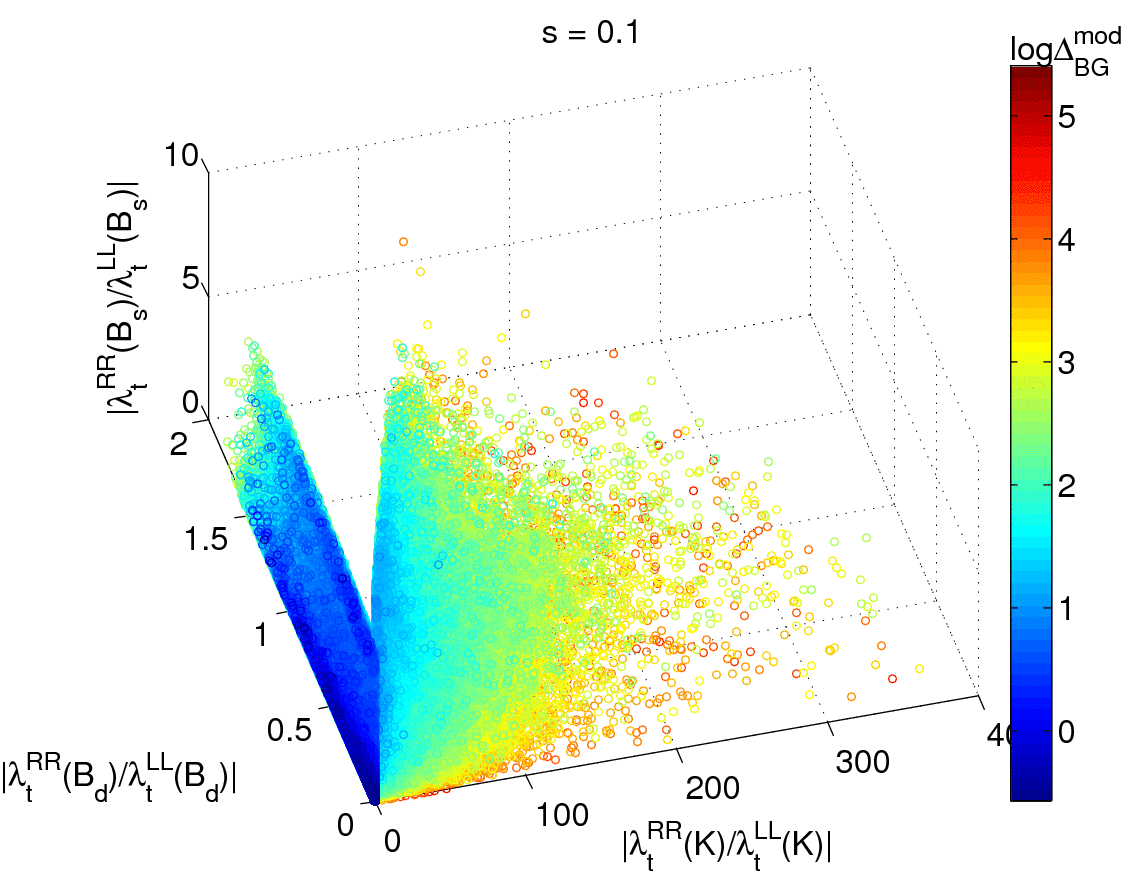}
\caption{The absolute values of ${\mathscr F}_{t}(q)$ for $q = K, B_d, B_s$ as functions of each other. The colour-code corresponds to the fine-tuning $\Delta_{\rm BG}^{\rm mod}$.}
\label{fig:possible_effects}
\end{figure}

{As discussed in section \ref{sec:normal_h} the relative size of LR contributions in the various meson systems is given by the quantity ${\mathscr F}_{t}(q)$.
In figure \ref{fig:possible_effects} we show the possible sizes of $|{\mathscr F}_{t}(q)|$ for the three meson systems as functions of each other for $s=0.1$.  We observe that in principle huge effects in the $K$ system are possible, while the effects in the $B_s$ system are much more moderate and the effects in the $B_d$ system are rather small. This indicates a rough hierarchy of NP effects $B_d < B_s \ll K$ also expected for rare $K$ and $B_{d,s}$ decays. Note that huge effects in the $K$ system are possible only for large fine-tuning, since in that case the large $tt$ contribution to $K^0-\bar K^0$ mixing has to be cancelled by other contributions.}

\subsection[The phase of $B_s$ mixing and $\Br(B\to X_q \gamma)$]{{\boldmath The phase of $B_s$ mixing and $\Br(B\to X_q \gamma)$}}
\begin{figure}
\centering
\includegraphics[width=0.48\textwidth]{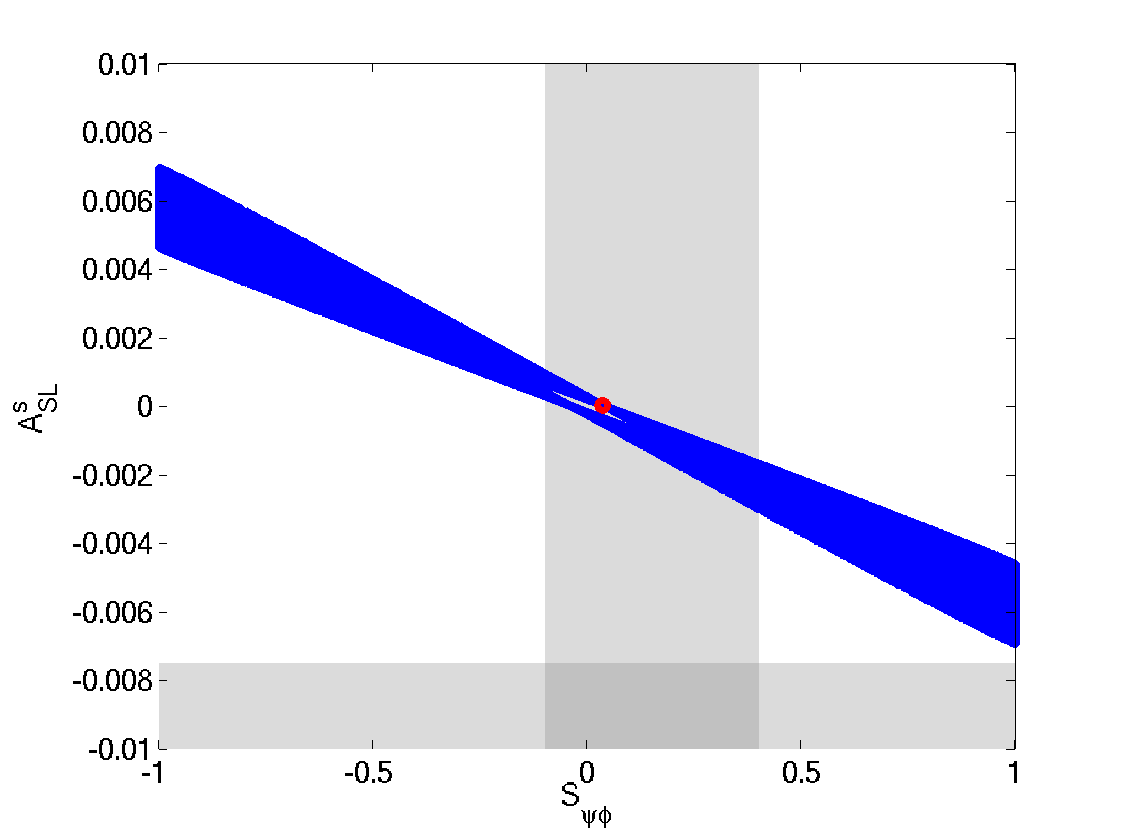}
\includegraphics[width=0.48\textwidth]{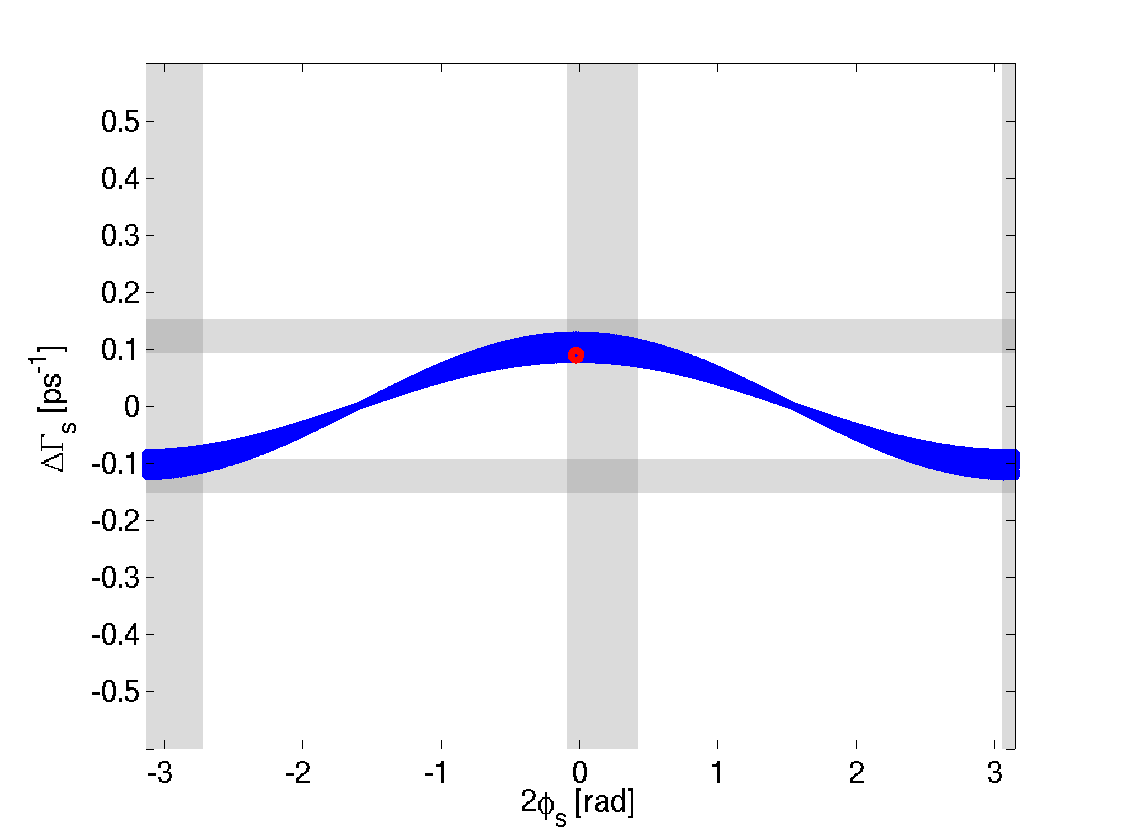}
\caption{The model independent correlation of $A_{\rm SL}^{s}$ and $S_{\psi\phi}$ (left panel) and the correlation of $\Delta\Gamma_s$ with $2\phi_s$ (right panel)}
\label{fig:asl_spsiphi}
\end{figure}

Now we want to briefly discuss the results of our analysis of observables related to the phase of $B_s$ mixing and $\Br(B\to X_q \gamma)$.
In the general case with and without fine-tuned points we observe no correlation between observables, except for model-independent ones such as $A_{\rm SL}^{s}$-$S_{\psi\phi}$ and $\Delta\Gamma_s$ versus $\phi_s$ shown in figure \ref{fig:asl_spsiphi}.

\begin{figure}
\centering
\includegraphics[width=0.48\textwidth]{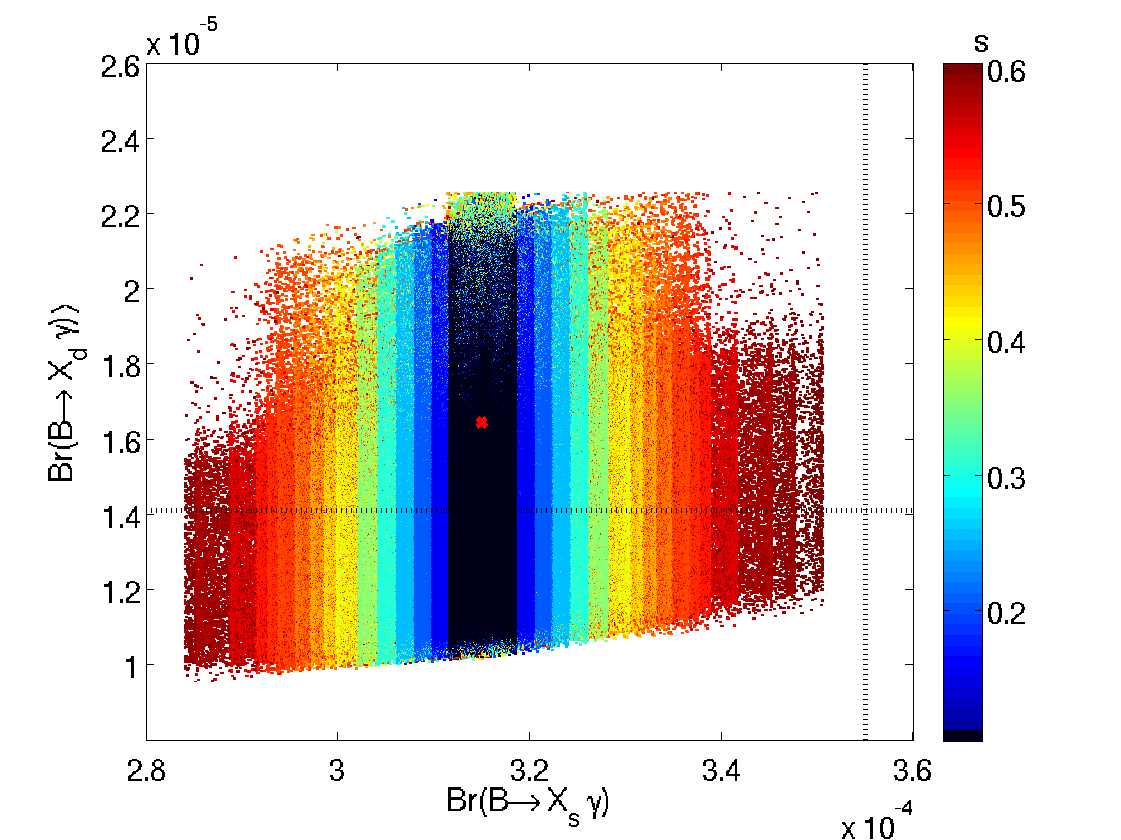}
\includegraphics[width=0.48\textwidth]{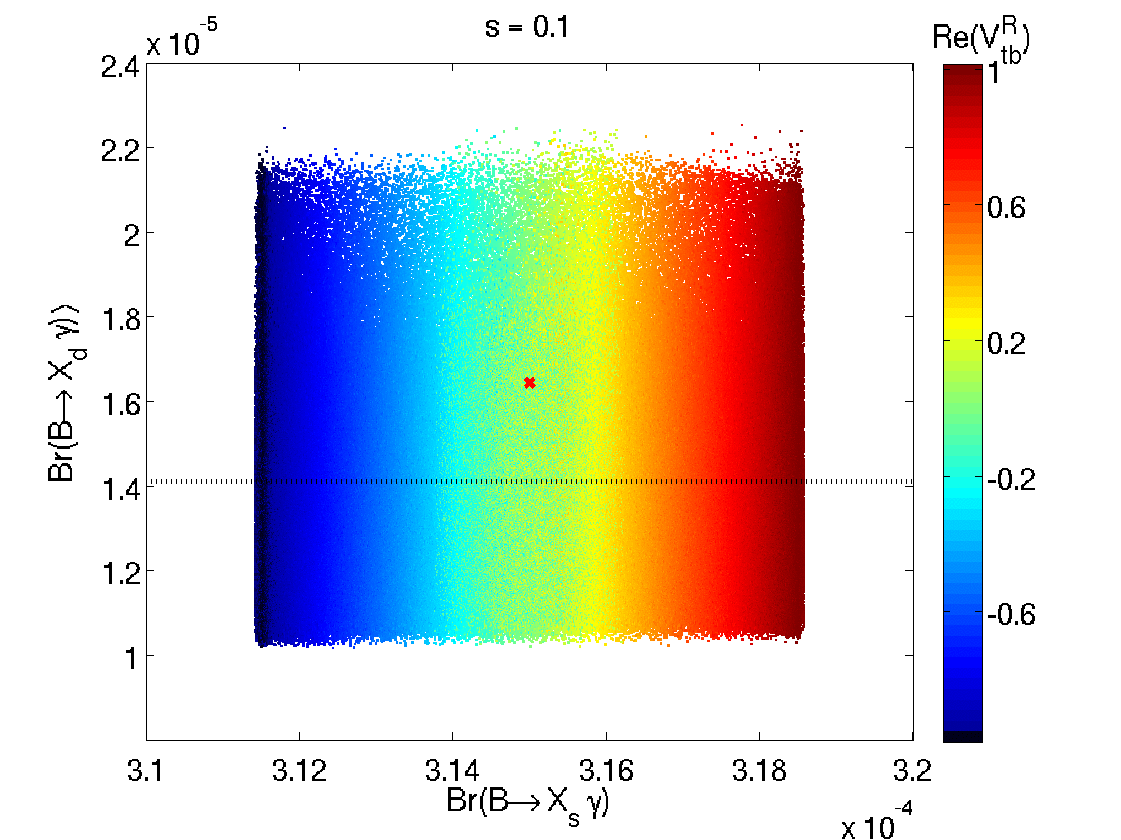}
\caption{The correlation of $\Br(B\to X_s\gamma)$ and  $\langle\Br(B\to X_d\gamma)\rangle$ showing the $s$ dependence (left panel) and $\Re(V^{\rm R}_{tb})$ dependence (right panel) in colour. The red cross indicates the SM values, while the experimental central values are given by the dashed lines.}
\label{fig:bdgbsg}
\end{figure}

{In figure \ref{fig:bdgbsg} we show the correlation between $\Br(B\to X_s\gamma)$ and $\Br(B\to X_d\gamma)$, encoding the $s$ dependence (left panel) and $\Re(V^{\rm R}_{tb})$ dependence (right panel, for $s=0.1$) in colour. 
For $s> 0.5$ the values of $\Br(B\to X_s\gamma)$ cover a larger range that originates in the divergence of the function $u(s)$ in the unphysical limit $s\to 1/\sqrt{2}$, see section \ref{eq:bsgamma-chargedHiggs}. For small $s$ the branching ratios depend only linearly on this parameter and the effects are much smaller.
As can be seen from the explicit formulae given in section \ref{sec:bsg_anatomy} 
$\Br(B\to X_s\gamma)$ depends linearly on the real part of $V^{\rm R}_{tb}$.

Combining the information of both plots in figure \ref{fig:bdgbsg} we conclude 
that in order to enhance $\Br(B\to X_s\gamma)$,
 we would need both $s>0.5$ and a dominantly real and positive $V^{\rm R}_{tb}$. As the SM prediction lies somewhat below the experimental value, albeit still in good agreement, a positive NP contribution to $\Br(B\to X_s\gamma)$ is welcome.}

Concerning the CP asymmetries in $b\to q \gamma$ we find moderate changes relative to the SM prediction
\begin{eqnarray}
A_{\rm CP}(b\to d\gamma)^{\rm SM} &=& -9.2\%\,,\\
A_{\rm CP}(b\to s\gamma)^{\rm SM} &=& 0.4\%\,.
\end{eqnarray}
Similar to the branching ratios also the CP asymmetries are $s$ dependent. For $s = 0.1$ we find
\begin{eqnarray}\label{eq:ACP-small-s}
-13.8\% <& A_{\rm CP}(b\to d\gamma) &< -7.9\%\,,\\
0.3\% <& A_{\rm CP}(b\to s\gamma) &< 0.6\%\,,
\end{eqnarray}
while for $s = 0.6$ the possible range is {slightly} larger
\begin{eqnarray}
-14.2\% <& A_{\rm CP}(b\to d\gamma) &< -8.3\%\,,\\
-0.18\% <& A_{\rm CP}(b\to s\gamma) &< 1.1\%\,.\label{eq:ACP-large-s}
\end{eqnarray}
{These effects will be difficult to disentangle from the SM contribution due to the large non-perturbative uncertainties present in these observables.}

{Note that in figure \ref{fig:bdgbsg} and in \eqref{eq:ACP-small-s}--\eqref{eq:ACP-large-s} we always used the maximal allowed Higgs mass (see section \ref{sec:higgsmass}). For lower Higgs mass the constraints on the parameter space become much stronger, predicting a much more hierarchical structure for $V^{\rm R}$. These constraints however do not suppress $\Br(B\to X_{s,d}\gamma)$ since the latter decays depend only on the diagonal element $V^{\rm R}_{tb}$. Therefore decreasing the heavy Higgs mass $M_H$ enhances the effects in $\Br(B\to X_{s,d}\gamma)$.}

\subsection[The $|V_{ub}|$ problem in the LRM]{\boldmath The $|V_{ub}|$ problem in the LRM}\label{sec:vub_in_lram}

\begin{figure}
\centering
\includegraphics[width=0.7\textwidth]{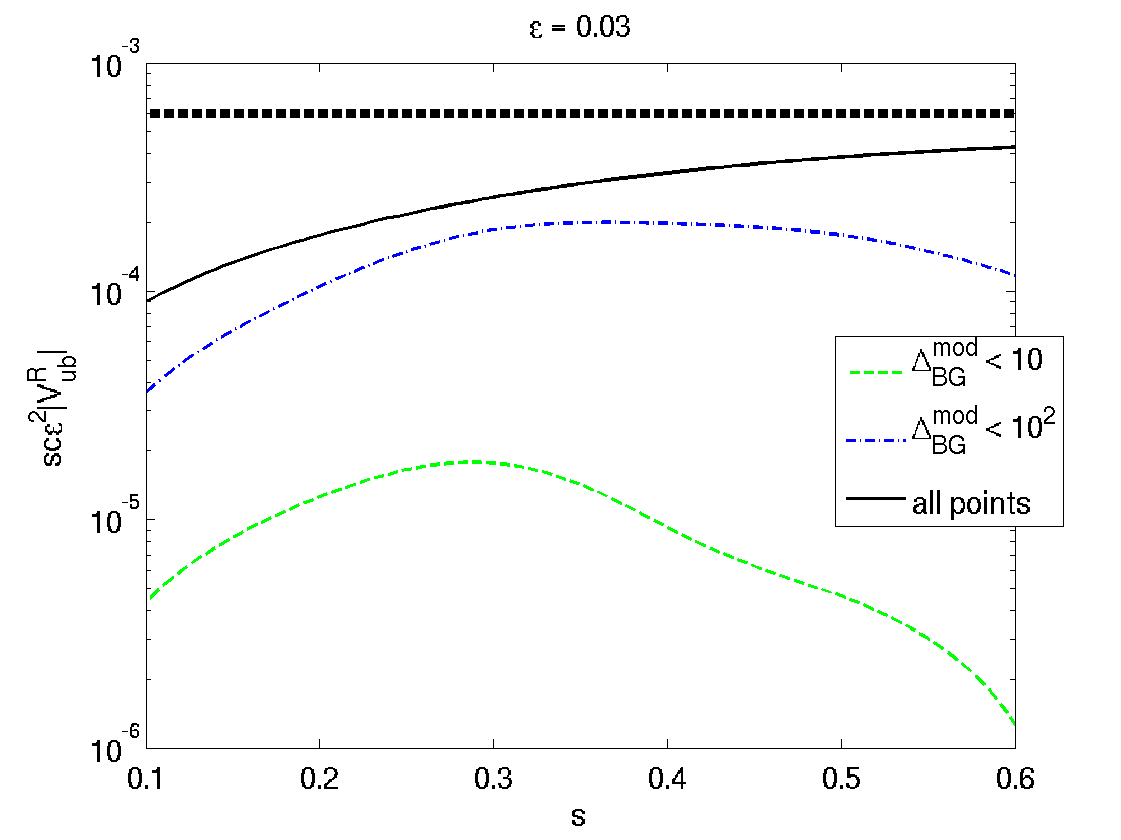}
\caption{Maximal value for $sc\epsilon^2|V^{\rm R}_{ub}|$ as a function of $s$, for different allowed fine-tunings. A complete solution to the $|V_{ub}|$ problem requires values for $sc\epsilon^2|V^{\rm R}_{ub}|$ above the dashed black line.}\label{fig:vub_problem}
\end{figure}
{As already discussed in section~\ref{sec:tree} in the LRM the true value of $|V_{ub}^{\rm L}|$ is given by the inclusive value of $\vub$ in (\ref{vub1}), which remains unaffected by the presence of RH currents. On the other hand 
the corrections from RH currents to the vector and  axial-vector couplings could in principle allow to explain the different values of $\vub$ found in exclusive decays and $B^+\to\tau^+\nu_\tau$ as given in (\ref{vub2}) and (\ref{vub3}), respectively 
\cite{Crivellin:2009sd,Chen:2008se,Feger:2010qc,Buras:2010pz}. 
A solution to the $|V_{ub}|$ problem can be provided if $sc\epsilon^2|V^{\rm R}_{ub}|$ is in the ballpark of $ 0.6 \times 10^{-3}$.
Since we can only restrict $sc\epsilon^2\le 10^{-3}$, while $|V^{\rm R}_{ub}|$ is not constrained if arbitrary fine-tuning is allowed, studying the $s$ dependence of the constraints is mandatory.

Figure \ref{fig:vub_problem} illustrates the situation in the LRM. In general the black dashed line, necessary for a complete solution of the $|V_{ub}|$ problem, cannot be reached in the LRM. The tension increases significantly if we consider only points with small fine-tuning. While we cannot exclude that there are some 
models with RH currents that could solve this problem, from the point of view of the LRM considered here, this is not the case and any value of $\vub$ among (\ref{vub1})--(\ref{vub2}) could be the true value.}

\section{The role of the heavy Higgs mass}\label{sec:higgsmass}
\begin{figure}
\centering
\includegraphics[width=0.78\textwidth]{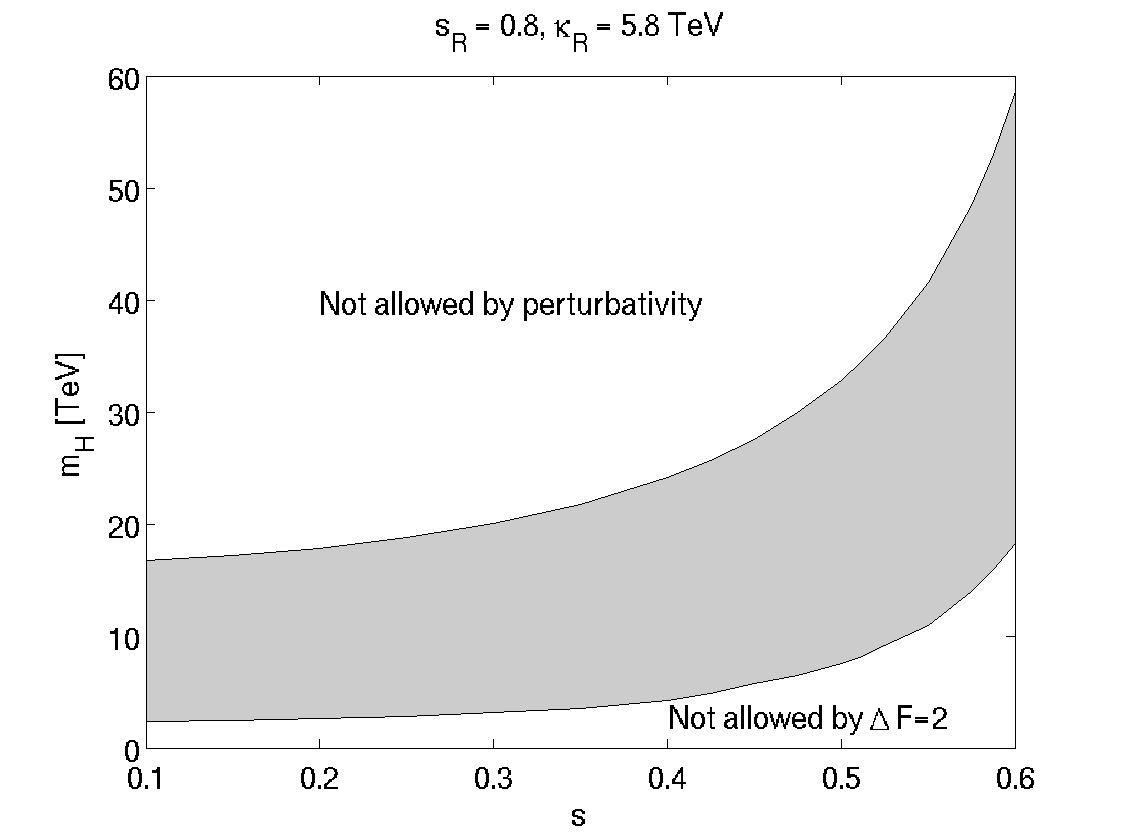}
\caption{The allowed range for the Higgs mass $M_H$ as a function of $s$ for $\kappa_R\approx 5.8\,{\rm TeV}$ and $s_R=0.8$.}
\label{fig:higgs_mass}
\end{figure}

Up to now in our analysis we have set $M_H = 16/\sqrt{1-2s^2}\,{\rm TeV}$, this ensures staying in the perturbative
regime of the coupling $\alpha_3$ in the Higgs potential while suppressing the Higgs contributions
in $\Delta F = 2$. In this section we want to investigate the lower bound on the Higgs mass from our
constraints. {Therefore we now allow $\alpha_3$ to vary between $0.1$ and $8$, or equivalently the heavy Higgs mass between $2/\sqrt{1-2s^2}\,{\rm TeV}$ and $16/\sqrt{1-2s^2}\,{\rm TeV}$. The result can be seen in figure \ref{fig:higgs_mass}, in particular we find} 
the `soft' lower limit on $M_H$ to be 
\begin{equation}
  2.4\,{\rm TeV} \lsim M_H\,.
\end{equation}
{This limit is 'soft' in the sense that lower values are not rigorously forbidden, but our parameter scan did not reveal any points which allow for lower masses.} One might suspect a very large fine-tuning for this low Higgs mass, but while the fine-tuning does increase 
with decreasing Higgs mass there are still points with fine-tuning $\Delta_\text{BG}^\text{mod} <10$  even for the lowest possible mass. This is of course
also a test of the idea of fine-tuning. Finally for low Higgs masses the matrix $V^{\rm R}$ shows  a very hierarchical structure
as can be expected. The mixing pattern is roughly given by
\begin{equation}
\tilde s_{12} \sim \ord(10^{-2})\,,\qquad \tilde s_{13} \sim \ord(10^{-4})\,,\qquad \tilde s_{23} \sim \ord(10^{-3})\,.
\end{equation}

\section{Comparison to other models}\label{CMODELS}

\subsection{Preliminaries}
{Now we compare the results found in the LRM to those obtained in other NP models.
The first part of this section is dedicated to the comparison with two specific versions of LR models discussed in the literature.} Subsequently we compare the pattern of flavour violation in the general LRM with patterns found by us in a number of extensions of the SM. A review of those analyses can be found in \cite{Buras:2010wr}. Our discussion below deals only with $\Delta F=2$ transitions and $B\to X_{s,d}\gamma$ decays.

In order to increase the transparency of  the comparison in question let us 
summarize the most characteristic features of NP contributions in the 
LRM:
\begin{itemize}
\item
The dominance of the operator $Q_2^{\rm LR}$ in all $\Delta F=2$ transitions: 
$K^0-\bar K^0$, $B_d^0-\bar B_d^0$ and $B_s^0-\bar B_s^0$, with the biggest 
impact on the $K^0-\bar K^0$ system.
\item
While both box diagrams with $W_L-W_R$ exchanges and tree level neutral 
Higgs contributions are the sources of the effects in $\Delta F = 2$ observables, the latter contributions 
are most important. Governed by the same operators and having the same quark mixing factors the trees always dominate over the boxes.
\item
The contributions of induced RH couplings of $W_L$ and of $H^\pm$ to the 
$B\to X_{s,d}\gamma$ decays are enhanced by the ratio $m_t/m_b$ due to the 
chirality flip on the top quark propagator. Moreover $H^\pm$ contributions 
receive an additional enhancement for large $s$.
\item
All effects listed above are beyond the usual MFV framework.
\end{itemize}

With this picture in mind we are in the position to make a transparent comparison with other versions of LR models and with a number of extensions of the SM.

\subsection{LR models with additional flavour symmetries}

Recently two different approaches have been presented to reconcile the LR models with flavour data even for relatively small symmetry breaking scales. In both cases this is achieved by introducing an additional symmetry that removes all flavour changing Higgs effects from the down quark sector.

In a first paper Guadagnoli and Mohapatra \cite{Guadagnoli:2010sd} suggested to extend the Higgs sector by a discrete $\mathbb{Z}_4$ symmetry. Eventually this leads to removing all flavour violating Higgs couplings from the down quark sector and the strongest constraints on the heavy Higgs mass then comes from $D^0-\bar D^0$ mixing. $\Delta S = 2$ and $\Delta B = 2$ processes are then governed by $W_L-W_R$ box diagrams. While in \cite{Guadagnoli:2010sd} the manifest version of the LRSM has been considered, it is straightforward to extend this model to the general LRM leading to a potentially interesting phenomenology. 

A different approach to avoid the stringent flavour constraints has been proposed in 
\cite{Guadagnoli:2011id}. Here a gauged LR symmetric flavour group $SU(3)_{Q_L} \times SU(3)_{Q_R}$ has been introduced following the idea of Grinstein et al. \cite{Grinstein:2010ve}. In this scenario due to a different structure of the Higgs sector only the SM Higgs boson is present and no dangerous tree level FCNH effects arise. The potentially dangerous tree level contributions from flavour gauge bosons are naturally suppressed by a see-saw-like mechanism, so that the most relevant new contributions to $\Delta F =2$ processes arise again from $W_L-W_R$ box diagrams. 

{The effects in the $B\to X_{s,d} \gamma$ decays turn out to be small in this scenario. This is related to both the absence of a heavy charged Higgs and the fact that in this model $W_L$ and $W_R$ do not mix with each other. Consequently no chiral enhancement $\propto m_t/m_b$ is possible in this case as in the 
scenario considered by us. On the other hand as pointed out in 
\cite{Buras:2011zb} in such models new flavour violating neutral gauge bosons 
could provide an enhancement of $B\to X_{s,d} \gamma$ through a chiral 
enhancement $\propto m_F/m_b$, where $F$ denotes a new vectorial neutral 
heavy fermion. 
Yet, as shown in \cite{Buras:2011zb} in models in which fermion masses are 
generated by a see-saw mechanism such contributions are strongly suppressed 
by the mixing between SM quarks and heavy fermions so that finally the 
NP contributions remain small.}

\subsection[2HDM with flavour blind phases: {${\rm 2HDM_{\overline{MFV}}}$}]{2HDM with flavour blind phases: \boldmath{${\rm 2HDM_{\overline{MFV}}}$}}
Concerning the $\Delta F=2$ transitions the most spectacular difference 
in  ${\rm 2HDM_{\overline{MFV}}}$  \cite{Buras:2010mh,Buras:2010zm} from the LRM 
considered here is the absence of relevant direct contributions to 
$\varepsilon_K$. The reason is that in 
${\rm 2HDM_{\overline{MFV}}}$ the flavour violating neutral Higgs couplings are 
effectively generated at one-loop level and are necessarily proportional to the 
masses of external quarks ($m_{d,s}$ in this case) and not to up-quark masses 
as in the case of the LRM, where flavour violating Higgs couplings are present in
the fundamental Lagrangian. 

Therefore in ${\rm 2HDM_{\overline{MFV}}}$ $\varepsilon_K$ can only be
 found close to the data if $\sin 2\beta\approx 0.80$ which corresponds 
to scenario 2 in section~\ref{sec:strategy}.
The interplay of flavour blind phases with the 
CKM phase allows then to bring $S_{\psi K_S}$ to its experimental value, while 
enhancing automatically $S_{\psi\phi}$ to values close to 0.3.  Having less 
parameters than the LRM, this prediction is rather unique. Finding experimentally $S_{\psi\phi}$ to be negative would rule out ${\rm 2HDM_{\overline{MFV}}}$. 
For LRM this would still not be a problem as one would move then to a different oasis in the parameter
space of this model than considered by us in \ref{sec:Matrix}.

It should also be emphasized that the neutral Higgs masses in the 
${\rm 2HDM_{\overline{MFV}}}$ are of the order of a few hundred GeV, 
whereas in the LRM they are by at least one order of magnitude larger.
Therefore LHC should easily distinguish between these two scenarios.

\subsection{Right-handed MFV}
Recently, in \cite{Buras:2010pz}, we have studied the effects of RH currents by means of an effective theory approach under incorporation of an extended MFV principle.

The similar symmetry pattern with respect to the LRM could lead to the premature assumption that both models can be matched by integrating out the heavy particles in the LRM. However as we have already mentioned in section \ref{sec:comops}, the operator structure of both models turns out to be complementary. Furthermore, due to the extended MFV mechanism the specific flavour structure  precludes this possibility.  

The differences display also in the phenomenology of both models. Even though the general parametrization for the RH mixing matrix is given by the same parameters, after imposing all constraints the matrices display a different pattern. This can be seen explicitly when comparing figure \ref{fig:absVR} to the structure of the matrix given in equation (105) of \cite{Buras:2010pz}. The fact that the $|V_{ub}|$ problem cannot be solved in the LRM whereas this was the case in RHMFV demonstrates once more the difference in the RH mixing matrices. 

Let us have a closer look at further flavour observables. While the RHMFV was originally designed to explain a large $S_{\psi\phi}$, a full theory like the LRM allows a more flexible analysis. Here no particular assumptions about $S_{\psi\phi}$ have been made and values within the full range are allowed. Since in the LRM we take into account the variation of all phases, the high number of parameters allows to find regions in parameters space where all SM tensions can be resolved. This was not the case in RHMFV where the $S_{\psi K_S}$ -$\sin(2\beta)$ tension persists under the assumption that $S_{\psi\phi}$ is significantly enhanced over its SM value.

On the other hand the general pattern of NP effects within the different meson sectors show similarities for both models. For example in both cases we find the  rough hierarchy $B_d < B_s \ll K$ for effects in the corresponding observables. The observable $\varepsilon_K$ acts in both cases as strongest constraint. Finally both frameworks cannot explain the anomalous $Z b \bar b$ coupling.

\subsection{Randall-Sundrum with custodial protection}

In the RS model with custodial protection (RSc), $\Delta F= 2$ operators are generated already at the tree level by 
the exchange of heavy Kaluza-Klein gluon and electroweak gauge boson modes (see \cite{Blanke:2008zb} for an extensive 
phenomenological analysis). Tree level flavour changing couplings of the SM Higgs generate $\Delta F = 2$ transitions as well, 
but they turn out to be subleading with respect to the aforementioned KK gauge boson contributions \cite{Buras:2009ka,Duling:2009pj}. 
In contrast to the LRM studied here, only $K-\bar K$ mixing is dominated by the chirally enhanced $Q_2^{\rm LR}$ operator 
generated by KK gluon exchange. Since the chiral enhancement is absent in the case of $B_{d,s}-\bar B_{d,s}$ mixing the 
latter processes are dominated by the operator $Q_1^{\rm VLL}$ and electroweak KK modes are relevant. Consequently while 
the constraint from $\eps_K$ generically puts strong bounds on the model in question \cite{Csaki:2008zd}, the contributions 
to $M_{12}^{d,s}$ are suppressed by the RS-GIM mechanism \cite{Agashe:2004cp}. Still non-zero effects are generally expected, 
so that a solution of the $\eps_K$ -- $S_{\psi K_S}$ tension is possible within this framework. Also $S_{\psi\phi}$ can receive large enhancements.

In contrast to $\Delta F= 2$ transitions, the dipole operators governing the $B\to X_{s,d}\gamma$ decays are generated first 
at the one loop level. As in the LRM the $m_b$ suppression can be overcome by a chirality flip inside the loop. However in 
contrast to that model, in the RSc the primed operators $C'_{7\gamma,8G}$ are generally dominant \cite{Agashe:2004cp,Isidori:2010kg,Blanke:2011xx}, {so that an enhanced branching ratio is generally expected.}

\subsection{Four generations: SM4} 
As opposed to the scenarios discussed so far, there are only SM operators 
contributing to $\Delta F=2$ processes and the NP effects are primarily 
enhanced through the non-decoupling effects of the $t'$-quark. The 
presence of new mixing angles and of new phases allows then to solve 
the $S_{\psi K_S}-\varepsilon_K$ anomaly in a straightforward manner and 
simultaneously enhance $S_{\psi\phi}$ if necessary. Moreover the agreement 
of theory and data can be improved in various observables.

{Concerning the $B\to X_s \gamma$ decay, as seen in figure 17 of \cite{Buras:2010pi}, the 
branching ratio can be enhanced up to $4\cdot 10^{-4}$ but also significantly 
suppressed below $3\cdot 10^{-4}$. Large enhancements like the ones from 
$H^\pm$ in the LRM are not possible in the SM4, in particular if one wants to
get interesting deviations from the SM3 predictions for $S_{\psi\phi}$ and 
$B_s\to\mu^+\mu^-$.}

\subsection{Littlest Higgs with T-parity}

As in the case of the SM4 no new operators enter in the LHT model and FCNC processes are generated first at the 
one-loop level \cite{Hubisz:2005bd,Blanke:2006sb}. Still due to the absence of an intrinsic flavour structure in 
the mirror fermion sector the effects on $\Delta F = 2$ transitions are generally large, and in particular $\eps_K$
 puts strong constraints on the new mixing parameters. At the same time the presence of new 
mixings and phases in the mirror sector allows to solve anomalies including a
possible enhancement of $S_{\psi\phi}$. However the effects are not as 
strong as in the case of SM4 \cite{Blanke:2008ac,Blanke:2009am}. 
Due to the absence of right-handed flavour violating currents, no chirality enhanced contribution to $B\to X_s\gamma$ arises and the effects remain very small.

\section{Summary}\label{sec:conc}
In this paper we have presented a complete study of $\Delta S=2$ and $\Delta B=2$ processes in the LRM. 
This includes $\varepsilon_K$, $\Delta M_K$,  $\Delta M_s$, $\Delta M_d$, $A_{\rm SL}^q$, $\Delta\Gamma_q$, 
and the mixing induced CP asymmetries $S_{\psi K_S}$ and  $S_{\psi \phi}$. 
We have included the new contributions from box diagrams with $W_R$ gauge boson and charged Higgs 
$(H^\pm)$  exchanges and tree level contributions from heavy neutral Higgs exchanges. We have also analysed the 
$B\to X_{s,d}\gamma$ decays that receive important new contributions from the 
$W_L-W_R$ mixing and $H^\pm$  exchanges. Compared to the existing literature the novel feature of our analysis 
is the search for correlations between various observables, simultaneous inclusion of all relevant contributions, 
in particular Higgs contributions and an improved treatment of QCD corrections. Our main findings are as follows:
\begin{itemize}
\item We find that the LRM is put under pressure by the 
$\varepsilon_K$ constraint. This is due to  the tree-level contributions of the neutral Higgs scalars and the 
related LR operators whose contributions are enhanced through renormalisation 
group effects and their chirally enhanced hadronic matrix elements. While 
this problem has been known for many years, we stress that even if the heavy Higgs
masses are of order\linebreak $15-20~\tev$, these contributions are not only important 
but even dominant. Increasing these masses much more would make the Higgs system non-perturbative. Leaving them out from 
phenomenological analysis, as done in some papers, is simply wrong. Figure~\ref{fig:lrcontrib} makes this point 
explicitly. On the other hand we show that there are structures (hierarchical or inverted) of the matrix $V^{\rm R} $ where the $\varepsilon_K$-constraint can be satisfied 
without large fine-tuning of parameters.
\item
The contributions of $W_R$ to $\Delta F=2$ observables, in particular 
$\varepsilon_K$,  is important but in view of the new improved lower experimental bound on its mass much less 
problematic than the tree level Higgs contributions, although also here
the LR operators mentioned above enter. Consequently, if only these 
contributions were present, the matrix $V^{\rm R}$ would be somewhat less 
hierarchical than the CKM matrix. 
\item
The charged Higgs $H^\pm$ being as heavy as the neutral ones plays a 
subdominant role in $\Delta F=2$ processes but can give significant 
contributions to $B\to X_{s,d}\gamma$ decays. These contributions 
are often neglected. We stress that this is not justified. 
Figure~\ref{fig:bsg_higgs_contrib}  makes this point 
explicitly. We find that the branching ratios for these decays can be significantly 
affected by new contributions but the CP-violating effects are 
too small to be distinguished from the SM results in view of 
hadronic uncertainties.
\item
{In our analysis of FCNC processes we have taken into account all 
existing constraints from electroweak precision observables and 
tree level decays. In this context we have found that in these 
models the SM problem with RH $Z\to b\bar b$ couplings and the 
so-called $\vub$-problem cannot be solved. The increased 
value of $M_{W_R}$ due to LHC combined with FCNC constraints does not 
allow to explain the difference in exclusive and inclusive determinations 
of $\vub$ with the help of RH currents within the context of the LRM considered 
here.}
\item
We have found that the NP effects are largest in the $K$ system but 
large effects can also be found in the $B_s$ system. Significant but 
smaller effects are found in the $B_d$ system.
\item
Guided by the persisting $S_{\psi K_S}$ - $\varepsilon_K$ tension present 
in the SM we have constructed a simple analytic expression for $V^{\rm R}$, which 
is given in (\ref{eq:golden Vr}). It is given  in 
terms of two mixing angles $\ts_{13}$ and $\ts_{23}$ and two corresponding 
phases $\phi_1=\phi_{13}$ and $\phi_2=\phi_{23}$ and allows to remove or 
soften the tension in question for scales 
$M_{W_R}\simeq 2-3\tev$ in the reach of the LHC.  In this scenario for 
$V^{\rm R}$ interesting correlations with CP violation in the $B_s$ system 
are present that depend on the value of $\vub$ chosen as well as the 
numerical values of the elements of $V^{\rm R}$. 
In two $\vub$ scenarios considered by us
 $S_{\psi\phi}$ can be significantly  enhanced over the SM value. When the data on 
$S_{\psi \phi}$ 
will improve one will be able to learn more about the structure of $V^{\rm R}$.
\item A more involved numerical analysis allows to find other structures 
of $V^{\rm R}$ with all its nine parameters entering the game 
that without large fine-tuning  of these parameters allows to obtain 
interesting results. {In particular as shown in figure~\ref{fig:absVR} three 
oases are found in the space $(|V_{us}^{\rm R}|,|V_{ub}^{\rm R}|,|V_{cb}^{\rm R}|)$, of which two 
have low fine-tuning. The pattern of flavour violation in the scenario
with ``normal'' hierarchy in $V^{\rm R}$ has been presented in 
several plots in section~\ref{sec:VR4S} and the corresponding Wolfenstein-like 
parametrization has been derived.} 
\item We find a soft lower limit for the mass of the heavy Higgses of $M_H \gsim 2.4\,{\rm TeV}$, see figure \ref{fig:higgs_mass}.
\end{itemize}

Our analysis of the LRM presented in this paper was dominated by the 
observables on which already good data exist. We have seen that already 
these data put stringent constraints on this NP scenario. 
{However there are other observables not discussed by us, that 
have not been measured yet but the experimental upper bounds on them put significant 
constraints on the LRM. Among them a prominent role is played by the 
neutron electric dipole moment (EDM). The most recent analysis in 
\cite{Zhang:2007da}
demonstrates that the neutron EDM provides lower bounds on $W^\prime$ and 
heavy neutral Higgs masses that are competitive with those coming from 
$\varepsilon_K$. Determining what impact  the present bounds on the
neutron EDM have on the LRM considered by us would require a detailed analysis of all contributions and taking into account the significant hadronic uncertainties.  
We leave such an analysis for  future work.
Similarly, in this decade} 
other observables, like rare $B$ and $K$ decay branching ratios and 
related CP violating observables,
 will be measured with sufficient precision so that we will be able to 
find out whether this NP scenario is viable. Of course the discovery of 
new gauge bosons $W^\prime$ and $Z^\prime$ at the LHC would be the most 
spectacular manifestation of the kind of new physics analysed in our paper.

\subsection*{Acknowledgements}
We would like to thank Andreas Crivellin, Jennifer Girrbach, Mikolaj Misiak, Kai Schmitz and Emmanuel Stamou for 
very useful discussions. This research was done in the context of the ERC Advanced Grant project ``FLAVOUR''(267104) and was
partially supported by the Graduiertenkolleg GRK 1054, the German Bundesministerium f{\"u}r Bildung und
Forschung under contract 05HT6WOA and by the U.S. National Science Foundation through grant PHY-0757868 and CAREER award PHY-0844667.

\begin{appendix}

\section{Higgs potential}\label{app:potential}
The most general renormalisable Higgs potential invariant under parity is given
by \cite{Deshpande:1990ip,Barenboim:2001vu,Kiers:2005gh,Zhang:2007da}
\begin{eqnarray}\label{eq:Higgspot}
&&V(\phi, \Delta_L, \Delta_R) = - \mu_1^2 {\rm Tr} (\phi^{\dag} \phi) - \mu_2^2
\left[ {\rm Tr} (\tilde{\phi} \phi^{\dag}) + {\rm Tr} (\tilde{\phi}^{\dag} \phi) \right]
- \mu_3^2 \left[ {\rm Tr} (\Delta_L \Delta_L^{\dag}) + {\rm Tr} (\Delta_R
\Delta_R^{\dag}) \right] \nonumber
\\
&&+ \lambda_1 \left[ {\rm Tr} (\phi^{\dag} \phi) \right]^2 + \lambda_2 \left\{ \left[
{\rm Tr} (\tilde{\phi} \phi^{\dag}) \right]^2 + \left[ {\rm Tr}
(\tilde{\phi}^{\dag} \phi) \right]^2 \right\} \nonumber \\
&&+ \lambda_3 {\rm Tr} (\tilde{\phi} \phi^{\dag}) {\rm Tr} (\tilde{\phi}^{\dag} \phi) +
\lambda_4 {\rm Tr} (\phi^{\dag} \phi) \left[ {\rm Tr} (\tilde{\phi} \phi^{\dag}) + {\rm
Tr}
(\tilde{\phi}^{\dag} \phi) \right]\nonumber \\
&& + \rho_1 \left\{ \left[ {\rm Tr} (\Delta_L \Delta_L^{\dag}) \right]^2 + \left[ {\rm
Tr} (\Delta_R \Delta_R^{\dag}) \right]^2 \right\} \nonumber \\ && + \rho_2 \left[ {\rm
Tr} (\Delta_L \Delta_L) {\rm Tr} (\Delta_L^{\dag} \Delta_L^{\dag}) + {\rm Tr} (\Delta_R
\Delta_R) {\rm Tr} (\Delta_R^{\dag} \Delta_R^{\dag}) \right] \nonumber
\\
&&+ \rho_3 {\rm Tr} (\Delta_L \Delta_L^{\dag}) {\rm Tr} (\Delta_R \Delta_R^{\dag})+
\rho_4 \left[ {\rm Tr} (\Delta_L \Delta_L) {\rm Tr} (\Delta_R^{\dag} \Delta_R^{\dag}) +
{\rm Tr} (\Delta_L^{\dag} \Delta_L^{\dag}) {\rm Tr} (\Delta_R
\Delta_R) \right]  \nonumber \\
&&+ \alpha_1 {\rm Tr} (\phi^{\dag} \phi) \left[ {\rm Tr} (\Delta_L \Delta_L^{\dag}) +
{\rm Tr} (\Delta_R \Delta_R^{\dag})  \right] \nonumber
\\
&&+ \left\{ \alpha_2 e^{i \delta_2} \left[ {\rm Tr} (\tilde{\phi} \phi^{\dag}) {\rm Tr}
(\Delta_L \Delta_L^{\dag}) + {\rm Tr} (\tilde{\phi}^{\dag} \phi) {\rm Tr} (\Delta_R
\Delta_R^{\dag}) \right] + {\rm h.c.}\right\} \nonumber
\\
&&+ \alpha_3 \left[ {\rm Tr}(\phi \phi^{\dag} \Delta_L \Delta_L^{\dag}) + {\rm
Tr}(\phi^{\dag} \phi \Delta_R \Delta_R^{\dag}) \right] + \beta_1 \left[ {\rm Tr}(\phi
\Delta_R \phi^{\dag} \Delta_L^{\dag}) +
{\rm Tr}(\phi^{\dag} \Delta_L \phi \Delta_R^{\dag}) \right] \nonumber \\
&&+ \beta_2 \left[ {\rm Tr}(\tilde{\phi} \Delta_R \phi^{\dag} \Delta_L^{\dag}) + {\rm
Tr}(\tilde{\phi}^{\dag} \Delta_L \phi \Delta_R^{\dag}) \right] + \beta_3 \left[ {\rm
Tr}(\phi \Delta_R \tilde{\phi}^{\dag} \Delta_L^{\dag}) + {\rm Tr}(\phi^{\dag} \Delta_L
\tilde{\phi} \Delta_R^{\dag}) \right] \,,
\end{eqnarray}
where there are a total of 18 parameters, $\mu^2_{1,2,3}$, $\lambda_{1,2,3,4}$,
$\rho_{1,2,3,4}$, $\alpha_{1,2,3}$, and $\beta_{1,2,3}$.  Due to the left-right symmetry, only one of them, the coupling involving $\alpha_2$ can become complex and all other couplings are real.

We have investigated the possibility of a more general Higgs potential, where no left-right symmetry is imposed. However as the parameters of the Higgs potential do not affect our flavour analysis, taking into account that leading order couplings of Higgs and quark fields are independent of these parameters, we find it sufficient to give here the left-right symmetric form. 
In fact among the parameters of the Higgs sector only $\alpha_3$ in 
(\ref{eq:Higgspot}), determining the heavy Higgs masses, is relevant for our 
analysis.

\section{Gauge boson masses and mixings}
\label{app:gauge-boson-masses}

The gauge boson mass matrices can straightforwardly be obtained from 
the relevant terms of equations (\ref{eq:Higgskin}) and (\ref{eq:covder}). Diagonalising the resulting mass matrix for the charged gauge bosons yields the mass eigenstates
\bea
W^\pm &=& W_L^\pm+sce^{\mp i\alpha} \frac{s_Rc_W}{s_W}\epsilon^2 W_R^\pm\,,\\
W'^\pm &=& W_R^\pm-sce^{\pm i\alpha} \frac{s_Rc_W}{s_W}\epsilon^2W_L^\pm\,,
\eea
where we defined $v=\sqrt{\kappa^2 + \kappa'^2}$, $c=\kappa/v$, $s=\kappa'/v$ and we neglected higher order terms in $\epsilon=v/\kappa_R$. We also introduced the mixing angles
\be
s_R = \frac{g'}{\sqrt{g'^2+g_R^2}}\,,\quad c_R = \sqrt{1-s_R^2}\,,\qquad  s_W = \frac{ s_R}{\sqrt{(g_L/g_R)^2+  s_R^2}}\,,\quad c_W = \sqrt{1-s_W^2}\,.
\ee
The corresponding masses read
\bea
(M_W)^2 &=& \frac{e^2 v^2}{2 s_W^2} \left( 1 - 2s^2c^2 \epsilon^2\right)
\,,\\
(M_{W_R})^2 &=& \frac{e^2 \kappa_R^2}{ c_W^2 s_R^2} \left(1+ \frac{1}{2}\epsilon^2\right)\,.
\eea

For the neutral gauge bosons we can write
\bea
A &=&  s_W W_L^3 +  s_R c_W W_R^3 +  c_R c_W B\,,\\
Z &=&  c_W W_L^3 - s_R s_W \left(1-\frac{ c_R^4}{4 s_W^2}\epsilon^2 \right) W_R^3 - c_R s_W \left(1+\frac{ s_R^2 c_R^2}{4 s_W^2}\epsilon^2 \right) B\,\\
Z' &=& - \frac{ s_R c_R^3c_W}{4s_W}\epsilon^2 W^3_L + 
c_R\left(1+ \frac{ s_R^2 c_R^2}{4}\epsilon^2\right)W_R^3  - s_R \left(1-\frac{ c_R^4}{4}\epsilon^2\right) B\,.
\eea
The masses are given by
\bea
(M_A)^2 &=& 0\,,\\
(M_Z)^2 &=& \frac{e^2 v^2}{2 s_W^2 c_W^2}\left(1-\frac{ c_R^4}{4}\epsilon^2\right)\,,\\
(M_{Z'})^2 &=& \frac{2e^2\kappa_R^2}{ s_R^2 c_R^2 c_W^2} \left(1+\frac{ c_R^4}{4}\epsilon^2\right)\,.
\eea

\section{Goldstone boson and Higgs mass eigenstates}
\label{app:goldhiggs} 

The doubly charged components $\delta_{L,R}^{++}$ lead to two physical doubly charged Higgses.

The singly charged fields ${\phi_{1,2}^\pm}$ and $\delta_{L,R}^\pm$ form the Goldstone bosons of $W^\pm$ and $W'^\pm$ and two charged Higgses. The Goldstone bosons are given by
\bea
G^\pm &=& \pm i \left[ c\left(1-s^4\epsilon^2\right)\phi_1^\pm - s e^{\mp i\alpha}\left(1-c^4\epsilon^2\right)\phi_2^\pm  - \sqrt{2} c s e^{\mp i\alpha} \epsilon \delta_R^\pm\right]\,,
\\
G'^\pm &=& \mp i\left[ \left(1-\frac{\epsilon^2}{4}\right) \delta_R^\pm + \frac{s e^{\pm i \alpha}}{\sqrt{2}}\epsilon \phi_1^\pm -\frac{c}{\sqrt{2}}\epsilon\phi_2^\pm\right]\,.
\eea
The physical singly charged Higgs states are linear combinations of $\delta_L^\pm$ and 
\be
h^\pm = se^{\pm i \alpha}  \left(1 - \frac{(c^2- s^2)^2}{4}\epsilon^2\right)\phi_1^\pm
+c  \left(1 - \frac{(c^2- s^2)^2}{4}\epsilon^2\right) \phi_2^\pm
+\frac{c^2-s^2}{\sqrt{2}}\epsilon\delta_R^\pm\,.
\ee

Finally the neutral Goldstone boson and Higgs fields are built out of $\phi_{1,2}^0$ and $\delta_{L,R}^0$. Defining $\pi^0 = c \Im \phi_1^0-s \Im (e^{-i\alpha}\phi_2^0)$, the Goldstone bosons of $Z$ and $Z'$ read
\bea
G^0 &=& {\sqrt{2}}\left(1-\frac{ c_R^4}{8}\epsilon^2\right)\pi^0-\frac{ c_R^2}{\sqrt{2}}\epsilon\Im \delta_R^0\,,\\
G'^0 &=& -{\sqrt{2}}\left(1-\frac{ c_R^4}{8}\epsilon^2\right)\Im \delta_R^0-\frac{ c_R^2}{\sqrt{2}}\epsilon\pi^0
\,.
\eea
In addition there are six neutral Higgs fields in the spectrum, that are linear combinations of $s \Im \phi_1^0+c \Im (e^{-i\alpha}\phi_2^0)$, $\Re
\phi_1^0$, $\Re(e^{-i\alpha}\phi_2^0)$, $\Re\delta_R^0$, $\Re\delta_L^0$ and $\Im\delta_L^0$.

In order to make a more detailed statement about the Higgs mass eigenstates one has to diagonalize the Higgs potential (\ref{eq:Higgspot}). Considering the neutral 8 by 8 mass matrix, its diagonalisation can be best done by using perturbation theory. {While in \cite{Zhang:2007da} a further hierarchy, namely $\kappa' \ll \kappa$, has been assumed, we do not restrict ourselves to this case but determine the Higgs mass eigenstates to leading order for arbitrary $s=\kappa'/v$.
It turns out that the leading order Higgs couplings are not sensitive to the detailed structure of the potential and in particular the parity invariant potential yields the same result. The leading order mass eigenstates of the Higgs fields of our interest are then given by
\begin{eqnarray}
h^0 &=& \sqrt{2}\left(c \Re \phi_1^0  + s \Re(e^{-i\alpha}\phi_2^0)  \right)\\
H_1^0 &=& \sqrt{2} \left(-s \Re \phi_1^0+c \Re(e^{-i\alpha}\phi_2^0)   \right) \\
H_2^0 &=& \sqrt{2} \left( s \Im \phi_1^0 + c \Im(e^{-i\alpha}\phi_2^0)   \right) 
\end{eqnarray}
where $h_0$ can be identified as the light Higgs with a mass of $\ord(v)$, while $H^0_1$ and $H^0_2$ are two new flavour-violating neutral Higgses with masses $\ord(\kappa_R)$.
To leading order their masses are equal to each other
\begin{equation}
M_H^2\equiv M_{H_1^0}^2 = M_{H_2^0}^2 = \frac{\alpha_3\kappa_R^2}{1-2s^2} = \alpha_3\kappa_R^2 \sqrt{u(s)}\,.
\end{equation}
In the course of the numerical analysis we regard $M_H$ as a free parameter.
These results agree with the ones given by the authors of \cite{Zhang:2007da} in the limit of small $s\ll 1$. A more explicit analysis of the  Higgs sector of this class of models can be also found in \cite{Zhang:2007da,Deshpande:1990ip,Kiers:2005gh}. Charged Higgs effects in LR models are often neglected in the literature with the 
argument that they have  to be small in FCNC processes as they have to take place at one loop order \cite{Mohapatra:1983ae}.  This is in fact true for the 
$\Delta F=2$ processes  but not for $B\to X_{s,d}\gamma$. This has already been pointed out in \cite{Asatrian:1989iu,Asatryan:1990na,Babu:1993hx,Fujikawa:1993zu,Asatrian:1996as,Frank:2010qv} in LR models and confirmed here by us. As in the case of the neutral Higgs sector we are only interested in leading order couplings and masses.
We find that the mass of the lightest charged Higgs,
\be
H^\pm = s e^{\pm i\alpha} \phi_1^\pm + c \phi_2^\pm \,,
\ee
is given to a very 
good approximation by
\bea
M_{H+}=M_{H} 
\eea
and hence cannot be chosen to be as light as $1\tev$ as done sometimes in the literature.}

\section{\boldmath Numerical details for $\Delta F=2$}\label{app:num_details_deltaf2}
Here we give the values of $R_{ij}$ as defined in (\ref{Rij}) and used
in the main part of our numerical analysis.
Including all contributions we find for $\epsilon=0.03$, $s_R=0.8$ and $s=0.1$
\be
\hat R(K) = \left(
\begin{array}{ccc}
  -4.3236\cdot 10^{-9} & -1.2585\cdot 10^{-6} & -7.8924\cdot 10^{-5}\\
  -1.2585\cdot 10^{-6} & -6.5793\cdot 10^{-4} & -4.1333\cdot 10^{-2}\\
  -7.8924\cdot 10^{-5} & -4.1333\cdot 10^{-2} & -9.1112
\end{array}
\right)\,,
\ee

\be
\hat R(B) = \left(
\begin{array}{ccc}
  -2.7079\cdot 10^{-10} & -7.8847\cdot 10^{-8} &-4.9574\cdot 10^{-6}\\
  -7.8847\cdot 10^{-8} & -4.1217\cdot 10^{-5} &-2.5958\cdot 10^{-3}\\
  -4.9574\cdot 10^{-6} & -2.5958\cdot 10^{-3} &-0.5727
\end{array}
\right)\,.
\ee
{For $s = 0.5$ the Higgs contributions get enhanced and the calculation of the $R_{ij}$ yields}
\be
\hat R(K) = \left(
\begin{array}{ccc}
  -4.6987\cdot 10^{-9} & -1.4447\cdot 10^{-6} & -1.3159\cdot 10^{-4}\\
  -1.4447\cdot 10^{-6} & -7.5036\cdot 10^{-4} & -6.7475\cdot 10^{-2}\\
  -1.3159\cdot 10^{-4} & -6.7475\cdot 10^{-2} & -16.546
\end{array}
\right)\,,
\ee

\be
\hat R(B) = \left(
\begin{array}{ccc}
  -2.9439\cdot 10^{-10} & -9.0565\cdot 10^{-8} & -8.2716\cdot 10^{-6}\\
  -9.0565\cdot 10^{-8} & -4.7034\cdot 10^{-5} & -4.2408\cdot 10^{-3}\\
  -8.2716\cdot 10^{-6} & -4.2408\cdot 10^{-3} & -1.0406
\end{array}
\right)\,.
\ee
Keeping only gauge boson contributions we find {(no $s$ dependence)}
\be
\hat R(K,{\rm gauge}) = \left(
\begin{array}{ccc}
    -3.9343\cdot 10^{-9} & -1.0653\cdot 10^{-6} & -2.4264\cdot 10^{-5} \\
    -1.0653\cdot 10^{-6} & -5.6199\cdot 10^{-4} & -1.4200\cdot 10^{-2} \\
    -2.4264\cdot 10^{-5} & -1.4200\cdot 10^{-2} & -1.3954
\end{array}
\right)\,,
\ee

\be
\hat R(B, {\rm gauge}) = \left(
\begin{array}{ccc}
  -2.4630\cdot 10^{-10} & -6.6689\cdot 10^{-8} & -1.5190\cdot 10^{-6} \\
  -6.6689\cdot 10^{-8} & -3.5182\cdot 10^{-5} & -8.8898\cdot 10^{-4} \\
  -1.5190\cdot 10^{-6} & -8.8898\cdot 10^{-4} & -8.7355\cdot 10^{-2}
\end{array}
\right)\,.
\ee
\noindent The $H^\pm$ contributions are significantly smaller than gauge contributions.

\section{Feynman rules}

In this part of the appendix we present the Feynman rules up to and 
including $\ord(\epsilon^2)$ corrections, except for the couplings to Higgs bosons where we
restrict ourselves to the leading order couplings\footnote{We would like to thank Jennifer Girrbach for checking the Feynman rules presented here.}. The Feynman rules are given in mass 
eigenstates of the particular fields. We want to stress that we keep here the correct notation 
$W$ and $W'$, as introduced in appendix \ref{app:gauge-boson-masses}, in spite of using throughout
the phenomenological analysis the simplified notation $W_L$ and $W_R$ for the mass eigenstates.

Concerning triple gauge couplings a comment is in order. The Dirac structure of all vertices is the same,
\FloatBarrier
\begin{figure}[h!]
\centering
\includegraphics[width=0.8\textwidth]{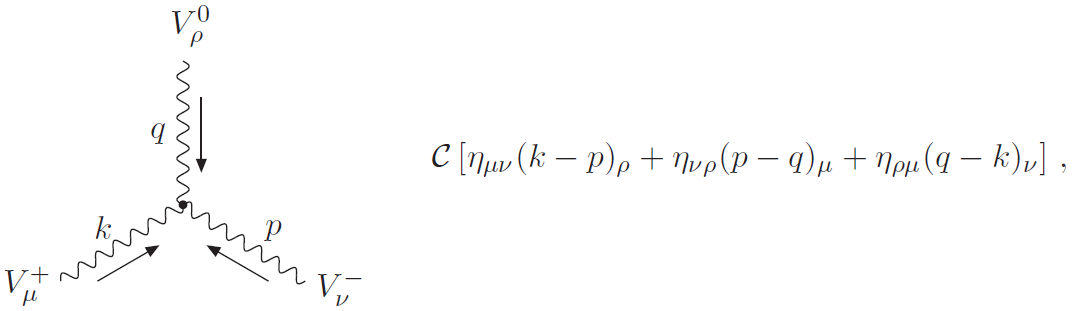}
\label{fig:tripgaugediag}
\end{figure}
\FloatBarrier
\noindent where $V_\mu^+ =W_\mu^+,W'^+_\mu$, $V_\nu^-=W_\nu^-, W'^-_\nu$, $V_\rho^0=A_\rho, Z_\rho,,Z_\rho'$, and $k,p,q$ are their incoming momenta. Therefore in table \ref{tab:WWZ} we collect only the coefficients $\mathcal{C}$ of the respective couplings.

\begin{table}[h!]
\renewcommand{\arraystretch}{1.5}
\begin{center}
\begin{tabular}{|c|c|c|} 
\hline
& $A_\mu$ & $G^a_\mu$\\ 
\hline\hline
$  \bar u_L^{i} u_L^{i} X_{\mu} $ &
$- i \frac{2}{3} e \gamma^{\mu}  $  &
$- i g_s \gamma^{\mu}\, t^a  $ \\\hline

$ \bar u_R^{i} u_R^{i} X_{\mu} $ &
$- i \frac{2}{3} e \gamma^{\mu} $ &
$- i g_s \gamma^{\mu}\, t^a $\\\hline

$  \bar d_L^{i} d_L^{i} X_{\mu}$ &
$- i  (-\frac{1}{3}) e \gamma^{\mu} $  &
$- i g_s \gamma^{\mu}\, t^a  $ \\\hline

$ \bar d_R^{i} d_R^{i} X_{\mu}$ &
$- i (-\frac{1}{3}) e \gamma^{\mu} $ &
$- i g_s \gamma^{\mu}\, t^a $ \\\hline
\end{tabular} 
\end{center}
\renewcommand{\arraystretch}{1.0}
\caption{Fermion couplings to the massless gauge bosons $X_\mu$: photon $A_\mu$ and gluons $G^{a}_{\mu}$}
\end{table}

\begin{table}[h!]
\renewcommand{\arraystretch}{1.5}
\begin{center}
\begin{tabular}{|c|c|c|} 
\hline
 & $W^+$ & $W^{\prime +}$\\
\hline\hline
$  \bar u_L^{i} d_L^{j} X^+_{\mu} $ &
$- \frac{i e}{\sqrt{2}  s_W} V^{\rm L}_{ij} \gamma^{\mu}  $  &
$+  \frac{i e c s e^{-i \alpha}  s_R c_W}{\sqrt{2}   s_W^2} \epsilon^2 V^{\rm L}_{ij} \gamma^{\mu}  $ \\\hline

$ \bar u_R^{i} d_R^{j} X^+_{\mu} $ &
$- \frac{i e c s e^{i \alpha}}{\sqrt{2}  s_W} \epsilon^2 V^{\rm R}_{ij} \gamma^{\mu}  $ &
$ -  \frac{i e}{\sqrt{2}  s_R  c_W}  V^{\rm R}_{ij} \gamma^{\mu} $ \\\hline
\end{tabular} 
\end{center}
\renewcommand{\arraystretch}{1.0}
\caption{Fermion couplings to $W^+$ and $W^{\prime +}$}
\end{table}

\begin{table}[h!]
\renewcommand{\arraystretch}{1.5}
\begin{center}
\begin{tabular}{|c|c|c|} 
\hline
& $Z$ & $Z^\prime$\\ 
\hline\hline
$  \bar u_L^{i} u_L^{i} X_{\mu} $ &
$-  \frac{i e}{ s_W  c_W} (\frac{1}{2} -\frac{2}{3}  s_W^2-\frac{1}{24} s_R^2  c_R^2 \epsilon^2) \gamma^{\mu}  $  &
$ie\, (\frac{1}{6} \frac{s_R}{c_R c_W}+\frac{1}{8}  (\frac{c_W c_R^3 s_R}{  s_W^2}-\frac{ c_R^3 s_R}{3  c_W})\epsilon^2) \gamma^{\mu}$ \\\hline

$  \bar d_L^{i} d_L^{i} X_{\mu}$ &
$-  \frac{i e}{ s_W  c_W} (-\frac{1}{2} + \frac{1}{3}  s_W^2-\frac{1}{24} s_R^2  c_R^2 \epsilon^2) \gamma^{\mu}  $  &
$ie\, (\frac{1}{6} \frac{s_R}{c_R c_W}-\frac{1}{8}  (\frac{c_W c_R^3 s_R}{ s_W^2}+\frac{ c_R^3 s_R}{3  c_W})\epsilon^2 ) \gamma^{\mu}$ \\\hline

$ \bar u_R^{i} u_R^{i} X_{\mu} $ &
$-  \frac{i e}{ s_W  c_W} (-\frac{2}{3}  s_W^2 +\frac{ c_R^2}{8} ( c_R^2-\frac{1}{3}  s_R^2) \epsilon^2) \gamma^{\mu}  $  &
$-  \frac{i e}{ c_W  s_R  c_R} (\frac{1}{2}-\frac{2}{3} s_R^2+\frac{ c_R^4  s_R^2}{6} \epsilon^2) \gamma^{\mu}  $ \\\hline

$ \bar d_R^{i} d_R^{i} X_{\mu}$ &
$-  \frac{i e}{ s_W  c_W} (\frac{1}{3}  s_W^2 -\frac{ c_R^2}{8} ( c_R^2+\frac{1}{3}  s_R^2) \epsilon^2) \gamma^{\mu}  $  &
$-  \frac{i e}{ c_W  s_R  c_R} (-\frac{1}{2}+\frac{1}{3} s_R^2-\frac{ c_R^4  s_R^2}{12} \epsilon^2) \gamma^{\mu}  $ \\\hline
\end{tabular} 
\end{center}
\renewcommand{\arraystretch}{1.0}
\caption{Fermion couplings to neutral gauge bosons $Z$ and $Z^{'}$}
\end{table}

\begin{table}[h!]
\renewcommand{\arraystretch}{1.5}
\begin{center}
\begin{tabular}{|c|c|c|} 
\hline
& $G^+$ & $G^{\prime +}$\\ 
\hline\hline
$  \bar u_L^{i} d_R^{j} X^+ $ & 
$ \frac{e}{\sqrt{2}M_W s_W}\left(m_d^j V^{\rm L}_{ij} - cse^{i\alpha}m_u^i \epsilon^2 V^{\rm R}_{ij}\right) $ & 
$ -\frac{e m_u^i}{\sqrt{2} c_W s_R M_{W_R}} V^{\rm R}_{ij} $
\\\hline

$ \bar u_R^{i} d_L^{j} X^+ $ & 
$ -\frac{e}{\sqrt{2}M_W s_W}\left(m_u^i V^{\rm L}_{ij} - cse^{i\alpha}m_d^j \epsilon^2 V^{\rm R}_{ij}\right) $ & 
$ \frac{e m_d^j}{\sqrt{2} c_W s_R M_{W_R}} V^{\rm R}_{ij} $
\\\hline
\end{tabular} 
\end{center}
\renewcommand{\arraystretch}{1.0}
\caption{Fermion couplings to charged Goldstone bosons}
\end{table}

\begin{table}[h!]
\renewcommand{\arraystretch}{1.5}
\begin{center}
\begin{tabular}{|c|c|c|} 
\hline
& $G^0$ & $G^{\prime 0}$ \\
\hline\hline
$  \bar u_L^{i} u_R^{i} X^0 $ & 
$ \frac{e m_u^i}{2 c_W s_W M_Z}\left(1-\frac{ c_R^4}{4}\epsilon^2\right) $ & 
$ -\frac{e  c_R m_u^i}{2 s_R c_W M_{Z'}} $ \\\hline

$  \bar d_L^{i} d_R^{i} X^0 $ & 
$ -\frac{e m_d^i}{2 c_W s_W M_Z}\left(1-\frac{ c_R^4}{4}\epsilon^2\right) $ & 
$ \frac{e  c_R m_d^i}{2 s_R c_W M_{Z'}} $ \\\hline
\end{tabular} 
\end{center}
\renewcommand{\arraystretch}{1.0}
\caption{Fermion couplings to neutral Goldstone bosons $G^0$ and $G^{\prime 0}$}
\end{table}

\begin{table}[htbp]
\renewcommand{\arraystretch}{1.5}
\centering
\begin{tabular}{|c|c|c|c|} 
\hline
& $Z$ & $Z^\prime$ & $\qquad A\qquad$\\ 
\hline\hline
$ W^+W^- X $ &
$ i e \frac{ c_W}{ s_W}$ &
$ -ie\frac{ c_R^3 s_R c_W}{4s_W^2}\epsilon^2 $ &
$i e $\\\hline

$  W^+W'^-X $ &
$-i e \frac{c s e^{i
\alpha} s_R}{ s_W^2}\epsilon^2 $ &
$ ie \frac{cse^{i\alpha} c_R}{ s_W}\epsilon^2 $ &
$0$ \\\hline

$  W'^+W^- X $ &
$ -i e \frac{c s e^{-i\alpha} s_R}{ s_W^2}\epsilon^2 $  &
$ ie \frac{cse^{-i\alpha} c_R}{ s_W}\epsilon^2 $ &
$0$ \\\hline

$ W^{\prime +}W'^- X  $ &
$- ie\left(\frac{s_W}{c_W}-\frac{ c_R^4}{4 c_W s_W}\epsilon^2\right) $ &
$ ie\frac{ c_R}{ c_W}\left(\frac{1}{ s_R} + \frac{ c_R^2 s_R}{4}  \epsilon^2\right) $ &
$ i e $\\\hline
\end{tabular} 
\renewcommand{\arraystretch}{1.0}
\caption{Triple gauge couplings, involving a $Z$ and $Z^\prime$ boson and the photon, respectively}\label{tab:WWZ}
\end{table}

\begin{table}[htbp]
\renewcommand{\arraystretch}{1.5}
\centering
\begin{tabular}{|c|c|c|} 
\hline
& $Z$ & $A$ \\
\hline\hline
$ G^+(p)G^-(q) X_\mu $ &
$-\frac{i e}{2 c_W s_W}(p-q)_\mu\left[1-2 s_W^2+\left(\frac{ c_R^4}{4}-2s^2c^2\right)\epsilon^2\right] $ &
$-i e(p-q)_\mu $ \\\hline
$ G'^+(p)G^-(q) X_\mu $ &
$\frac{i e cse^{-i\alpha}}{\sqrt{2} c_W s_W}\epsilon(p-q)_\mu $ &
$0 $ \\\hline
$ G'^+(p)G'^-(q) X_\mu $ &
$i e \frac{s_W}{c_W} (p-q)_\mu\left[1-\frac{1- s_R^2 c_R^2}{4 s_W^2}\epsilon^2\right] $ &
$-i e(p-q)_\mu $ \\\hline
\end{tabular} 
\renewcommand{\arraystretch}{1.0}
\caption{Charged Goldstone couplings to the photon and the $Z$ boson}\label{tab:GGZ}
\end{table}

\begin{table}[htbp]
\renewcommand{\arraystretch}{1.5}
\centering
\begin{tabular}{|c|c|} 
\hline
$ G^+(p)G^-(q) Z'_\mu $ &
$ -\frac{ie c_R}{2 s_R c_W}(p-q)_\mu\left[1-\frac{ s_R^2 c_R^4(1-2 s_W^2) +8s^2c^2 s_W^2(1+ s_R^2)}{4 c_R^2 s_W^2} \epsilon^2 \right] $ \\\hline
$ G'^+(p)G^-(q) Z'_\mu $ &
$ \frac{iecse^{-i\alpha}(1+ s_R^2)}{\sqrt{2} c_R s_R c_W} \epsilon(p-q)_\mu $ \\\hline
$ G'^+(p)G'^-(q) Z'_\mu $ &
$ \frac{ie s_R}{ c_R c_W}(p-q)_\mu\left[1-\frac{1+ s_R^2(1+ c_R^4)}{4 s_R^2}\epsilon^2\right] $ \\\hline
\end{tabular} 
\renewcommand{\arraystretch}{1.0}
\caption{Charged Goldstone couplings to the $Z'$ boson}\label{tab:GGZp}
\end{table}

\begin{table}[htbp]
\renewcommand{\arraystretch}{1.5}
\centering
\begin{tabular}{|c|c|c|} 
\hline
& $W$ & $W^\prime$ \\ 
\hline\hline
$ G^0(p)G^-(q) X^+_\mu $ &
$ \frac{ie}{2 s_W}(p-q)_\mu\left[1-\frac{1}{8}{( c_R^4-8s^2c^2)}\epsilon^2\right] $ &
$ \frac{iecse^{-i\alpha}}{ c_W s_R}(p-q)_\mu\left[1-\frac{1}{2}\left(\frac{ c_R^4}{4}+\frac{ s_R^2}{ s_W^2}-2s^2c^2\right)\epsilon^2\right]$ \\\hline

$ G'^0(p)G^-(q) X^+_\mu $ &
$ -\frac{ie c_R^2}{4 s_W}\epsilon (p-q)_\mu$ &
$  \frac{iecse^{-i\alpha}(2- c_R^2)}{2 c_W s_R}\epsilon(p-q)_\mu$ \\\hline

$ G^0(p)G'^-(q) X^+_\mu $ &
$ -\frac{iecse^{i\alpha}}{\sqrt{2} s_W}\epsilon(p-q)_\mu$ &
$ -\frac{ie s_R}{2\sqrt{2} c_W}\epsilon(p-q)_\mu$ \\\hline

$ G'^0(p)G'^-(q) X^+_\mu $ &
$ \frac{iecse^{i\alpha}(2+ c_R^2)}{2\sqrt{2} s_W}\epsilon^2(p-q)_\mu  $ &
$ \frac{ie}{\sqrt{2} c_W s_R}(p-q)_\mu\left[1-\frac{1}{8}(1+s^4_R)\epsilon^2\right]  $ \\\hline
\end{tabular} 
\renewcommand{\arraystretch}{1.0}
\caption{Couplings of charged and neutral Goldstone Bosons to the $W$ and $W^\prime$ boson}
\end{table}

\begin{table}[htbp]
\renewcommand{\arraystretch}{1.5}
\centering
\begin{tabular}{|c|c|c|} 
\hline
& $G^0$ & $G^{\prime 0}$\\ 
\hline\hline
$ W^+W^- X^0 $ &
$ 0 $ &
$ 0 $ \\\hline

$  W^+W'^- X^0 $ &
$-\frac{2ecse^{i\alpha}M_Z}{ s_R}g_{\mu\nu} $ &
$ \frac{e c_R^3 cse^{i\alpha}M_{Z'} }{2 s_W}\epsilon^2 g_{\mu\nu} $\\\hline

$ W^{\prime +}W'^- X^0  $ &
$ 0 $ &
$ 0 $ \\\hline
\end{tabular} 
\renewcommand{\arraystretch}{1.0}
\caption{Couplings of charged gauge bosons to the $G^0$ and $G'^0$ boson}
\end{table}

\begin{table}[htbp]
\renewcommand{\arraystretch}{1.5}
\centering
\begin{tabular}{|c|c|c|} 
\hline
& $G^+$ & $G^{\prime +}$ \\ 
\hline\hline
$ W_\mu^- A_\nu X^+ $ &
$ -e M_W g_{\mu\nu} $ &
$ 0 $ \\\hline

$ W^{\prime -}_\mu A_\nu X^+ $ &
$ 0 $ &
$ -eM_{W_R}g_{\mu\nu} $ \\\hline

$ W_\mu^-Z_\nu X^+ $ &
$ \frac{e s_WM_W}{ c_W}g_{\mu\nu}\left(1 - \frac{c_R^4 - 8 s^2 c^2}{4 s_W^2} \epsilon^2\right) $ &
$ 0 $\\\hline

$ W_\mu^{\prime -}Z_\nu X^+ $ &
$ \frac{2ecse^{i\alpha} M_W}{ c_W^2 s_R}g_{\mu\nu} $ &
$ \frac{e s_WM_{W_R}}{c_W}\left(1-\frac{1+s_R^4}{4 s_W^2}\epsilon^2\right)g_{\mu\nu} $\\\hline

$ W_\mu^{ -}Z'_\nu X^+ $ &
$ -\frac{e c_R M_W}{ c_W s_R} g_{\mu\nu} $ &
$  \frac{2ecse^{-i\alpha}M_{W_R}}{ c_R s_W}\epsilon^2g_{\mu\nu} $\\\hline

$ W_\mu^{\prime -}Z'_\nu X^+ $ &
$ \frac{2ecse^{i\alpha} s_W(1+ s_R^2) M_W}{ c_W^2 s_R^2 c_R} g_{\mu\nu} $ &
$ \frac{e(1+ s_R^2)M_{W_R}}{ c_R s_R c_W}\left(1-\left(\frac{1}{2}+\frac{ s_R^2 c_R^4}{4(1+ s_R^2)}\right)\epsilon^2\right) g_{\mu\nu} $ \\\hline
\end{tabular} 
\renewcommand{\arraystretch}{1.0}
\caption{Couplings of charged and neutral gauge bosons to the $G^+$ and $G^{\prime +}$ boson}
\end{table}

\begin{table}[htbp]
\renewcommand{\arraystretch}{1.5}
\centering
\begin{tabular}{|c|c|c|} 
\hline
 & $H_1^0$ & $H_2^0$ \\\hline\hline

$  \bar d_L^{i} d_R^{j} X^0 $  & 
$ -\frac{i}{\sqrt{2}(1-2s^2)v}\left( e^{i\alpha} m_u^{a} V_{ai}^{\rm L*}  V^{\rm R}_{aj} - 2cs m_d^i \delta_{ij} \right) $ & 
$\frac{1}{\sqrt{2}(1-2s^2)v}\left( e^{i\alpha} m_u^{a} V_{ai}^{\rm L*}  V^{\rm R}_{aj} - 2cs m_d^i \delta_{ij} \right) $
\\\hline

$  \bar d_R^{i} d_L^{j} X^0$  & 
$ -\frac{i}{\sqrt{2}(1-2s^2)v}\left( e^{-i\alpha} m_u^{a} V^{\rm R *}_{ai}V_{aj}^{\rm L} - 2cs m_d^i \delta_{ij} \right) $ & 
$ -\frac{1}{\sqrt{2}(1-2s^2)v}\left( e^{-i\alpha} m_u^{a} V^{\rm R *}_{ai} V_{aj}^{\rm L} - 2cs m_d^i \delta_{ij} \right) $\\\hline

$  \bar u_L^{i} u_R^{j} X^0 $  & 
$ -\frac{i}{\sqrt{2}(1-2s^2)v}\left( e^{-i\alpha} m_d^{a} V_{ai}^{\rm L}  V^{\rm R *}_{aj} - 2cs m_u^i \delta_{ij} \right) $ & 
$  -\frac{1}{\sqrt{2}(1-2s^2)v}\left( e^{-i\alpha} m_d^{a} V_{ai}^{\rm L}  V^{\rm R *}_{aj} - 2cs m_u^i \delta_{ij} \right) $\\\hline

$  \bar u_R^{i} u_L^{j} X^0 $  &
$  -\frac{i}{\sqrt{2}(1-2s^2)v}\left( e^{i\alpha} m_d^{a} V^{\rm R }_{ai}V_{aj}^{\rm L *}   - 2cs m_u^i \delta_{ij} \right) $ &
$  \frac{1}{\sqrt{2}(1-2s^2)v}\left( e^{i\alpha} m_d^{a} V^{\rm R }_{ai}V_{aj}^{\rm L *}  - 2cs m_u^i \delta_{ij} \right) $ \\\hline
\end{tabular} 
\renewcommand{\arraystretch}{1.0}
\caption{Fermion couplings to the flavour violating neutral Higgses. Here $m_u^{a}$ and $m_d^{a}$ denote the $a$th up and down quark mass, respectively. Summation over $a$ is understood.}\label{tab:H01}
\end{table}

\begin{table}[htbp]
\renewcommand{\arraystretch}{1.5}
\centering
\begin{tabular}{|c|c|} 
\hline
$  \bar u_L^{i} d_R^{j} H^+ $  & 
$ -\frac{i}{(1-2s^2)v}\left( m_u^i V^{\rm R}_{ij} - 2cse^{-i\alpha}V^{\rm L}_{ij} m_d^j \right) $ \\\hline

$  \bar u_R^{i} d_L^{j} H^+ $  &
$ \frac{i}{(1-2s^2)v}\left( V^{\rm R}_{ij} m_d^j - 2cse^{-i\alpha}m_u^iV^{\rm L}_{ij}\right) $  \\\hline
\end{tabular} 
\renewcommand{\arraystretch}{1.0}
\caption{Charged Higgs couplings to fermions. Here $m_u^{a}$ and $m_d^{a}$ denote the $a$th up and down quark mass, respectively.}\label{tab:Hch}
\end{table}

\FloatBarrier

\end{appendix}

\bibliographystyle{JHEP}
\bibliography{lrsm}

\providecommand{\href}[2]{#2}\begingroup\raggedright\begin{thebibliography}{10%
0}

\bibitem{Glashow:1970gm}
S.~L. Glashow, J.~Iliopoulos, and L.~Maiani, {\it {Weak Interactions with
  Lepton-Hadron Symmetry}},  {\em Phys. Rev.} {\bf D2} (1970) 1285--1292.

\bibitem{Bevan:2011zz}
A.~Bevan, M.~Bona, M.~Ciuchini, D.~Derkach, A.~Stocchi, {\em et.~al.}, {\it
  {The unitarity triangle analysis within and beyond the standard model}},
  {\em PoS} {\bf HQL2010} (2011) 019. {Updates available on
  \texttt{http://www.utfit.org}.}

\bibitem{Charles:2004jd}
{\bf CKMfitter} Collaboration, J.~Charles {\em et.~al.}, {\it {CP violation and
  the CKM matrix: Assessing the impact of the asymmetric B factories}},  {\em
  Eur. Phys. J.} {\bf C41} (2005) 1--131,
  [\href{http://xxx.lanl.gov/abs/hep-ph/0406184}{{\tt hep-ph/0406184}}].
  {Updates available on \texttt{http://ckmfitter.in2p3.fr/}.}

\bibitem{Lunghi:2008aa}
E.~Lunghi and A.~Soni, {\it {Possible Indications of New Physics in
  $B_d$-mixing and in $\sin(2 \beta)$ Determinations}},  {\em Phys. Lett.} {\bf
  B666} (2008) 162--165, [\href{http://xxx.lanl.gov/abs/0803.4340}{{\tt
  arXiv:0803.4340}}].

\bibitem{Buras:2008nn}
A.~J. Buras and D.~Guadagnoli, {\it {Correlations among new CP violating
  effects in $\Delta F = 2$ observables}},  {\em Phys. Rev.} {\bf D78} (2008)
  033005, [\href{http://xxx.lanl.gov/abs/0805.3887}{{\tt arXiv:0805.3887}}].

\bibitem{Buras:2010wr}
A.~J. Buras, {\it {Minimal flavour violation and beyond: Towards a flavour code
  for short distance dynamics}},  {\em Acta Phys. Polon.} {\bf B41} (2010)
  2487--2561, [\href{http://xxx.lanl.gov/abs/1012.1447}{{\tt
  arXiv:1012.1447}}].

\bibitem{Isidori:2010zz}
G.~Isidori, {\it {The challenges of flavour physics}},  {\em PoS} {\bf
  ICHEP2010} (2010) 543, [\href{http://xxx.lanl.gov/abs/1012.1981}{{\tt
  arXiv:1012.1981}}].

\bibitem{Lunghi:2010gv}
E.~Lunghi and A.~Soni, {\it {Possible evidence for the breakdown of the
  CKM-paradigm of CP-violation}},  {\em Phys. Lett.} {\bf B697} (2011)
  323--328, [\href{http://xxx.lanl.gov/abs/1010.6069}{{\tt arXiv:1010.6069}}].

\bibitem{Lenz:2010gu}
A.~Lenz, U.~Nierste, J.~Charles, S.~Descotes-Genon, A.~Jantsch, {\em et.~al.},
  {\it {Anatomy of New Physics in $B - \bar{B}$ mixing}},  {\em Phys.Rev.} {\bf
  D83} (2011) 036004, [\href{http://xxx.lanl.gov/abs/1008.1593}{{\tt
  arXiv:1008.1593}}].

\bibitem{Lenz:2011ti}
A.~Lenz and U.~Nierste, {\it {Numerical updates of lifetimes and mixing
  parameters of B mesons}},  \href{http://xxx.lanl.gov/abs/1102.4274}{{\tt
  arXiv:1102.4274}}.

\bibitem{Laiho:2011nz}
J.~Laiho, E.~Lunghi, and R.~Van De~Water, {\it {Lessons for new physics from
  CKM studies}},  {\em PoS} {\bf FPCP2010} (2010) 040,
  [\href{http://xxx.lanl.gov/abs/1102.3917}{{\tt arXiv:1102.3917}}].

\bibitem{Lunghi:2011xy}
E.~Lunghi and A.~Soni, {\it {Demise of CKM and its aftermath}},
  \href{http://xxx.lanl.gov/abs/1104.2117}{{\tt arXiv:1104.2117}}.

\bibitem{Pati:1974yy}
J.~C. Pati and A.~Salam, {\it {Lepton Number as the Fourth Color}},  {\em Phys.
  Rev.} {\bf D10} (1974) 275--289.

\bibitem{Mohapatra:1974gc}
R.~N. Mohapatra and J.~C. Pati, {\it {A Natural Left-Right Symmetry}},  {\em
  Phys. Rev.} {\bf D11} (1975) 2558.

\bibitem{Mohapatra:1974hk}
R.~N. Mohapatra and J.~C. Pati, {\it {Left-Right Gauge Symmetry and an
  Isoconjugate Model of CP Violation}},  {\em Phys. Rev.} {\bf D11} (1975)
  566--571.

\bibitem{Senjanovic:1975rk}
G.~Senjanovic and R.~N. Mohapatra, {\it {Exact Left-Right Symmetry and
  Spontaneous Violation of Parity}},  {\em Phys. Rev.} {\bf D12} (1975) 1502.

\bibitem{Senjanovic:1978ev}
G.~Senjanovic, {\it {Spontaneous Breakdown of Parity in a Class of Gauge
  Theories}},  {\em Nucl. Phys.} {\bf B153} (1979) 334.

\bibitem{Mohapatra:1977mj}
R.~N. Mohapatra, F.~E. Paige, and D.~P. Sidhu, {\it {Symmetry Breaking and
  Naturalness of Parity Conservation in Weak Neutral Currents in Left-Right
  Symmetric Gauge Theories}},  {\em Phys. Rev.} {\bf D17} (1978) 2462.

\bibitem{Chang:1982dp}
D.~Chang, {\it {A Minimal Model of Spontaneous CP Violation with the Gauge
  Group $SU(2)_L \times SU(2)_R \times U(1)_{B-L}$}},  {\em Nucl. Phys.} {\bf
  B214} (1983) 435.

\bibitem{Branco:1982wp}
G.~Branco, J.~Frere, and J.~Gerard, {\it {The value of $\epsilon' / \epsilon$
  in models based on $SU(2)_L\times SU(2)_R\times U(1)$}},  {\em Nucl.Phys.}
  {\bf B221} (1983) 317.

\bibitem{Harari:1983gq}
H.~Harari and M.~Leurer, {\it {Left-Right Symmetry and the Mass Scale of a
  Possible Right-Handed Weak Boson}},  {\em Nucl. Phys.} {\bf B233} (1984) 221.

\bibitem{Kiers:2002cz}
K.~Kiers, J.~Kolb, J.~Lee, A.~Soni, and G.-H. Wu, {\it {Ubiquitous CP violation
  in a top inspired left-right model}},  {\em Phys. Rev.} {\bf D66} (2002)
  095002, [\href{http://xxx.lanl.gov/abs/hep-ph/0205082}{{\tt
  hep-ph/0205082}}].

\bibitem{Beall:1981ze}
G.~Beall, M.~Bander, and A.~Soni, {\it {Constraint on the Mass Scale of a
  Left-Right Symmetric Electroweak Theory from the $K_L - K_S$ Mass
  Difference}},  {\em Phys. Rev. Lett.} {\bf 48} (1982) 848.

\bibitem{Ecker:1985rr}
G.~Ecker and W.~Grimus, {\it {$\epsilon$, $\epsilon'$ in a model with
  spontaneous $P$ and $CP$ violation}},  {\em Phys. Lett.} {\bf B153} (1985)
  279--285.

\bibitem{Frere:1991db}
J.~M. Frere {\em et.~al.}, {\it {$K^0 -\bar K^0$ in the $SU(2)_L \times SU(2)_R
  \times U(1)$ model of CP violation}},  {\em Phys. Rev.} {\bf D46} (1992)
  337--353.

\bibitem{Barenboim:1996wz}
G.~Barenboim, J.~Bernabeu, and M.~Raidal, {\it {Spontaneous CP-violation in the
  left-right model and the kaon system}},  {\em Nucl. Phys.} {\bf B478} (1996)
  527--543, [\href{http://xxx.lanl.gov/abs/hep-ph/9608450}{{\tt
  hep-ph/9608450}}].

\bibitem{Mohapatra:1983ae}
R.~N. Mohapatra, G.~Senjanovic, and M.~D. Tran, {\it {Strangeness changing
  processes and the limit on the right-handed gauge boson mass}},  {\em Phys.
  Rev.} {\bf D28} (1983) 546.

\bibitem{Zhang:2007da}
Y.~Zhang, H.~An, X.~Ji, and R.~N. Mohapatra, {\it {General CP Violation in
  Minimal Left-Right Symmetric Model and Constraints on the Right-Handed
  Scale}},  {\em Nucl. Phys.} {\bf B802} (2008) 247--279,
  [\href{http://xxx.lanl.gov/abs/0712.4218}{{\tt arXiv:0712.4218}}].

\bibitem{Barenboim:2001vu}
G.~Barenboim, M.~Gorbahn, U.~Nierste, and M.~Raidal, {\it {Higgs sector of the
  minimal left-right symmetric model}},  {\em Phys. Rev.} {\bf D65} (2002)
  095003, [\href{http://xxx.lanl.gov/abs/hep-ph/0107121}{{\tt
  hep-ph/0107121}}].

\bibitem{Ball:1999mb}
P.~Ball, J.~M. Frere, and J.~Matias, {\it {Anatomy of Mixing-Induced CP
  Asymmetries in Left-Right-Symmetric Models with Spontaneous CP Violation}},
  {\em Nucl. Phys.} {\bf B572} (2000) 3--35,
  [\href{http://xxx.lanl.gov/abs/hep-ph/9910211}{{\tt hep-ph/9910211}}].

\bibitem{Langacker:1989xa}
P.~Langacker and S.~Uma~Sankar, {\it {Bounds on the Mass of $W_R$ and the
  $W_L-W_R$ Mixing Angle $\xi$ in General $SU(2)_L \times SU(2)_R \times U(1)$
  Models}},  {\em Phys. Rev.} {\bf D40} (1989) 1569--1585.

\bibitem{Barenboim:1996nd}
G.~Barenboim, J.~Bernabeu, J.~Prades, and M.~Raidal, {\it {Constraints on the
  $W_R$ mass and CP violation in left-right models}},  {\em Phys. Rev.} {\bf
  D55} (1997) 4213--4221, [\href{http://xxx.lanl.gov/abs/hep-ph/9611347}{{\tt
  hep-ph/9611347}}].

\bibitem{Zhang:2007fn}
Y.~Zhang, H.~An, X.~Ji, and R.~N. Mohapatra, {\it {Right-handed quark mixings
  in minimal left-right symmetric model with general CP violation}},  {\em
  Phys. Rev.} {\bf D76} (2007) 091301,
  [\href{http://xxx.lanl.gov/abs/0704.1662}{{\tt arXiv:0704.1662}}].

\bibitem{Maiezza:2010ic}
A.~Maiezza, M.~Nemevsek, F.~Nesti, and G.~Senjanovic, {\it {Left-Right Symmetry
  at LHC}},  {\em Phys. Rev.} {\bf D82} (2010) 055022,
  [\href{http://xxx.lanl.gov/abs/1005.5160}{{\tt arXiv:1005.5160}}].

\bibitem{Hsieh:2010zr}
K.~Hsieh, K.~Schmitz, J.-H. Yu, and C.~P. Yuan, {\it {Global Analysis of
  General $SU(2) \times SU(2) \times U(1)$ Models with Precision Data}},  {\em
  Phys. Rev.} {\bf D82} (2010) 035011,
  [\href{http://xxx.lanl.gov/abs/1003.3482}{{\tt arXiv:1003.3482}}].

\bibitem{Crivellin:2011ba}
A.~Crivellin and L.~Mercolli, {\it {$B\to X_d \gamma$ and constraints on new
  physics}},  \href{http://xxx.lanl.gov/abs/1106.5499}{{\tt arXiv:1106.5499}}.

\bibitem{Csaki:2003zu}
C.~Csaki, C.~Grojean, L.~Pilo, and J.~Terning, {\it {Towards a realistic model
  of Higgsless electroweak symmetry breaking}},  {\em Phys. Rev. Lett.} {\bf
  92} (2004) 101802, [\href{http://xxx.lanl.gov/abs/hep-ph/0308038}{{\tt
  hep-ph/0308038}}].

\bibitem{Nomura:2003du}
Y.~Nomura, {\it {Higgsless theory of electroweak symmetry breaking from warped
  space}},  {\em JHEP} {\bf 11} (2003) 050,
  [\href{http://xxx.lanl.gov/abs/hep-ph/0309189}{{\tt hep-ph/0309189}}].

\bibitem{Barbieri:2003pr}
R.~Barbieri, A.~Pomarol, and R.~Rattazzi, {\it {Weakly coupled Higgsless
  theories and precision electroweak tests}},  {\em Phys. Lett.} {\bf B591}
  (2004) 141--149, [\href{http://xxx.lanl.gov/abs/hep-ph/0310285}{{\tt
  hep-ph/0310285}}].

\bibitem{Georgi:2004iy}
H.~Georgi, {\it {Fun with Higgsless theories}},  {\em Phys. Rev.} {\bf D71}
  (2005) 015016, [\href{http://xxx.lanl.gov/abs/hep-ph/0408067}{{\tt
  hep-ph/0408067}}].

\bibitem{Crivellin:2009sd}
A.~Crivellin, {\it {Effects of right-handed charged currents on the
  determinations of $|V_{ub}|$ and $|V_{cb}|$}},  {\em Phys. Rev.} {\bf D81}
  (2010) 031301, [\href{http://xxx.lanl.gov/abs/0907.2461}{{\tt
  arXiv:0907.2461}}].

\bibitem{Chen:2008se}
C.-H. Chen and S.-h. Nam, {\it {Left-right mixing on leptonic and semileptonic
  $b\to u$ decays}},  {\em Phys. Lett.} {\bf B666} (2008) 462--466,
  [\href{http://xxx.lanl.gov/abs/0807.0896}{{\tt arXiv:0807.0896}}].

\bibitem{Feger:2010qc}
R.~Feger, T.~Mannel, V.~Klose, H.~Lacker, and T.~Luck, {\it {Limit on a
  Right-Handed Admixture to the Weak $b \to c$ Current from Semileptonic
  Decays}},  {\em Phys. Rev.} {\bf D82} (2010) 073002,
  [\href{http://xxx.lanl.gov/abs/1003.4022}{{\tt arXiv:1003.4022}}].

\bibitem{Buras:2010pz}
A.~J. Buras, K.~Gemmler, and G.~Isidori, {\it {Quark flavour mixing with
  right-handed currents: an effective theory approach}},  {\em Nucl. Phys.}
  {\bf B843} (2011) 107--142, [\href{http://xxx.lanl.gov/abs/1007.1993}{{\tt
  arXiv:1007.1993}}].

\bibitem{Nemevsek:2011hz}
M.~Nemevsek, F.~Nesti, G.~Senjanovic, and Y.~Zhang, {\it {First Limits on
  Left-Right Symmetry Scale from LHC Data}},  {\em Phys.Rev.} {\bf D83} (2011)
  115014, [\href{http://xxx.lanl.gov/abs/1103.1627}{{\tt arXiv:1103.1627}}].

\bibitem{Grojean:2011vu}
C.~Grojean, E.~Salvioni, and R.~Torre, {\it {A weakly constrained W' at the
  early LHC}},  {\em JHEP} {\bf 1107} (2011) 002,
  [\href{http://xxx.lanl.gov/abs/1103.2761}{{\tt arXiv:1103.2761}}].

\bibitem{Ecker:1983uh}
G.~Ecker, W.~Grimus, and H.~Neufeld, {\it {Higgs induced flavor changing
  neutral interactions in $SU(2)_L \times SU(2)_R \times U(1)$}},  {\em Phys.
  Lett.} {\bf B127} (1983) 365.

\bibitem{Gilman:1983ce}
F.~J. Gilman and M.~H. Reno, {\it {Restrictions from the neutral $K$ and $B$
  meson systems on left-right symmetric gauge theories}},  {\em Phys. Rev.}
  {\bf D29} (1984) 937.

\bibitem{Ecker:1985vv}
G.~Ecker and W.~Grimus, {\it {CP Violation and Left-Right Symmetry}},  {\em
  Nucl. Phys.} {\bf B258} (1985) 328--360.

\bibitem{Hou:1985ur}
W.-S. Hou and A.~Soni, {\it {Gauge invariance of the $K_L -\bar K_S$ mass
  difference in left-right symmetric model}},  {\em Phys. Rev.} {\bf D32}
  (1985) 163.

\bibitem{London:1989cf}
D.~London and D.~Wyler, {\it {Left-right symmetry and CP violation in the $B$
  system}},  {\em Phys. Lett.} {\bf B232} (1989) 503.

\bibitem{Ball:1999yi}
P.~Ball and R.~Fleischer, {\it {An Analysis of $B_s$ decays in the left-right
  symmetric model with spontaneous CP violation}},  {\em Phys.Lett.} {\bf B475}
  (2000) 111--119, [\href{http://xxx.lanl.gov/abs/hep-ph/9912319}{{\tt
  hep-ph/9912319}}].

\bibitem{Sahoo:2005wb}
S.~Sahoo, L.~Maharana, A.~Roul, and S.~Acharya, {\it {The masses of $W_R$,
  triplet Higgs, and $Z'$ bosons in the Left-Right Symmetric Model}},  {\em
  Int. J. Mod. Phys.} {\bf A20} (2005) 2625--2638.

\bibitem{Asatrian:1989iu}
G.~M. Asatrian and A.~N. Ionnisian, {\it {Rare B meson decays in $SU(2)_L\times
  SU(2)_R\times U(1)$ model}},  {\em Mod. Phys. Lett.} {\bf A5} (1990)
  1089--1096.

\bibitem{Asatryan:1990na}
G.~M. Asatryan and A.~N. Ioannisyan, {\it {The decay $b \to s \gamma$ in the
  $SU(2)_L \times SU(2)_R \times U(1)$ model}},  {\em Sov. J. Nucl. Phys.} {\bf
  51} (1990) 858--860.

\bibitem{Cocolicchio:1988ac}
D.~Cocolicchio, G.~Costa, G.~L. Fogli, J.~H. Kim, and A.~Masiero, {\it {Rare
  $B$ decays in left-right symmetric models}},  {\em Phys. Rev.} {\bf D40}
  (1989) 1477.

\bibitem{Cho:1993zb}
P.~L. Cho and M.~Misiak, {\it {$b \to s \gamma$ decay in $SU(2)_L \times
  SU(2)_R \times U(1)$ extensions of the Standard Model}},  {\em Phys. Rev.}
  {\bf D49} (1994) 5894--5903,
  [\href{http://xxx.lanl.gov/abs/hep-ph/9310332}{{\tt hep-ph/9310332}}].

\bibitem{Babu:1993hx}
K.~S. Babu, K.~Fujikawa, and A.~Yamada, {\it {Constraints on left-right
  symmetric models from the process $b \to s \gamma$}},  {\em Phys. Lett.} {\bf
  B333} (1994) 196--201, [\href{http://xxx.lanl.gov/abs/hep-ph/9312315}{{\tt
  hep-ph/9312315}}].

\bibitem{Fujikawa:1993zu}
K.~Fujikawa and A.~Yamada, {\it {Test of the chiral structure of the top -
  bottom charged current by the process $b \to s \gamma$}},  {\em Phys. Rev.}
  {\bf D49} (1994) 5890--5893.

\bibitem{Asatrian:1996as}
G.~M. Asatrian and A.~Ioannisian, {\it {CP-Violation in the Decay $b \to
  s\gamma$ in the Left-Right Symmetric Model}},  {\em Phys. Rev.} {\bf D54}
  (1996) 5642--5646, [\href{http://xxx.lanl.gov/abs/hep-ph/9603318}{{\tt
  hep-ph/9603318}}].

\bibitem{Frank:2010qv}
M.~Frank, A.~Hayreter, and I.~Turan, {\it {$B$ Decays in an Asymmetric
  Left-Right Model}},  {\em Phys. Rev.} {\bf D82} (2010) 033012,
  [\href{http://xxx.lanl.gov/abs/1005.3074}{{\tt arXiv:1005.3074}}].

\bibitem{Guadagnoli:2011id}
D.~Guadagnoli, R.~N. Mohapatra, and I.~Sung, {\it {Gauged Flavor Group with
  Left-Right Symmetry}},  {\em JHEP} {\bf 04} (2011) 093,
  [\href{http://xxx.lanl.gov/abs/1103.4170}{{\tt arXiv:1103.4170}}].

\bibitem{Mohapatra:1979ia}
R.~N. Mohapatra and G.~Senjanovic, {\it {Neutrino mass and spontaneous parity
  nonconservation}},  {\em Phys. Rev. Lett.} {\bf 44} (1980) 912.

\bibitem{Mohapatra:1980yp}
R.~N. Mohapatra and G.~Senjanovic, {\it {Neutrino Masses and Mixings in Gauge
  Models with Spontaneous Parity Violation}},  {\em Phys. Rev.} {\bf D23}
  (1981) 165.

\bibitem{Khasanov:2001tu}
O.~Khasanov and G.~Perez, {\it {On neutrino masses and a low breaking scale of
  left-right symmetry}},  {\em Phys. Rev.} {\bf D65} (2002) 053007,
  [\href{http://xxx.lanl.gov/abs/hep-ph/0108176}{{\tt hep-ph/0108176}}].

\bibitem{Aranda:2009ut}
A.~Aranda, J.~Diaz-Cruz, E.~Ma, R.~Noriega, and J.~Wudka, {\it {Asymmetric
  Higgs Sector and Neutrino Mass in an $SU(2)_R$ Model}},  {\em Phys.Rev.} {\bf
  D80} (2009) 115003, [\href{http://xxx.lanl.gov/abs/0909.1754}{{\tt
  arXiv:0909.1754}}].

\bibitem{Blanke:2006sb}
M.~Blanke {\em et.~al.}, {\it {Particle antiparticle mixing, $\varepsilon_K$,
  $\Delta\Gamma_q$, $A^q_\text{SL}$, $A_\text{CP}(B_d \to \psi K_S)$,
  $A_\text{CP}(B_s \to \psi \phi)$ and $B \to X_{s,d}\gamma$ in the Littlest
  Higgs model with T-parity}},  {\em JHEP} {\bf 12} (2006) 003,
  [\href{http://xxx.lanl.gov/abs/hep-ph/0605214}{{\tt hep-ph/0605214}}].

\bibitem{Blanke:2008zb}
M.~Blanke, A.~J. Buras, B.~Duling, S.~Gori, and A.~Weiler, {\it {$\Delta F=2$
  Observables and Fine-Tuning in a Warped Extra Dimension with Custodial
  Protection}},  {\em JHEP} {\bf 03} (2009) 001,
  [\href{http://xxx.lanl.gov/abs/0809.1073}{{\tt arXiv:0809.1073}}].

\bibitem{Buras:2010pi}
A.~J. Buras {\em et.~al.}, {\it {Patterns of Flavour Violation in the Presence
  of a Fourth Generation of Quarks and Leptons}},  {\em JHEP} {\bf 09} (2010)
  106, [\href{http://xxx.lanl.gov/abs/1002.2126}{{\tt arXiv:1002.2126}}].

\bibitem{Buras:2000if}
A.~J. Buras, M.~Misiak, and J.~Urban, {\it {Two-loop QCD anomalous dimensions
  of flavour-changing four-quark operators within and beyond the standard
  model}},  {\em Nucl. Phys.} {\bf B586} (2000) 397--426,
  [\href{http://xxx.lanl.gov/abs/hep-ph/0005183}{{\tt hep-ph/0005183}}].

\bibitem{Chang:1984hr}
D.~Chang, J.~Basecq, L.-F. Li, and P.~B. Pal, {\it {Comment on the $K_L - K_S$
  mass difference in left-right model}},  {\em Phys. Rev.} {\bf D30} (1984)
  1601.

\bibitem{Basecq:1985cr}
J.~Basecq, L.-F. Li, and P.~B. Pal, {\it {Gauge invariant calculation of the
  $K_L-K_S$ mass difference in the left-right model}},  {\em Phys. Rev.} {\bf
  D32} (1985) 175.

\bibitem{Buras:2001ra}
A.~J. Buras, S.~Jager, and J.~Urban, {\it {Master formulae for $\Delta F = 2$
  NLO-QCD factors in the standard model and beyond}},  {\em Nucl. Phys.} {\bf
  B605} (2001) 600--624, [\href{http://xxx.lanl.gov/abs/hep-ph/0102316}{{\tt
  hep-ph/0102316}}].

\bibitem{Gorbahn:2009pp}
M.~Gorbahn, S.~Jager, U.~Nierste, and S.~Trine, {\it {The supersymmetric Higgs
  sector and $B-\bar{B}$ mixing for large tan $\beta$}},  {\em Phys.Rev.} {\bf
  D84} (2011) 034030, [\href{http://xxx.lanl.gov/abs/0901.2065}{{\tt
  arXiv:0901.2065}}].

\bibitem{Buras:2010mh}
A.~J. Buras, M.~V. Carlucci, S.~Gori, and G.~Isidori, {\it {Higgs-mediated
  FCNCs: Natural Flavour Conservation vs. Minimal Flavour Violation}},  {\em
  JHEP} {\bf 10} (2010) 009, [\href{http://xxx.lanl.gov/abs/1005.5310}{{\tt
  arXiv:1005.5310}}].

\bibitem{Buras:2010zm}
A.~J. Buras, G.~Isidori, and P.~Paradisi, {\it {EDMs vs. CPV in $B_{s,d}$
  mixing in two Higgs doublet models with MFV}},  {\em Phys. Lett.} {\bf B694}
  (2011) 402--409, [\href{http://xxx.lanl.gov/abs/1007.5291}{{\tt
  arXiv:1007.5291}}].

\bibitem{Becirevic:2001xt}
D.~Becirevic, V.~Gimenez, G.~Martinelli, M.~Papinutto, and J.~Reyes, {\it
  {$B$-parameters of the complete set of matrix elements of $\Delta B = 2$
  operators from the lattice}},  {\em JHEP} {\bf 04} (2002) 025,
  [\href{http://xxx.lanl.gov/abs/hep-lat/0110091}{{\tt hep-lat/0110091}}].

\bibitem{Babich:2006bh}
R.~Babich {\em et.~al.}, {\it {$K^0 -\bar K^0$ mixing beyond the standard model
  and CP- violating electroweak penguins in quenched QCD with exact chiral
  symmetry}},  {\em Phys. Rev.} {\bf D74} (2006) 073009,
  [\href{http://xxx.lanl.gov/abs/hep-lat/0605016}{{\tt hep-lat/0605016}}].

\bibitem{Ciuchini:1997bw}
M.~Ciuchini, E.~Franco, V.~Lubicz, G.~Martinelli, I.~Scimemi, {\em et.~al.},
  {\it {Next-to-leading order QCD corrections to $\Delta F = 2$ effective
  Hamiltonians}},  {\em Nucl.Phys.} {\bf B523} (1998) 501--525,
  [\href{http://xxx.lanl.gov/abs/hep-ph/9711402}{{\tt hep-ph/9711402}}].

\bibitem{D'Agostini:2004yu}
G.~D'Agostini, {\it {Asymmetric uncertainties: Sources, treatment and potential
  dangers}},  \href{http://xxx.lanl.gov/abs/physics/0403086}{{\tt
  physics/0403086}}.

\bibitem{Altmannshofer:2009ne}
W.~Altmannshofer, A.~J. Buras, S.~Gori, P.~Paradisi, and D.~M. Straub, {\it
  {Anatomy and Phenomenology of FCNC and CPV Effects in SUSY Theories}},  {\em
  Nucl.Phys.} {\bf B830} (2010) 17--94,
  [\href{http://xxx.lanl.gov/abs/0909.1333}{{\tt arXiv:0909.1333}}].

\bibitem{Grinstein:2010ve}
B.~Grinstein, M.~Redi, and G.~Villadoro, {\it {Low Scale Flavor Gauge
  Symmetries}},  {\em JHEP} {\bf 11} (2010) 067,
  [\href{http://xxx.lanl.gov/abs/1009.2049}{{\tt arXiv:1009.2049}}].

\bibitem{Buras:2010pza}
A.~J. Buras, D.~Guadagnoli, and G.~Isidori, {\it {On $\varepsilon_K$ beyond
  lowest order in the Operator Product Expansion}},  {\em Phys. Lett.} {\bf
  B688} (2010) 309--313, [\href{http://xxx.lanl.gov/abs/1002.3612}{{\tt
  arXiv:1002.3612}}].

\bibitem{Buras:2009pj}
A.~J. Buras and D.~Guadagnoli, {\it {On the consistency between the observed
  amount of CP violation in the $K$ and $B_d$ systems within minimal flavor
  violation}},  {\em Phys.Rev.} {\bf D79} (2009) 053010,
  [\href{http://xxx.lanl.gov/abs/0901.2056}{{\tt arXiv:0901.2056}}].

\bibitem{Bona:2005eu}
{\bf UTfit} Collaboration, M.~Bona {\em et.~al.}, {\it {The UTfit collaboration
  report on the status of the unitarity triangle beyond the standard model. I.
  Model-independent analysis and minimal flavor violation}},  {\em JHEP} {\bf
  0603} (2006) 080, [\href{http://xxx.lanl.gov/abs/hep-ph/0509219}{{\tt
  hep-ph/0509219}}].

\bibitem{Beneke:1998sy}
M.~Beneke, G.~Buchalla, C.~Greub, A.~Lenz, and U.~Nierste, {\it
  {Next-to-leading order QCD corrections to the lifetime difference of $B_s$
  mesons}},  {\em Phys.Lett.} {\bf B459} (1999) 631--640,
  [\href{http://xxx.lanl.gov/abs/hep-ph/9808385}{{\tt hep-ph/9808385}}].

\bibitem{Beneke:2002rj}
M.~Beneke, G.~Buchalla, C.~Greub, A.~Lenz, and U.~Nierste, {\it {The $B^+ -
  B^0_d$ lifetime difference beyond leading logarithms}},  {\em Nucl.Phys.}
  {\bf B639} (2002) 389--407,
  [\href{http://xxx.lanl.gov/abs/hep-ph/0202106}{{\tt hep-ph/0202106}}].

\bibitem{Beneke:2003az}
M.~Beneke, G.~Buchalla, A.~Lenz, and U.~Nierste, {\it {CP asymmetry in flavor
  specific $B$ decays beyond leading logarithms}},  {\em Phys.Lett.} {\bf B576}
  (2003) 173--183, [\href{http://xxx.lanl.gov/abs/hep-ph/0307344}{{\tt
  hep-ph/0307344}}].

\bibitem{Ciuchini:2001vx}
M.~Ciuchini, E.~Franco, V.~Lubicz, and F.~Mescia, {\it {Next-to-leading order
  QCD corrections to spectator effects in lifetimes of beauty hadrons}},  {\em
  Nucl.Phys.} {\bf B625} (2002) 211--238,
  [\href{http://xxx.lanl.gov/abs/hep-ph/0110375}{{\tt hep-ph/0110375}}].

\bibitem{Ciuchini:2003ww}
M.~Ciuchini, E.~Franco, V.~Lubicz, F.~Mescia, and C.~Tarantino, {\it {Lifetime
  differences and CP violation parameters of neutral $B$ mesons at the
  next-to-leading order in QCD}},  {\em JHEP} {\bf 08} (2003) 031,
  [\href{http://xxx.lanl.gov/abs/hep-ph/0308029}{{\tt hep-ph/0308029}}].

\bibitem{Lenz:2011zz}
A.~Lenz, {\it {A simple relation for $B_s$-mixing}},
  \href{http://xxx.lanl.gov/abs/1106.3200}{{\tt arXiv:1106.3200}}.

\bibitem{Giurgiu:2010is}
{\bf For the CDF} Collaboration, G.~Giurgiu, {\it {New Measurement of the $B_s$
  Mixing Phase at CDF}},  {\em PoS} {\bf ICHEP2010} (2010) 236,
  [\href{http://xxx.lanl.gov/abs/1012.0962}{{\tt arXiv:1012.0962}}].

\bibitem{Abazov:2011ry}
{\bf D0} Collaboration, V.~M. Abazov {\em et.~al.}, {\it {Measurement of the
  CP-violating phase $\phi_s^{J/\psi \phi}$ using the flavor-tagged decay
  $B_s^0 \rightarrow J/\psi \phi$ in 8 fb$^{-1}$ of $p \overline p$
  collisions}},  \href{http://xxx.lanl.gov/abs/1109.3166}{{\tt
  arXiv:1109.3166}}.

\bibitem{Raven:1378074}
{\bf LHCb} Collaboration, G.~Raven, ``{B Physics Results from the LHC}.''
  Review talk, rather than LHCb specific, but featuring new LHCb results. {\tt
  http://cdsweb.cern.ch/record/1378074?ln=en}, Aug, 2011.

\bibitem{Asner:2010qj}
{\bf Heavy Flavor Averaging Group} Collaboration, D.~Asner {\em et.~al.}, {\it
  {Averages of b-hadron, c-hadron, and tau-lepton Properties}},
  \href{http://xxx.lanl.gov/abs/1010.1589}{{\tt arXiv:1010.1589}}. {Updates
  available on \texttt{http://www.slac.stanford.edu/xorg/hfag/}.}

\bibitem{Abazov:2011yk}
{\bf D0} Collaboration, V.~M. Abazov {\em et.~al.}, {\it {Measurement of the
  anomalous like-sign dimuon charge asymmetry with $9 fb^{-1}$ of $p \bar p$
  collisions}},  \href{http://xxx.lanl.gov/abs/1106.6308}{{\tt
  arXiv:1106.6308}}.

\bibitem{Ligeti:2006pm}
Z.~Ligeti, M.~Papucci, and G.~Perez, {\it {Implications of the measurement of
  the $B^0_s - \bar B^0_s$ mass difference}},  {\em Phys. Rev. Lett} {\bf 97}
  (2006) 101801, [\href{http://xxx.lanl.gov/abs/hep-ph/0604112}{{\tt
  hep-ph/0604112}}].

\bibitem{Blanke:2006ig}
M.~Blanke, A.~J. Buras, D.~Guadagnoli, and C.~Tarantino, {\it {Minimal Flavour
  Violation Waiting for Precise Measurements of $\Delta M_s$, $S_{\psi \phi}$,
  $A^s_\text{SL}$, $|V_{ub}|$, $\gamma$ and $B^0_{s,d} \to \mu^+ \mu^-$}},
  {\em JHEP} {\bf 10} (2006) 003,
  [\href{http://xxx.lanl.gov/abs/hep-ph/0604057}{{\tt hep-ph/0604057}}].

\bibitem{Grossman:2009mn}
Y.~Grossman, Y.~Nir, and G.~Perez, {\it {Testing New Indirect CP Violation}},
  {\em Phys. Rev. Lett.} {\bf 103} (2009) 071602,
  [\href{http://xxx.lanl.gov/abs/0904.0305}{{\tt arXiv:0904.0305}}].

\bibitem{Bobeth:1999ww}
C.~Bobeth, M.~Misiak, and J.~Urban, {\it {Matching conditions for $b \to s
  \gamma$ and $b \to s \text{gluon}$ in extensions of the standard model}},
  {\em Nucl. Phys.} {\bf B567} (2000) 153--185,
  [\href{http://xxx.lanl.gov/abs/hep-ph/9904413}{{\tt hep-ph/9904413}}].

\bibitem{Misiak:2006zs}
M.~Misiak, H.~Asatrian, K.~Bieri, M.~Czakon, A.~Czarnecki, {\em et.~al.}, {\it
  {Estimate of $B(\bar B \to X_s \gamma)$ at $\ord(\alpha_s^2)$}},  {\em
  Phys.Rev.Lett.} {\bf 98} (2007) 022002,
  [\href{http://xxx.lanl.gov/abs/hep-ph/0609232}{{\tt hep-ph/0609232}}].

\bibitem{Buras:2011zb}
A.~J. Buras, L.~Merlo, and E.~Stamou, {\it {The Impact of Flavour Changing
  Neutral Gauge Bosons on $\bar{B}\to X_s \gamma$}},  {\em JHEP} {\bf 1108}
  (2011) 124, [\href{http://xxx.lanl.gov/abs/1105.5146}{{\tt
  arXiv:1105.5146}}].

\bibitem{Misiak:2006ab}
M.~Misiak and M.~Steinhauser, {\it {NNLO QCD corrections to the $B \to X_s
  \gamma$ matrix elements using interpolation in $m_c$}},  {\em Nucl. Phys.}
  {\bf B764} (2007) 62--82, [\href{http://xxx.lanl.gov/abs/hep-ph/0609241}{{\tt
  hep-ph/0609241}}].

\bibitem{Benzke:2010js}
M.~Benzke, S.~J. Lee, M.~Neubert, and G.~Paz, {\it {Factorization at Subleading
  Power and Irreducible Uncertainties in $\bar B\to X_s\gamma$ Decay}},  {\em
  JHEP} {\bf 08} (2010) 099, [\href{http://xxx.lanl.gov/abs/1003.5012}{{\tt
  arXiv:1003.5012}}].

\bibitem{Hurth:2003dk}
T.~Hurth, E.~Lunghi, and W.~Porod, {\it {Untagged $B\to X_{s,d}\gamma$ CP
  asymmetry as a probe for new physics}},  {\em Nucl. Phys.} {\bf B704} (2005)
  56--74, [\href{http://xxx.lanl.gov/abs/hep-ph/0312260}{{\tt
  hep-ph/0312260}}].

\bibitem{Soares:1991te}
J.~M. Soares, {\it {CP violation in radiative $b$ decays}},  {\em Nucl.Phys.}
  {\bf B367} (1991) 575--590.

\bibitem{Kagan:1998bh}
A.~L. Kagan and M.~Neubert, {\it {Direct CP violation in $B \to X_s\gamma$
  decays as a signature of new physics}},  {\em Phys.Rev.} {\bf D58} (1998)
  094012, [\href{http://xxx.lanl.gov/abs/hep-ph/9803368}{{\tt
  hep-ph/9803368}}].

\bibitem{Kagan:1998ym}
A.~L. Kagan and M.~Neubert, {\it {QCD anatomy of $B \to X_s\gamma$ decays}},
  {\em Eur.Phys.J.} {\bf C7} (1999) 5--27,
  [\href{http://xxx.lanl.gov/abs/hep-ph/9805303}{{\tt hep-ph/9805303}}].

\bibitem{Benzke:2010tq}
M.~Benzke, S.~J. Lee, M.~Neubert, and G.~Paz, {\it {Long-Distance Dominance of
  the CP Asymmetry in $B\to X_{s,d}\gamma$ Decays}},  {\em Phys. Rev. Lett.}
  {\bf 106} (2011) 141801, [\href{http://xxx.lanl.gov/abs/1012.3167}{{\tt
  arXiv:1012.3167}}].

\bibitem{Chen:2011de}
M.-C. Chen and J.~Huang, {\it {TeV Scale Models of Neutrino Masses and Their
  Phenomenology}},  \href{http://xxx.lanl.gov/abs/1105.3188}{{\tt
  arXiv:1105.3188}}.

\bibitem{Czakon:2002wm}
M.~Czakon, J.~Gluza, and J.~Hejczyk, {\it {Muon decay to one loop order in the
  left-right symmetric model}},  {\em Nucl. Phys.} {\bf B642} (2002) 157--172,
  [\href{http://xxx.lanl.gov/abs/hep-ph/0205303}{{\tt hep-ph/0205303}}].

\bibitem{Nakamura:2010zzi}
{\bf Particle Data Group} Collaboration, K.~Nakamura {\em et.~al.}, {\it
  {Review of particle physics}},  {\em J. Phys.} {\bf G37} (2010) 075021.
  {Updates available on \texttt{http://pdg.lbl.gov/}.}

\bibitem{Antonelli:2010yf}
M.~Antonelli, V.~Cirigliano, G.~Isidori, F.~Mescia, M.~Moulson, {\em et.~al.},
  {\it {An Evaluation of $|V_{us}|$ and precise tests of the Standard Model
  from world data on leptonic and semileptonic kaon decays}},  {\em
  Eur.Phys.J.} {\bf C69} (2010) 399--424,
  [\href{http://xxx.lanl.gov/abs/1005.2323}{{\tt arXiv:1005.2323}}].

\bibitem{Laiho:2009eu}
J.~Laiho, E.~Lunghi, and R.~S. Van~de Water, {\it {Lattice QCD inputs to the
  CKM unitarity triangle analysis}},  {\em Phys. Rev.} {\bf D81} (2010) 034503,
  [\href{http://xxx.lanl.gov/abs/0910.2928}{{\tt arXiv:0910.2928}}]. Updates
  available on {\tt http://latticeaverages.org/}.

\bibitem{Bailey:2010gb}
{\bf Fermilab Lattice and MILC} Collaboration, J.~A. Bailey {\em et.~al.}, {\it
  {$B \to D^* l \nu$ at zero recoil: an update}},  {\em PoS} {\bf LATTICE2010}
  (2010) 311, [\href{http://xxx.lanl.gov/abs/1011.2166}{{\tt
  arXiv:1011.2166}}].

\bibitem{Okamoto:2004xg}
M.~Okamoto, C.~Aubin, C.~Bernard, C.~E. DeTar, M.~Di~Pierro, {\em et.~al.},
  {\it {Semileptonic $D \to \pi/K$ and $B \to \pi/D$ decays in 2+1 flavor
  lattice QCD}},  {\em Nucl.Phys.Proc.Suppl.} {\bf 140} (2005) 461--463,
  [\href{http://xxx.lanl.gov/abs/hep-lat/0409116}{{\tt hep-lat/0409116}}].

\bibitem{Finkemeier:1994ev}
M.~Finkemeier, {\it {Radiative corrections to $\pi_{l2}$ and $K_{l2}$ decays}},
   \href{http://xxx.lanl.gov/abs/hep-ph/9501286}{{\tt hep-ph/9501286}}.

\bibitem{Abazov:2011zk}
{\bf D0} Collaboration, V.~Abazov {\em et.~al.}, {\it {Precision measurement of
  the ratio ${\rm B}(t \to Wb)/{\rm B}(t \to Wq)$ and Extraction of $V_{tb}$}},
   {\em Phys.Rev.Lett.} {\bf 107} (2011) 121802,
  [\href{http://xxx.lanl.gov/abs/1106.5436}{{\tt arXiv:1106.5436}}].

\bibitem{:2005ema}
{\bf ALEPH, DELPHI, L3, OPAL, SLD, LEP Electroweak Working Group, SLD
  Electroweak Group, SLD Heavy Flavour Group} Collaboration, {\it {Precision
  electroweak measurements on the $Z$ resonance}},  {\em Phys.Rept.} {\bf 427}
  (2006) 257--454, [\href{http://xxx.lanl.gov/abs/hep-ex/0509008}{{\tt
  hep-ex/0509008}}].

\bibitem{Giri:2003ty}
A.~Giri, Y.~Grossman, A.~Soffer, and J.~Zupan, {\it {Determining gamma using
  $B^\pm \to D K^\pm$ with multibody $D$ decays}},  {\em Phys.Rev.} {\bf D68}
  (2003) 054018, [\href{http://xxx.lanl.gov/abs/hep-ph/0303187}{{\tt
  hep-ph/0303187}}].

\bibitem{Belle-Moriond-2011}
{\bf Belle} Collaboration, A.~Poluektov, ``{Recent EW results from Belle}.''
  Talk given at Rencontres de Moriond, March 13-20, 2011, slides available on
  \texttt{http://indico.in2p3.fr/conferenceOtherViews.py?view=standard\&confId%
=4403}.

\bibitem{Buras:2005xt}
A.~J. Buras, {\it {Flavor physics and CP violation}},
  \href{http://xxx.lanl.gov/abs/hep-ph/0505175}{{\tt hep-ph/0505175}}.

\bibitem{Faller:2008zc}
S.~Faller, M.~Jung, R.~Fleischer, and T.~Mannel, {\it {The Golden Modes $B^0
  \rightarrow J/\psi K_{S,L}$ in the Era of Precision Flavour Physics}},  {\em
  Phys.Rev.} {\bf D79} (2009) 014030,
  [\href{http://xxx.lanl.gov/abs/0809.0842}{{\tt arXiv:0809.0842}}].

\bibitem{Flacher:2008zq}
H.~Flacher, M.~Goebel, J.~Haller, A.~Hocker, K.~Monig, {\em et.~al.}, {\it
  {Revisiting the Global Electroweak Fit of the Standard Model and Beyond with
  Gfitter}},  {\em Eur.Phys.J.} {\bf C60} (2009) 543--583,
  [\href{http://xxx.lanl.gov/abs/0811.0009}{{\tt arXiv:0811.0009}}].

\bibitem{Arbuzov:2005ma}
A.~Arbuzov, M.~Awramik, M.~Czakon, A.~Freitas, M.~Grunewald, {\em et.~al.},
  {\it {ZFITTER: A Semi-analytical program for fermion pair production in e+ e-
  annihilation, from version 6.21 to version 6.42}},  {\em Comput.Phys.Commun.}
  {\bf 174} (2006) 728--758,
  [\href{http://xxx.lanl.gov/abs/hep-ph/0507146}{{\tt hep-ph/0507146}}].
  {Updates available on \texttt{http://zfitter.desy.de/}.}

\bibitem{Peskin:1991sw}
M.~E. Peskin and T.~Takeuchi, {\it {Estimation of oblique electroweak
  corrections}},  {\em Phys.Rev.} {\bf D46} (1992) 381--409.

\bibitem{Baak:2011ze}
M.~Baak {\em et.~al.}, {\it {Updated Status of the Global Electroweak Fit and
  Constraints on New Physics}},  \href{http://xxx.lanl.gov/abs/1107.0975}{{\tt
  arXiv:1107.0975}}.

\bibitem{Z-Pole}
{{\bf The ALEPH, DELPHI, L3, OPAL, SLD Collaborations, the LEP Electroweak
  Working Group, the SLD Electroweak and Heavy Flavour Groups}}, {\it
  {Precision Electroweak Measurements on the Z Resonance}},  {\em Phys. Rept.}
  {\bf 427} (2006) 257, [\href{http://xxx.lanl.gov/abs/hep-ex/0509008}{{\tt
  hep-ex/0509008}}].

\bibitem{Bayes:2011zz}
{\bf TWIST} Collaboration, R.~Bayes {\em et.~al.}, {\it {Experimental
  Constraints on Left-Right Symmetric Models from Muon Decay}},  {\em Phys.
  Rev. Lett.} {\bf 106} (2011) 041804.

\bibitem{Aad:2011yg}
{\bf ATLAS} Collaboration, G.~Aad {\em et.~al.}, {\it {Search for a heavy gauge
  boson decaying to a charged lepton and a neutrino in $1\,\text{fb}^{-1}$ of
  $pp$ collisions at $\sqrt{s} = 7\,\text{TeV}$ using the ATLAS detector}},
  \href{http://xxx.lanl.gov/abs/1108.1316}{{\tt arXiv:1108.1316}}.

\bibitem{CMS-PAS-EXO-11-024}
{\bf CMS} Collaboration, ``{Search for $W'$ in the leptonic channels in pp
  Collisions at $\sqrt{s} = 7\,\text{TeV}$}.'' {\tt
  http://cdsweb.cern.ch/record/1369201?ln=en}, 2011.

\bibitem{CMS-PAS-EXO-11-002}
{\bf CMS} Collaboration, ``{Search for a heavy neutrino and right-handed $W$ of
  the left-right symmetric model in $pp$ collisions at $\sqrt{s} =
  7\,\text{TeV}$}.'' {\tt http://cdsweb.cern.ch/record/1369255?ln=en}, 2011.

\bibitem{Chatrchyan:1370086}
{\bf CMS} Collaboration, S.~Chatrchyan {\em et.~al.}, {\it {Search for
  Resonances in the Dijet Mass Spectrum from 7 TeV $pp$ Collisions at CMS}}, .
  {\tt http://cdsweb.cern.ch/record/1370086?ln=en}.

\bibitem{Aad:2011fq}
{\bf ATLAS} Collaboration, G.~Aad {\em et.~al.}, {\it {Search for New Physics
  in the Dijet Mass Distribution using $1 \text{fb}^{-1}$ of $pp$ Collision
  Data at $\sqrt{s} = 7 \text{TeV}$ collected by the ATLAS Detector}},
  \href{http://xxx.lanl.gov/abs/1108.6311}{{\tt arXiv:1108.6311}}.

\bibitem{Lunghi:2009sm}
E.~Lunghi and A.~Soni, {\it {Hints for the scale of new CP-violating physics
  from B-CP anomalies}},  {\em JHEP} {\bf 0908} (2009) 051,
  [\href{http://xxx.lanl.gov/abs/0903.5059}{{\tt arXiv:0903.5059}}].

\bibitem{Lunghi:2009ke}
E.~Lunghi and A.~Soni, {\it {Unitarity Triangle Without Semileptonic Decays}},
  {\em Phys.Rev.Lett.} {\bf 104} (2010) 251802,
  [\href{http://xxx.lanl.gov/abs/0912.0002}{{\tt arXiv:0912.0002}}].

\bibitem{Antonio:2007pb}
{\bf RBC} Collaboration, D.~J. Antonio {\em et.~al.}, {\it {Neutral kaon mixing
  from 2+1 flavor domain wall QCD}},  {\em Phys. Rev. Lett.} {\bf 100} (2008)
  032001, [\href{http://xxx.lanl.gov/abs/hep-ph/0702042}{{\tt
  hep-ph/0702042}}].

\bibitem{Aubin:2009jh}
C.~Aubin, J.~Laiho, and R.~S. Van~de Water, {\it {The Neutral kaon mixing
  parameter B(K) from unquenched mixed-action lattice QCD}},  {\em Phys.Rev.}
  {\bf D81} (2010) 014507, [\href{http://xxx.lanl.gov/abs/0905.3947}{{\tt
  arXiv:0905.3947}}].

\bibitem{Bae:2010ki}
T.~Bae, Y.-C. Jang, C.~Jung, H.-J. Kim, J.~Kim, {\em et.~al.}, {\it {$B_K$
  using HYP-smeared staggered fermions in $N_f=2+1$ unquenched QCD}},  {\em
  Phys.Rev.} {\bf D82} (2010) 114509,
  [\href{http://xxx.lanl.gov/abs/1008.5179}{{\tt arXiv:1008.5179}}].

\bibitem{Constantinou:2010qv}
{\bf ETM} Collaboration, M.~Constantinou {\em et.~al.}, {\it {$B_K$-parameter
  from $N_f$ = 2 twisted mass lattice QCD}},  {\em Phys.Rev.} {\bf D83} (2011)
  014505, [\href{http://xxx.lanl.gov/abs/1009.5606}{{\tt arXiv:1009.5606}}].

\bibitem{Aoki:2010pe}
Y.~Aoki, R.~Arthur, T.~Blum, P.~Boyle, D.~Brommel, {\em et.~al.}, {\it
  {Continuum Limit of $B_K$ from 2+1 Flavor Domain Wall QCD}},  {\em Phys.Rev.}
  {\bf D84} (2011) 014503, [\href{http://xxx.lanl.gov/abs/1012.4178}{{\tt
  arXiv:1012.4178}}].

\bibitem{Brod:2010mj}
J.~Brod and M.~Gorbahn, {\it {$\varepsilon_K$ at Next-to-Next-to-Leading Order:
  The Charm-Top-Quark Contribution}},  {\em Phys. Rev.} {\bf D82} (2010)
  094026, [\href{http://xxx.lanl.gov/abs/1007.0684}{{\tt arXiv:1007.0684}}].

\bibitem{Brod:2011ty}
J.~Brod and M.~Gorbahn, {\it {The NNLO Charm-Quark Contribution to $\epsilon_K$
  and $\Delta M_K$}},  \href{http://xxx.lanl.gov/abs/1108.2036}{{\tt
  arXiv:1108.2036}}.

\bibitem{Barbieri:2011ci}
R.~Barbieri, G.~Isidori, J.~Jones-Perez, P.~Lodone, and D.~M. Straub, {\it
  {U(2) and Minimal Flavour Violation in Supersymmetry}},  {\em Eur.Phys.J.}
  {\bf C71} (2011) 1725, [\href{http://xxx.lanl.gov/abs/1105.2296}{{\tt
  arXiv:1105.2296}}].

\bibitem{Aushev:2010bq}
T.~Aushev, W.~Bartel, A.~Bondar, J.~Brodzicka, T.~Browder, {\em et.~al.}, {\it
  {Physics at Super B Factory}},  \href{http://xxx.lanl.gov/abs/1002.5012}{{\tt
  arXiv:1002.5012}}.

\bibitem{Bona:2007qt}
{\bf SuperB} Collaboration, M.~Bona {\em et.~al.}, {\it {SuperB: A
  High-Luminosity Asymmetric $e^+ e^-$ Super Flavor Factory. Conceptual Design
  Report}},  \href{http://xxx.lanl.gov/abs/0709.0451}{{\tt arXiv:0709.0451}}.

\bibitem{O'Leary:2010af}
{\bf SuperB} Collaboration, B.~O'Leary {\em et.~al.}, {\it {SuperB Progress
  Reports -- Physics}},  \href{http://xxx.lanl.gov/abs/1008.1541}{{\tt
  arXiv:1008.1541}}.

\bibitem{Meadows:2011bk}
B.~Meadows, M.~Blanke, A.~Stocchi, A.~Drutskoy, A.~Cervelli, {\em et.~al.},
  {\it {The impact of SuperB on flavour physics}},
  \href{http://xxx.lanl.gov/abs/1109.5028}{{\tt arXiv:1109.5028}}.

\bibitem{Altmannshofer:2011iv}
W.~Altmannshofer and M.~Carena, {\it {B Meson Mixing in Effective Theories of
  Supersymmetric Higgs Bosons}},  \href{http://xxx.lanl.gov/abs/1110.0843}{{\tt
  arXiv:1110.0843}}.

\bibitem{Faller:2008gt}
S.~Faller, R.~Fleischer, and T.~Mannel, {\it {Precision Physics with $B^0_s \to
  J/\psi \phi$ at the LHC: The Quest for New Physics}},  {\em Phys.Rev.} {\bf
  D79} (2009) 014005, [\href{http://xxx.lanl.gov/abs/0810.4248}{{\tt
  arXiv:0810.4248}}].

\bibitem{Herrlich:1996vf}
S.~Herrlich and U.~Nierste, {\it {The Complete $|\Delta S|=2$ Hamiltonian in
  the Next-To-Leading Order}},  {\em Nucl. Phys.} {\bf B476} (1996) 27--88,
  [\href{http://xxx.lanl.gov/abs/hep-ph/9604330}{{\tt hep-ph/9604330}}].

\bibitem{Buras:1990fn}
A.~J. Buras, M.~Jamin, and P.~H. Weisz, {\it {Leading and next-to-leading QCD
  corrections to $\varepsilon$ parameter and $B^0 - \bar B^0$ mixing in the
  presence of a heavy top quark}},  {\em Nucl. Phys.} {\bf B347} (1990)
  491--536.

\bibitem{Urban:1997gw}
J.~Urban, F.~Krauss, U.~Jentschura, and G.~Soff, {\it {Next-to-leading order
  QCD corrections for the $B^0 -\bar B^0$ mixing with an extended Higgs
  sector}},  {\em Nucl. Phys.} {\bf B523} (1998) 40--58,
  [\href{http://xxx.lanl.gov/abs/hep-ph/9710245}{{\tt hep-ph/9710245}}].

\bibitem{Allison:2008xk}
{\bf HPQCD} Collaboration, I.~Allison {\em et.~al.}, {\it {High-Precision
  Charm-Quark Mass from Current-Current Correlators in Lattice and Continuum
  QCD}},  {\em Phys. Rev.} {\bf D78} (2008) 054513,
  [\href{http://xxx.lanl.gov/abs/0805.2999}{{\tt arXiv:0805.2999}}].

\bibitem{Csaki:2008zd}
C.~Csaki, A.~Falkowski, and A.~Weiler, {\it {The Flavor of the Composite
  Pseudo-Goldstone Higgs}},  {\em JHEP} {\bf 0809} (2008) 008,
  [\href{http://xxx.lanl.gov/abs/0804.1954}{{\tt arXiv:0804.1954}}].

\bibitem{Barbieri:1987fn}
R.~Barbieri and G.~F. Giudice, {\it {Upper Bounds on Supersymmetric Particle
  Masses}},  {\em Nucl. Phys.} {\bf B306} (1988) 63.

\bibitem{Athron:2007ry}
P.~Athron and D.~J. Miller, {\it {A New Measure of Fine Tuning}},  {\em Phys.
  Rev.} {\bf D76} (2007) 075010, [\href{http://xxx.lanl.gov/abs/0705.2241}{{\tt
  arXiv:0705.2241}}].

\bibitem{Guadagnoli:2010sd}
D.~Guadagnoli and R.~N. Mohapatra, {\it {TeV Scale Left Right Symmetry and
  Flavor Changing Neutral Higgs Effects}},  {\em Phys.Lett.} {\bf B694} (2011)
  386--392, [\href{http://xxx.lanl.gov/abs/1008.1074}{{\tt arXiv:1008.1074}}].

\bibitem{Buras:2009ka}
A.~J. Buras, B.~Duling, and S.~Gori, {\it {The Impact of Kaluza-Klein Fermions
  on Standard Model Fermion Couplings in a RS Model with Custodial
  Protection}},  {\em JHEP} {\bf 0909} (2009) 076,
  [\href{http://xxx.lanl.gov/abs/0905.2318}{{\tt arXiv:0905.2318}}].

\bibitem{Duling:2009pj}
B.~Duling, {\it {A Comparative Study of Contributions to $\epsilon_K$ in the RS
  Model}},  {\em JHEP} {\bf 1005} (2010) 109,
  [\href{http://xxx.lanl.gov/abs/0912.4208}{{\tt arXiv:0912.4208}}].

\bibitem{Agashe:2004cp}
K.~Agashe, G.~Perez, and A.~Soni, {\it {Flavor structure of warped extra
  dimension models}},  {\em Phys.Rev.} {\bf D71} (2005) 016002,
  [\href{http://xxx.lanl.gov/abs/hep-ph/0408134}{{\tt hep-ph/0408134}}].

\bibitem{Isidori:2010kg}
G.~Isidori, Y.~Nir, and G.~Perez, {\it {Flavor Physics Constraints for Physics
  Beyond the Standard Model}},  {\em Ann.Rev.Nucl.Part.Sci.} {\bf 60} (2010)
  355, [\href{http://xxx.lanl.gov/abs/1002.0900}{{\tt arXiv:1002.0900}}].

\bibitem{Blanke:2011xx}
M.~Blanke, B.~Shakya, P.~Tanedo, and Y.~Tsai, {\it {The birds and the $B$s in
  RS: the $b \to s \gamma$ penguin in a warped extra dimension}},
  \href{http://xxx.lanl.gov/abs/1203.6650}{{\tt arXiv:1203.6650}}.

\bibitem{Hubisz:2005bd}
J.~Hubisz, S.~J. Lee, and G.~Paz, {\it {The Flavor of a little Higgs with
  T-parity}},  {\em JHEP} {\bf 0606} (2006) 041,
  [\href{http://xxx.lanl.gov/abs/hep-ph/0512169}{{\tt hep-ph/0512169}}].

\bibitem{Blanke:2008ac}
M.~Blanke, A.~J. Buras, S.~Recksiegel, and C.~Tarantino, {\it {The Littlest
  Higgs Model with T-Parity Facing CP-Violation in $B_s - \bar B_s$ Mixing}},
  \href{http://xxx.lanl.gov/abs/0805.4393}{{\tt arXiv:0805.4393}}.

\bibitem{Blanke:2009am}
M.~Blanke, A.~J. Buras, B.~Duling, S.~Recksiegel, and C.~Tarantino, {\it {FCNC
  Processes in the Littlest Higgs Model with T-Parity: an Update}},  {\em Acta
  Phys.Polon.} {\bf B41} (2010) 657--683,
  [\href{http://xxx.lanl.gov/abs/0906.5454}{{\tt arXiv:0906.5454}}].

\bibitem{Deshpande:1990ip}
N.~G. Deshpande, J.~F. Gunion, B.~Kayser, and F.~I. Olness, {\it {Left-right
  symmetric electroweak models with triplet Higgs}},  {\em Phys. Rev.} {\bf
  D44} (1991) 837--858.

\bibitem{Kiers:2005gh}
K.~Kiers, M.~Assis, and A.~A. Petrov, {\it {Higgs sector of the left-right
  model with explicit CP violation}},  {\em Phys. Rev.} {\bf D71} (2005)
  115015, [\href{http://xxx.lanl.gov/abs/hep-ph/0503115}{{\tt
  hep-ph/0503115}}].

\end{thebibliography}\endgroup

\end{document}